\documentclass[aps,prc,nofootinbib,preprintnumbers,12pt,tightenlines,superscriptaddress]{revtex4}

\usepackage{graphicx}
\usepackage{rotating}
\usepackage{latexsym}
\usepackage{axodraw}
\usepackage{slashed}

\newcommand{\be}{\begin{equation}}
\newcommand{\ee}{\end{equation}}
\newcommand{\ba}{\begin{eqnarray}}
\newcommand{\ea}{\end{eqnarray}}
\newcommand{\fr}[2]{{\frac{#1}{#2}\,}}
\newcommand{\tr}{{\rm Tr\,}}
\newcommand{\rmi}[1]{{\mbox{\scriptsize #1}}}
\newcommand{\e}{\epsilon}
\renewcommand{\(}{\left(}
\renewcommand{\)}{\right)}
\newcommand{\msbar}{{\overline{\mbox{\rm MS}}}}
\newcommand{\nn}{\nonumber \\}
\newcommand{\pmq}{a}

\newcommand{\lpic}[1]{\;\parbox[c]{30pt}{\begin{picture}(15,0)(0,0)
\SetWidth{1.0}\SetScale{1.0} #1 \end{picture}}\;}

\def\Lwidth{1}

\def\Lgl(#1,#2)(#3,#4){\Photon(#1,#2)(#3,#4){\Lwidth}
{#1 #3 sub #1 #3 sub mul #2 #4 sub #2 #4 sub mul add sqrt Ldensity mul}}

\def\Lsc(#1,#2)(#3,#4){\ArrowLine(#1,#2)(#3,#4)}

\begin{document}

\hspace{0.2cm}

\vspace{2.0cm}

\title{Cold Quark Matter}

\preprint{\hskip4.0in \vbox{BI-TP 2009/30 \\ CERN-PH-TH-2009-229 \\ INT-PUB-09-060 \\ TUW-09-19}}

\author{Aleksi Kurkela}
\affiliation{Institute for Theoretical Physics, ETH Zurich, CH-8093 Zurich, Switzerland}
\author{Paul Romatschke}
\affiliation{Institute for Nuclear Theory, University of Washington, Box 351550, Seattle, WA, 98195}
\author{Aleksi Vuorinen}
\affiliation{Faculty of Physics, University of Bielefeld, D-33501 Bielefeld, Germany}
\affiliation{CERN, Physics Department, TH Unit, CH-1211 Geneva 23, Switzerland}
\affiliation{ITP, TU Vienna, Wiedner Hauptstr. 8-10, A-1040 Vienna, Austria}

\begin{abstract}
We perform an ${\mathcal O}(\alpha_s^2)$ perturbative calculation of the equation of state of cold but dense QCD matter with two massless and one massive quark flavor, finding that perturbation theory converges reasonably well for quark chemical potentials above 1 GeV. Using a running coupling constant and strange quark mass, and allowing for further non-perturbative effects, our results point to a narrow range where absolutely stable strange quark matter may exist. Absent stable strange quark matter, our findings suggest that quark matter in compact star cores becomes confined to hadrons only slightly above the density of atomic nuclei. Finally, we show that equations of state including quark matter lead to hybrid star masses up to $M\sim2M_{\odot}$, in agreement with current observations. For strange stars, we find maximal masses of $M\sim2.75M_{\odot}$ and conclude that confirmed observations of compact stars with $M>2M_{\odot}$ would strongly favor the existence of stable strange quark matter.
\end{abstract}

\maketitle

\newpage

\tableofcontents

\section{Introduction}

The properties of cold nuclear matter at densities above that of atomic nuclei, in particular its equation of state (EoS) and the location of the phase transition to deconfined quark matter, remain poorly known to this day. The difficulty in performing first principles calculations in such systems can be traced back to the complicated non-linear and non-perturbative nature of Quantum Chromodynamics (QCD). These properties have precluded an analytic solution describing confinement, while non-perturbative numerical techniques, such as lattice QCD, are inapplicable at large baryon densities and small temperatures due to the so-called sign problem. This should be contrasted with the situation at small baryon density and large temperatures, where close to the deconfinement transition region lattice QCD has provided controlled results for the EoS as well as the nature of the transition \cite{Aoki:2006we,Bazavov:2009zn}, while at temperatures much above the transition, the system is well described by analytic results from resummed perturbation theory \cite{Kajantie:2002wa,Blaizot:2003iq,Vuorinen:2003fs,Andersen:2004fp}.

Experimentally, the high temperature / low baryon density regime of QCD can be studied in relativistic heavy-ion collisions at the Relativistic Heavy Ion Collider (RHIC) \cite{Adcox:2004mh,Back:2004je,Arsene:2004fa,Adams:2005dq} and in the future at the Large Hadron Collider (LHC) \cite{Carminati:2004fp}. Collisions at lower energy, \textit{e.g.~}at the Alternating Gradient Synchrotron (AGS) and Super Proton Synchrotron (SPS) \cite{Barrette:1994xr,Abreu:2000ni}, as well as those planned at the Facility for Antiproton and Ion Research (FAIR) and RHIC \cite{Stephans:2006tg,Senger:2008zz}, study QCD matter at somewhat higher baryon density, and may give some insight into the EoS of cold nuclear matter. However, at truly low temperatures and supra-nuclear densities, QCD matter exists only in somewhat inconveniently located 'laboratories': Compact stars.

In the cores of compact stars, nuclear matter is expected to reach densities several times that of atomic nuclei $n_{\rm sat}\sim 0.16\ {\rm fm}^{-3}$, so that astrophysical observations may be able to provide critical information about the EoS of strongly interacting matter in a regime inaccessible to terrestrial experiments. Theoretically, the bulk properties of nuclear matter at or close to $n_{\rm sat}$ have been studied using microscopic calculations \cite{Heiselberg:2000dn} as well as phenomenological mean-field theory \cite{Bender:2003jk}. While giving matching results for symmetric nuclear matter (equal number of protons and neutrons) \cite{Weber:2004kj}, in neutron rich matter, relevant for compact star cores, the theoretical predictions differ amongst themselves by more than $100$ per cent for basic quantities such as the pressure at $n\sim n_{\rm sat}$ \cite{Lattimer:2000nx}. Extrapolating to higher densities further increases these differences, on top of which new phenomena such as pion and kaon condensation \cite{pionco,Kaplan:1986yq} and the presence of hyperons (such as $\Lambda,\Sigma^{\pm}$) only add to enlarge the uncertainties in the EoS.

At some critical density, nuclear matter is expected to undergo a phase transition to deconfined quark matter, which is theoretically well understood only at asymptotically high densities, where the QCD coupling $\alpha_s$ is small \cite{Alford:2007xm}. There, the stable ground state of quark matter is that of a color superconductor, but it is not known whether such a state persists to densities closer to the deconfinement transition, or whether normal unpaired quark matter, or some novel phase, becomes favored. Even the possibility of the normal quark phase being the fundamental ground state of nuclear matter (with atomic nuclei being only metastable) has been suggested in the so-called strange quark matter hypothesis \cite{Bodmer:1971we,Witten:1984rs,Farhi:1984qu}. This opened up the possibility for entire stars being made up of self-bound quark matter ('strange stars') \cite{Alcock:1986hz}.

Interestingly, despite all the advances in our understanding of QCD, most of the analysis of cold but dense quark matter still continues to be performed using the MIT bag model dating back 35 years \cite{Chodos:1974pn}. In this model, the interactions between quarks are absorbed into a phenomenological 'bag constant', which is not calculable within the model but effectively generated by the QCD interactions (see Ref.~\cite{Fraga:2001id} for a discussion of this issue), and is simply added to the pressure of a non-interacting system. We believe that a refinement of this model, using a perturbative EoS for quark matter evaluated with a running $\alpha_s$ and strange quark mass, should be quite superior to the plain bag model, and --- absent advances in truly non-perturbative methods --- should replace the latter whenever aiming for at least semi-quantitative results.

To this end, in this paper we consider the perturbative evaluation of the QCD pressure at zero temperature, where the state-of-the-art result is still the pioneering order $\alpha_s^2$ calculation of Freedman and McLerran \cite{Freedman:1976xs,Freedman:1976ub} and Baluni \cite{Baluni:1977ms}. These authors, however, only included effects of the strange quark mass up to order $\alpha_s$, dropping the mass entirely at order $\alpha_s^2$. As present day knowledge suggests a strange quark mass of about 100 MeV \cite{Amsler:2008zzb}, with atomic nuclei corresponding to a quark chemical potential of roughly 300 MeV, one can expect non-negligible strange quark mass effects in the EoS (\textit{cf.~}Ref.~\cite{Fraga:2004gz}). In view of the situation, we believe that a perturbative calculation of the cold QCD EoS to order $\alpha_s^2$ --- including the complete strange quark mass effects --- is long overdue. This provides the motivation for us to take on this challenge in the present work.

Our paper is organized as follows. In Section II, we introduce our notation, explain how renormalization is performed, and outline the general structure of the computation. In Section III, we then go through all the different parts of the calculation, presenting the results for the individual terms and in the end assembling the entire grand canonical potential of the system. Section IV is devoted to a detailed analysis of our result, covering aspects such as the choice of the renormalization scale and the dependence of the result on the strange quark mass. Having gained control of the perturbative EoS, in Section V we consider various applications of it, studying the scenarios of stable strange quark matter and a phase transition between ordinary quark matter and the hadronic phase. In Section VI, we finally consider the implications of our work on astrophysical systems, while in Section VII we draw our conclusions. Several technical details, as well as most of the partial results of our computation, are left to Appendices A--E.

\section{Setup}

The equation of state of a thermodynamic system is dictated by the functional relation between some fundamental quantity, such as the pressure or energy density, and various (usually intensive) parameters, such as the temperature and different chemical potentials. In the grand canonical ensemble, it can be solved from the grand potential, or Landau free energy,
\ba
\Omega&=&E-\mu N\;=\;-T\ln\,Z\;=\;-PV,
\ea
where $E$ is the (microcanonical) energy, and $Z$ the partition function of the system. In this paper, we set out to perform a perturbative evaluation of the grand potential of QCD to order $g^4=(4 \pi \alpha_s)^2$ in the strong coupling constant, keeping the temperature at zero but assuming the quark chemical potentials to be high enough so that, due to asymptotic freedom, the expansion converges to a satisfactory degree. Various thermodynamic quantities can then be obtained from the grand potential, which itself is determined by computing (minus) the sum of all the one-particle-irreducible (1PI) vacuum graphs of the theory.

Due to the infrared (IR) sensitivity of $\Omega$, at order $g^4$ it is no longer sufficient to only consider the strict loop expansion of the grand potential. In addition, the so-called plasmon sum must be performed to its full leading order, which implies resumming all gluonic ring diagrams containing an arbitrary number of insertions of the one-loop gluon polarization tensor. Upon performing renormalization, the expansion of the grand potential can be written as the IR and ultraviolet (UV) finite sum of (at least semi-) analytically obtainable one-, two- and three-loop two-gluon-irreducible (2GI) diagrams, as well as a numerically computable plasmon integral.

The aim of this section is to introduce our notation, explain how the theory is renormalized, and demonstrate how the computation can be divided into several distinct pieces, which will be addressed one by one in Section III.

\subsection{Notation and Conventions}

Let us consider a system of $N_l$ flavors of massless quarks and gluons, to which we add one massive flavor with a renormalized mass $m$, having in mind $N_l=2$ and $m=m_s$, the mass of the strange quark. We use the Euclidean metric, $g_{\mu\nu}=\delta_{\mu\nu}$, and consequently define QCD (in the presence of finite quark number chemical potentials $\mu_i$) through the action $S_{\rmi{QCD}}=\int {\rm d} ^dx \,\mathcal{L}_{\rm{QCD}}$, with
\be
\mathcal{L}_\rmi{QCD}=\frac{1}{4}F^a_{\mu\nu}F^a_{\mu\nu}+\bar{\psi}_i(\gamma_\mu D_\mu+m_\rmi{B}^i
-\mu_i \gamma_0)\psi_i. \label{LQCD}
\ee
As usual, the subscript $\rm{B}$ denotes bare quantities, and we have defined
\ba
F_{\mu\nu}^a &\equiv& \partial_\mu A_\nu^a-\partial_\nu A_\mu^a+g_\rmi{B}f^{abc}A^b_\mu A^c_\nu,\nonumber\\
D_\mu&=&\partial_\mu-ig_\rmi{B} A_\mu,\nonumber\\
A_\mu&=&A_\mu^a T^a.\nonumber
\ea

In accordance with the metric used, our gamma matrices are all Hermitian, $\gamma_\mu=\gamma_\mu^\dagger$, and satisfy $\{\gamma_\mu,\gamma_\nu\}=2\delta_{\mu\nu}$. The flavor index $i$ runs from 1 to $N_f\equiv N_l+1$, with $m_i=0$ for $1\leq i\leq N_l$ and $m_{N_f}=m$. Throughout the paper, dimensional regularization in $d=4-2\epsilon$ dimensions will be used to regulate divergent integrals, and hence $\mu=0,\ldots,d-1$. As working at zero temperature implies dealing with $d=4-2\e$ dimensional integrals rather than discrete sum-integrals, the integration measure becomes
$$
\int_{-\infty}^{\infty}\!\fr{{\rm d}p_0}{2\pi} \int\! \fr{{\rm d}^{d-1} p}{(2\pi)^{d-1}}
= \Lambda^{-2\e}
\Bigg[\(\fr{{\rm e}^{\gamma_E} \bar\Lambda^2}{4\pi}\)^{\e} \int_{
-\infty}^{\infty}\!\fr{{\rm d}p_0}{2\pi} \int\! \fr{{\rm d}^{d-1} p}{(2\pi)^{d-1}}\Bigg],
$$
where $\gamma_E=0.5772\ldots$ is the Euler-Mascheroni constant, and the renormalization scales $\Lambda$ and $\bar\Lambda$ of the MS and $\msbar$ schemes have been introduced for later convenience. The chemical potential is accommodated in the Feynman rules by shifting the zeroth components of the fermionic momenta by $p_0\rightarrow p_0+i\mu_i$, where $\mu_i$ is the chemical potential of the quark flavor in question. We choose to denote the chemical potential corresponding to the massive quark by  $\mu_{N_f}\equiv \mu$.

Our notation for the Feynman diagrams is as follows: Solid lines with an arrow correspond to quarks and wiggly lines to gluons. We work in the Feynman gauge throughout the calculation, and hence our propagators have the form
\ba
\lpic{\Lsc(0,0)(30,0)}&=&\fr{1}{i\slashed{P}+m}\;\;\,=\;\;\,\fr{-i\slashed{P}+m}{P^2+m^2},\nonumber\\
\lpic{\Lgl(0,0)(30,0)}&=&\fr{\delta_{\mu\nu}}{P^2},\nonumber
\ea
consistent with the choice to employ the Euclidean metric. Here and in the following, capital letters are used denote Euclidean four-momenta such as $P=(p_0,{\bf p})$, with $P^2=p_0^2+{\bf p}^2$. We will also frequently use the abbreviations
\ba
E(p)&=&\sqrt{m^2+p^2},\qquad
u=\sqrt{\mu^2-m^2}, \qquad
z \equiv \hat{u}-\hat{m}^2\,\ln\bigg[\fr{1+\hat{u}}{\hat{m}}\bigg], \label{freqdef}
\ea
where we have introduced the dimensionless parameters $\hat{m}=m/\mu, \hat{u}=u/\mu$. In addition, the group theory factors that appear in the calculation read
\ba
d_A&\equiv&\delta^{aa}\;=\; N_c^2-1,\nonumber\\
C_A \delta^{cd} & \equiv & f^{abc}f^{abd} \;=\; N_c \delta^{cd},\nonumber \\
C_F \delta_{ij} & \equiv & (T^a T^a)_{ij} \;=\; \fr{N_c^2-1}{2N_c}\delta_{ij},\nonumber
\ea
where the number of colors $N_c$ will later be set to three, and we have normalized the fundamental representation generators according to $\tr T^a T^b =\fr{1}{2}\delta^{ab}$.

Finally, all expressions in this work are given using natural units, in which $\hbar=c=k_B=1$, except when explicitly stated otherwise.

\subsection{Renormalization}
\label{sec:renorm}

We perform all our calculations in terms of the bare parameters appearing in the Lagrangian of Eq.~(\ref{LQCD}), and only in the end express them in terms of the physical, renormalized ones. By doing this, one avoids dealing with explicit counter terms, but on the other hand cannot set $\e$ to zero until the end of the computation. The relations between the bare and renormalized mass and coupling constant can be obtained from the literature (see \textit{e.g.~}Ref.~\cite{Fleischer:1998dw}),
\ba
m_\rmi{B}&=&Z_m m \; \equiv \; \(1+\delta_1 \fr{g^2}{(4\pi)^2} +\delta_2 \fr{g^4}{(4\pi)^4}
+\mathcal{O}(g^6) \)m,\\
(g_\rmi{B})^2&=&Z_g g^2 \;\, \equiv \; \(1+\delta_3 \fr{g^2}{(4\pi)^2}+\mathcal{O}(g^4)\)g^2,
\ea
where we have defined
\ba
&\delta_1=-\gamma_0 \e^{-1}\,\qquad
\delta_2= C_F \left( \frac{11}{2}C_A+\frac{9}{2}C_F-N_f \right)\e^{-2}-\frac{\gamma_1}{2} \e^{-1},\qquad
\delta_3 =-\beta_0 \e^{-1},&
\nonumber
\ea
with $\beta_0$ and $\gamma_i$ being constants that will be given below. As noted above, all of our calculations are performed in the Feynman gauge, which is defined by the unrenormalized gauge parameter $\xi_B$ taking the value $\xi_B=1$. The gauge parameter renormalization constant then becomes equal to that of the gauge field wave function, but neither of them enters the calculation. The same applies to the fermion wave function renormalization constant, and hence we refrain from quoting the corresponding results here.

Closely related to the above quantities are the renormalization group equations for the finite renormalized parameters $g$ and $m$, which --- given to higher order than would be required by our computation --- read \cite{Amsler:2008zzb}
\ba
\fr{\partial g^2}{\partial\, \ln\,{\bar\Lambda}^2}&=&-\beta_0\fr{g^4}{(4\pi)^2}-2 \beta_1 \fr{g^6}{(4\pi)^4}-4 \beta_2 \fr{g^8}{(4\pi)^6},\\
\fr{\partial m}{\partial\, \ln\,{\bar\Lambda}^2}&=&-m \left[\gamma_0
\fr{g^2}{(4\pi)^2}+\gamma_1 \fr{g^4}{(4\pi)^4}+\gamma_2 \fr{g^6}{(4\pi)^6}
\right].
\ea
All constants appearing here are available from Refs.~\cite{vanRitbergen:1997va,Vermaseren:1997fq,Chetyrkin:1997dh,Czakon:2004bu} and have the forms
\ba
&\beta_0 =\fr{11C_A-2N_f}{3},\qquad
\beta_1 =\frac{17}{3}C_A^2- C_F N_f-\frac{5}{3} C_A N_f,&\nn
&\beta_2 = \fr{2857}{216}C_A^3+\fr{1}{4}C_F^2N_f-\fr{205}{72}C_A C_F N_f-\fr{1415}{216}C_A^2 N_f+\fr{11}{36}C_F N_f^2+\fr{79}{216}C_AN_f^2,&\nn
&\gamma_0 =3C_F,\qquad
\gamma_1 =C_F\left(\frac{97}{6}C_A+\frac{3}{2}C_F-\frac{5}{3}N_f\right),&\nonumber\\
&\gamma_2 =C_F\left(\frac{129}{2}C_F^2-\frac{129}{4}C_F C_A+\frac{11413}{108}C_A^2+C_F N_F(-23+24 \zeta(3))
\right.&\nonumber\\
&\left.\quad+C_A N_F \(-\frac{278}{27}-24 \zeta(3)\)-\frac{35}{27} N_f^2\right).&\label{betafunc}
\ea
The renormalization group equations may furthermore be integrated to give the running coupling constant $\alpha_s(\bar\Lambda)=g^2(\bar\Lambda)/(4\pi)$ and strange quark mass $m(\bar\Lambda)$ as functions of the renormalization scale. Following the treatment of Ref.~\cite{Vermaseren:1997fq} by expanding the results in powers of $\alpha_s$, we obtain
\ba
\alpha_s(\bar\Lambda)&=&\frac{4 \pi}{\beta_0 L}\left(1-\frac{2 \beta_1}{\beta_0^2}
\frac{\ln\,L}{L}\right),\qquad
L=\ln\,\left(\bar\Lambda^2/\Lambda_{\tiny \msbar}^2\right);\label{runningalpha}\\
m(\bar\Lambda)&=&m(2\, {\rm GeV})\(\fr{\alpha_s(\bar\Lambda)}{\alpha_s(2\,{\rm GeV})}\)^{\!\gamma_0/\beta_0} \fr{1+A_1\fr{\alpha_s(\bar\Lambda)}{\pi}+ \fr{A_1^2+A_2}{2}\(\fr{\alpha_s(\bar\Lambda)}{\pi}\)^2}{1+A_1\fr{\alpha_s(2\,{\rm GeV})}{\pi}+ \fr{A_1^2+A_2}{2}\(\fr{\alpha_s(2\,{\rm GeV})}{\pi}\)^2},
\label{runningms}
\ea
with
\ba
A_1 &\equiv& -\fr{\beta_1 \gamma_0}{2\beta_0^2}+\fr{\gamma_1}{4\beta_0},\nn
A_2 &\equiv&\fr{\gamma_0}{4\beta_0^2}\(\fr{\beta_1^2}{\beta_0}-\beta_2\)-\fr{\beta_1 \gamma_1}{8\beta_0^2}+\fr{\gamma_2}{16\beta_0}.
\ea
Here, $\Lambda_{\tiny \msbar}$ denotes the $\msbar$ renormalization point, and we have chosen a fiducial scale of $2$ GeV where the strange quark mass takes the value $m(2\, {\rm GeV})\simeq 100\pm30$ MeV \cite{Amsler:2008zzb}.

\begin{figure}[t]
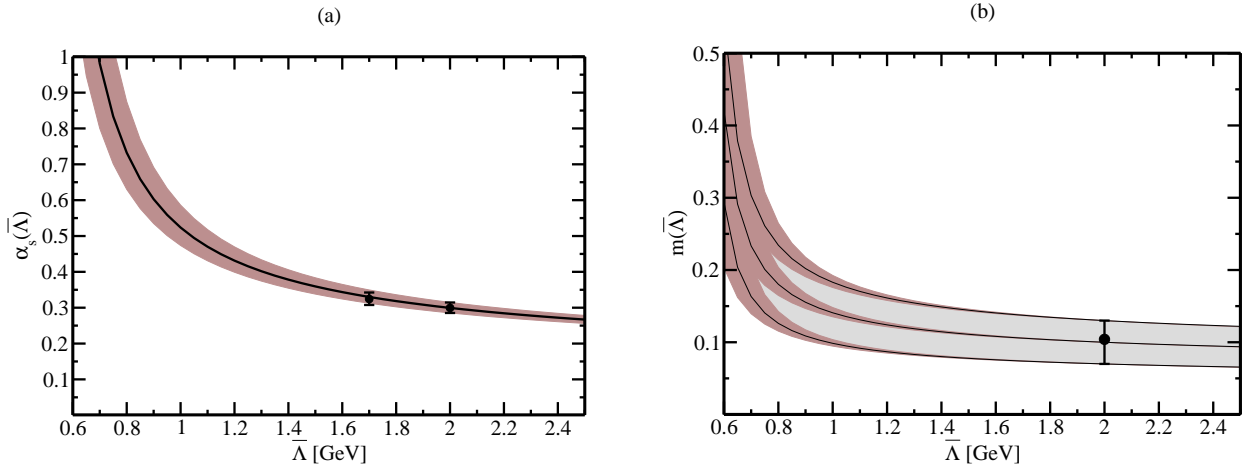

\center
\includegraphics[width=.47\linewidth]{fig1a.eps}
\hfill
\includegraphics[width=.47\linewidth]{fig1b.eps}
\caption{Renormalization scale dependence of $\alpha_s$ and $m$ from Eqs.~(\ref{runningalpha}), (\ref{runningms}), using values of $\bar\Lambda$ relevant for compact stars. The full lines and dark shaded regions correspond to the central value and uncertainty of $\Lambda_{\tiny \msbar}=0.378+0.034-0.032$ GeV,
respectively. The light shaded region in $m$ corresponds to the uncertainty in $m(2\ {\rm GeV})$. The circles with error bars correspond to reference points obtained from Ref.~\cite{Amsler:2008zzb}.}
\label{fig:as}
\end{figure}

One way to fix the renormalization point is by requiring that the coupling of Eq.~(\ref{runningalpha}) agrees with the current best experimental value for the parameter at the fiducial scale, $\alpha_s(2\ {\rm GeV})=0.2994+0.0152-0.0141$ \cite{Amsler:2008zzb}, leading to $\Lambda_{\tiny \msbar}=0.378+0.034-0.032$ GeV. For completeness, we note that an alternative to this would be to use a lattice QCD determination of the deconfinement temperature at small baryonic densities. In one lattice study, this was found to give $\Lambda_{\tiny \msbar}\simeq T_c/0.49$ \cite{Gupta:2000hr}, which --- depending on the result for $T_c$ one uses \cite{Cheng:2006qk,Aoki:2006br} --- results in $\Lambda_{\tiny \msbar}\sim0.375\pm0.016$ GeV. As this result happens to lie within the uncertainties of the first method, we thus adopt $\Lambda_{\tiny \msbar}=0.378+0.034-0.032$ GeV for the remainder of this work.

In Fig.~\ref{fig:as}, we display the functions $\alpha_s(\bar\Lambda)$ and $m(\bar\Lambda)$, as given by Eqs.~(\ref{runningalpha}) and (\ref{runningms}), to provide an indication of the value and uncertainties of the quantities. From here, it is evident that below $\bar\Lambda\sim 0.8$ GeV, the results become dominated by the various uncertainties discussed above, setting a lower limit for the values of $\bar\Lambda$ for which one can even optimistically expect quantitative results from any perturbative calculation (even in the hypothetical case that one was able to determine the quantity in question to infinitely many orders in $\alpha_s$).\footnote{In principle, one should adjust $N_f$ in Eq.~(\ref{betafunc}) when crossing the strange quark mass threshold (\textit{cf.~}Ref.~\cite{Rodrigo:1993hc} for a pedagogical review on heavy quark decoupling). However, from Eq.~(\ref{runningms}) one finds that the decoupling of the strange quark occurs at $\bar\Lambda\sim0.6$ GeV, where our results are not quantitatively correct in any case. Hence, we will simply ignore this effect in the following. In contrast, note that we will find that setting $m=0$ would change our results for the partition function at $\bar\Lambda\gg0.6$ GeV, so the strange quark mass effect cannot be altogether ignored.} Reducing these uncertainties would require a better determination of both $\alpha_s$ and $m$ at the fiducial scale, which is not the aim of our work.

\subsection{Organizing the Calculation}

\begin{figure}[t]

\centerline{\epsfxsize=13cm \epsfbox{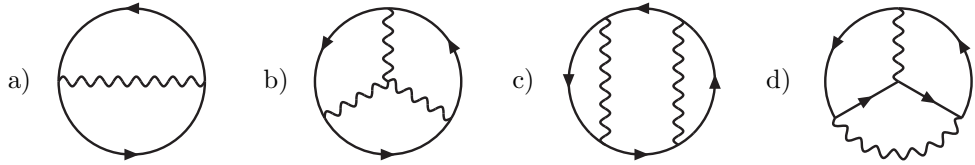}}

\caption[a]{The two- and three-loop 2GI diagrams contributing to the grand potential of QCD.
    \label{fig:graphs}
    }
\end{figure}

\begin{figure}[t]

\centerline{\epsfxsize=15cm \epsfbox{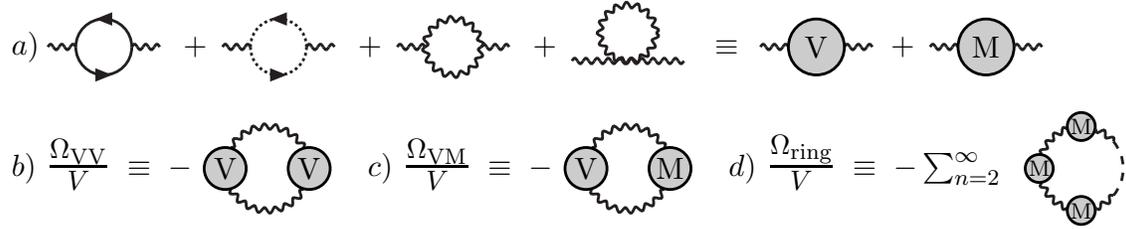}}

\caption[a]{The generic form of the ring or plasmon sum.
    \label{fig:ring}
    }
\end{figure}

To the perturbative order we are working in, the grand potential of QCD obtains contributions from the (renormalized) single quark loop $\Omega_\rmi{1L}$, the two- and three-loop two-gluon irreducible (2GI) vacuum diagrams of the theory $\Omega_\rmi{2GI}$, as well as the plasmon ring sum $\Omega_\rmi{plas}$,
\ba
\Omega&=&\Omega_\rmi{1L}+\Omega_\rmi{2GI}+\Omega_\rmi{plas}. \label{presexp1}
\ea
The 2GI graphs are displayed in Fig.~\ref{fig:graphs}, while the plasmon sum $\Omega_\rmi{plas}$ corresponds to the sum of all gluonic ring diagrams containing at least two insertions of the one-loop gluon polarization tensor, depicted in Fig.~\ref{fig:ring}a (\textit{cf.~}Appendix \ref{app:pimunu}). As explained in more detail in Section~\ref{sec:plasmon} as well as Ref.~\cite{Vuorinen:2003fs}, the benign IR behavior of the vacuum ($T=\mu=0$) part of the tensor allows the division of the ring sum into three parts. These include the three-loop vacuum-vacuum (VV) and vacuum-matter (VM) graphs of Fig.~\ref{fig:ring}b and c, as well as the 'matter' (vacuum subtracted) ring sum of Fig.~\ref{fig:ring}d,
\ba
\Omega_\rmi{plas}=\Omega_\rmi{VV}+\Omega_\rmi{VM}+\Omega_\rmi{ring}.
\ea
Here, the VV diagram only produces a chemical potential independent contribution to the grand potential, and is thus neglected. The matter part of the polarization tensor merely includes the fermion loop, which behaves like $1/P^2$ in the UV, so one finds that the ring sum $\Omega_\rmi{ring}$ is both IR and UV finite (\textit{cf.~}Section \ref{sec:plasmon}).

To further organize the calculation, one may separate the contributions of the massless quarks to both $\Omega_\rmi{1L}$ and $\Omega_\rmi{2GI}$ from the rest by writing
\ba
\Omega_\rmi{1L} &=& \Omega_\rmi{1L}^{m=0}+\Omega_\rmi{1L}^m,\;\;\;\;\Omega_\rmi{1L}^{m=0} \;\equiv\; \Omega_\rmi{1L}(\mu=m),\\
\Omega_\rmi{2GI} &=& \Omega_\rmi{2GI}^{m=0}+\Omega_\rmi{2GI}^m,\;\;\;\;\Omega_\rmi{2GI}^{m=0} \;\equiv\; \Omega_\rmi{2GI}(\mu=m),
\ea
and defining $\Omega_\rmi{1L}^m$ and $\Omega_\rmi{2GI}^m$ as the respective differences. With the vacuum-matter diagram, the issue is slightly more subtle, and we in fact have to write it in three parts,
\ba
\Omega_\rmi{VM}&=&\Omega_\rmi{VM}^{m=0} + \Omega_\rmi{VM}^m + \Omega_\rmi{VM}^x.
\ea
Here, $\Omega_\rmi{VM}^{m=0}$ is defined as the VM diagram composed of the massless part of the matter polarization tensor and the $m\rightarrow 0$ limit of the vacuum polarization tensor (\textit{i.e.~}containing contributions from $N_f=N_l+1$ massless quarks), while $\Omega_\rmi{VM}^m$ consists of a massive matter loop and the entire, $m$ dependent vacuum polarization tensor. Finally, $\Omega_\rmi{VM}^x$ is the contribution to the grand potential originating from a diagram composed of the massless part of the matter polarization tensor coupled to the difference of a massive and massless vacuum quark loop. It is easy to verify that the sum of these three functions indeed equals $\Omega_\rmi{VM}$.

Collecting the various pieces of the above expansion, the result for the grand potential may be written in the form
\ba
\Omega &=& \Omega^{m=0}+ \Omega^{m} + \Omega_\rmi{VM}^x + \Omega_\rmi{ring}, \label{lnZparts}
\ea
where we have denoted
\ba
\Omega^{m=0} &=& \Omega_\rmi{1L}^{m=0} + \Omega_\rmi{2GI}^{m=0} + \Omega_\rmi{VM}^{m=0},\\
\Omega^{m} &=& \Omega_\rmi{1L}^m + \Omega_\rmi{2GI}^m + \Omega_\rmi{VM}^m.\label{lnZm}
\ea
The reason for the above definitions now becomes clear: Having written $\Omega$ in terms of altogether eight different parts, we have managed to separate the contributions of the massless quark flavors (up to their contribution to the ring sum) into the function $\Omega^{m=0}$, the value of which may be directly taken from Ref.~\cite{Vuorinen:2003fs}. In the massive part, the contributions from the single quark loop $\Omega_\rmi{1L}^m$ (including renormalization corrections to it), the sum of the two- and three-loop 2GI diagrams $\Omega_\rmi{2GI}^m$, as well as the VM part $\Omega_\rmi{VM}^m$ have also been separated. In the following section, we will proceed to evaluate these various parts of the grand potential one by one.

\section{The Computation}
\label{sec:computation}

This section contains a discussion of the evaluation of each of the independent contributions to the grand potential, as defined in the 'master formula' of Eq.~(\ref{lnZparts}). For brevity of presentation, many of the calculational details as well as nearly all intermediate results are left to be listed in the Appendices. Readers not interested in the calculational details may wish to skip this section altogether and proceed straight to Section \ref{sec:QCDphaset}, where the final result is analyzed.

\subsection{$\Omega^{m=0}$: Massless Quarks}

The contribution of the massless quark flavors to the grand potential, including the massless single quark loops $\Omega_\rmi{1L}^{m=0}$, the sum of the two- and three-loop 2GI graphs containing massless quarks $\Omega_\rmi{2GI}^{m=0}$, and the massless VM graph $\Omega_\rmi{VM}^{m=0}$ can be extracted directly from Ref.~\cite{Vuorinen:2003fs}. One finds
\ba
\frac{\Omega_\rmi{1L}^{m=0}}{V} &=& -\fr{N_c}{12\pi^2}\sum_{i=1}^{N_l}\mu_i^4,\\
\frac{\Omega_\rmi{2GI}^{m=0}}{V}&=&\fr{d_A}{4\pi^2}\bigg(\!\fr{g(\bar{\Lambda})}{4\pi}\!\!\bigg)^{\!\!2}\sum_{i=1}^{N_l}\mu_i^4\bigg\{
1-\bigg[\fr{5C_A-2N_f}{3\e}+\fr{8}{3}\(C_A-N_f\)\ln\,\fr{\bar{\Lambda}}{2\mu_i}\nn
&-& \fr{2}{3}C_A+\fr{17}{2}C_F-\fr{10}{3}N_f\bigg]
\fr{g^2(\bar{\Lambda})}{(4\pi)^2}\bigg\},\\
\frac{\Omega_\rmi{VM}^{m=0}}{V}&=&\fr{d_A}{4\pi^2}\((5/2+\e)C_A-N_f\)\sum_{i=1}^{N_l}\mu_f^4\bigg\{\fr{2}{3\e}
+4\,\ln\,\fr{\bar{\Lambda}}{2\mu_i}+\fr{52}{9}\bigg\}\fr{g^4(\bar{\Lambda})}{(4\pi)^4},
\ea
where $V$ is the volume of the system, renormalization corrections have been taken into account, and the result is expressed in terms of the renormalized gauge coupling constant $g$.

Adding the three pieces together, one obtains for the function $\Omega^{m=0}$
\ba
\frac{\Omega^{m=0}}{V}&=&-\fr{1}{4\pi^2}\sum_{i=1}^{N_l}\mu_i^4\Bigg\{\fr{N_c}{3}-d_A\fr{g^2(\bar{\Lambda})}{(4\pi)^2}\nonumber\\
&+&d_A\bigg[\fr{4}{3}\(N_f-\fr{11C_A}{2}\)\ln\,\fr{\bar{\Lambda}}{2\mu_i}- \fr{142}{9}C_A + \fr{17}{2}C_F +\fr{22}{9}N_f\bigg]
\fr{g^4(\bar{\Lambda})}{(4\pi)^4}\Bigg\}.
\label{masslessp}
\ea

\subsection{$\Omega^{m}$: Massive Quark}

In this subsection, the contribution of the massive quark to the grand potential of QCD is evaluated by separately considering all the parts of the function $\Omega^{m}$, as listed in Eq.~(\ref{lnZm}). The final result for the function will be given at the end.

\vspace{0.4cm}
\paragraph{The Massive Quark Loop}
\label{sec:mqloop}

\mbox{} \vspace{0.1cm} \\
The grand potential of one free massive quark flavor at zero temperature but finite chemical potential can be written in the form
\ba
\frac{\Omega_\rmi{1L}^m}{V} &=& -2N_c\,\Lambda^{2\e}\int\! \fr{{\rm d}^{3-2\e} p}{(2\pi)^{3-2\e}}
(\mu-E(p))\,\theta(\mu-E(p)), \label{freethy}
\ea
where the regulator $\e$ has been kept non-zero because of the bare mass parameter appearing inside of $E(p)$. Writing the integral in the form
\ba
\frac{\Omega_\rmi{1L}^m}{V}&=&-2N_c\,\Lambda^{2\e}\fr{S_{3-2\e}}{(2\pi)^{3-2\e}}
\int_0^{\sqrt{\mu^2-m_\rmi{B}^2}}{\rm d}p\, p^{2-2\e}\(\mu-\sqrt{p^2+m_\rmi{B}^2}\),
\ea
with $S_n$ denoting the area of an $n$-sphere, one can perform a power series expansion in $g$ and $\e$. Keeping terms up to and including $O(g^4)$ and $O(\e^2)$, the integral is evaluated to be
\ba
&&\frac{\Omega_\rmi{1L}^m}{V}=-2 N_c\,\Lambda^{2\e}\Bigg\{\int\! \fr{{\rm d}^{3} p}{(2\pi)^{3}} (\mu-E(p))\,
\theta(\mu-E(p)) \label{freethy2}\\
&-&\left(\delta_1 m^2\fr{g^2(\bar{\Lambda})}{(4\pi)^2}+\left[\delta_2m^2+\fr{\delta_1^2m^2}{2}\(1+2m^2\fr{\partial}{\partial m^2}\)\right]\fr{g^4(\bar{\Lambda})}{(4\pi)^4}\right)\int\!
\fr{{\rm d}^{3-2\e} p}{(2\pi)^{3-2\e}}\fr{\theta(\mu-E(p))}{E(p)}\Bigg\},\nonumber
\ea
in which the mass and coupling constant are now the renormalized ones.

A straightforward evaluation of the remaining integrals finally gives
\ba
\fr{\Omega_\rmi{1L}^m}{V}&=&-\fr{N_c}{12\pi^2}\Bigg\{u\mu \(\mu^2-\fr{5}{2}m^2\)+\fr{3m^4}{2}\,\ln\bigg[\fr{\mu+u}{m}\bigg]\Bigg\} - d_A m^2\,\fr{6}{\e}I_1\,\fr{g^2(\bar{\Lambda})}{(4\pi)^2}\nn
&+& d_A m^2\Bigg\{C_A\Bigg[\fr{11}{\e^2}-\fr{97}{6\e}\Bigg]I_1
+C_F\Bigg[\bigg\{\fr{18}{\e^2}-\fr{3}{2\e}\bigg\}I_1+18 I_{1b}\Bigg]\nn
&-&N_f
\Bigg[\fr{2}{\e^2}-\fr{5}{3\e}\Bigg]I_1 \Bigg\}\,\fr{g^4(\bar{\Lambda})}{(4\pi)^4}, \label{freethy3}
\ea
where the $I_n(\hat{m})$'s denote integral functions defined in Appendix \ref{app:idefs}. Here and in the following, we have suppressed the arguments of these functions, except when they are necessary to avoid confusion.

\vspace{0.4cm}
\paragraph{The Massive Two- and Three-loop Skeleton Graphs}
\label{sec:2gi}

\mbox{} \vspace{0.1cm} \\
The four massive 2GI graphs of Fig.~\ref{fig:graphs} contribute to the grand potential as
\ba
-\frac{\Omega_\rmi{2GI}^m}{V}=D_a+D_b+D_c+D_d,
\label{2GIfirstdef}
\ea
where the functions $D_i$ appearing on the right hand side correspond to the diagrams a--d. Their evaluation proceeds in several steps. First, we write down expressions for the graphs in terms of the Feynman rules and contract all color and Lorentz indices, reducing the original functions to sums of scalar integrals. The finiteness of the chemical potential prevents the application of integration-by-parts identities at this point, so instead we perform the $p_0$ energy integrations using the residue theorem. Collecting the outcome of this step, one observes that each graph hereby reduces to a simple form that can be viewed as the result of having performed all possible 'cuts' on the fermionic lines of the diagram, their number ranging from zero to the number of loops.

Each time a line is cut, the corresponding propagator is placed on shell and the remaining part of the diagram integrated with respect to the three-momentum in question with the weight
\ba
-\int\fr{{\rm d}^3p}{(2\pi)^3}\fr{\theta(\mu-E({p}))}{2E({p})},
\ea
while $\mu$ is set to zero in all other propagators. This implies that the integrands of the phase space integrals become $d=4-2\e$ dimensional vacuum ($T=\mu=0$) two-, four- and six-point functions, which may be evaluated using integration-by-parts relations, conveniently implemented \textit{e.g.~}in the Mathematica package FIRE \cite{Smirnov:2008iw} and results available in the literature (see \textit{e.g.~}Refs.~\cite{Laporta:2004rb,Bonciani:2003cj,Bonciani:2003hc,Schroder:2005va}).

In the end, the cutting procedure separates each $n$-loop graph into $n+1$ parts
\ba
D_i&=&D_i^{0c} + D_i^{1c}  + D_i^{2c}  + D_i^{3c}+\cdots,
\ea
corresponding to the number of cuts performed. Thus the entire ${\mathcal O}(g^4)$ 2GI contribution to the grand potential separates into three pieces
\ba
\Omega_\rmi{2GI}^m&=&\Omega_\rmi{2GI}^{m,1c}+\Omega_\rmi{2GI}^{m,2c}+\Omega_\rmi{2GI}^{m,3c}.
\label{2GIsecdefinition}
\ea
Here, we have taken advantage of the fact that the zero-cut contribution is always independent of the chemical potential and may thus be dropped. We also note that the computationally most complicated triple-cut part of the grand potential is automatically UV finite and thus in no need of regularization, which simplifies the computations significantly.

As the detailed evaluation of the function $\Omega_\rmi{2GI}^m$ is quite lengthy and almost void of physical content, we leave it to Appendix \ref{app:m2GI}. There, we first go through the scalarization and cutting procedures for each graph, and in the end list the results for the functions $\Omega_\rmi{2GI}^{m,1c},\;\Omega_\rmi{2GI}^{m,2c}$ and $\Omega_\rmi{2GI}^{m,3c}$ in Eqs.~(\ref{lnZ2GI1c})--(\ref{lnZ2GI3c}).

\vspace{0.4cm}
\paragraph{The VM Graph}
\label{sec:vm}

\mbox{}\vspace{0.1cm}  \\
The evaluation of the massive vacuum-matter graph $\Omega_\rmi{VM}^m$ proceeds in a way highly analogous to that of the 2GI graphs, \textit{i.e.~}via first performing the Lorentz and color contractions and then taking one and two cuts in the remaining scalarized diagrams (the number of fermion propagators limits the maximum number of cuts to two here). The result can be written in the form
\ba
\Omega_\rmi{VM}^{m} &=& \Omega_\rmi{VM}^{m,1c}+\Omega_\rmi{VM}^{m,2c},
\label{VMmfirstdef}
\ea
where the functions $\Omega_\rmi{VM}^{m,1c},\Omega_\rmi{VM}^{m,2c}$ are given in Eqs.~(\ref{lnZVMm1c}) and (\ref{lnZVMm2c}).

\vspace{0.4cm}
\paragraph{Assembling the Result for $\Omega^{m}$}

\mbox{} \vspace{0.1cm} \\
Collecting the results of this section and Appendices \ref{app:m2GI} and \ref{app:VMgraph}, we see that the contribution of the massive quark flavor to the grand potential is of the form
\ba
-\fr{\Omega^m}{V}&=&{\mathcal M}_1+{\mathcal M}_2\,\fr{g^2(\bar{\Lambda})}{(4\pi)^2} + {\mathcal M}_3\,\fr{g^4(\bar{\Lambda})}{(4\pi)^4},
\label{massivepcollect}
\ea
where the two first terms of the result have been calculated before, \textit{e.g.~}in Ref.~\cite{Fraga:2004gz}. In our presentation, the function ${\mathcal M}_1$, corresponding to the grand potential of one free massive quark flavor, can be read off from Eq.~(\ref{freethy3}),
\ba
{\mathcal M}_1&=&
\fr{N_c \mu^4}{24\pi^2}\bigg\{2\hat{u}^3-3z \hat{m}^2\bigg\},\label{m1}
\ea
where we recall the abbreviations introduced in Eq.~(\ref{freqdef}). At the next order ${\mathcal O}(g^2)$, the function $\Omega^m$ on the other hand obtains contributions from the two-loop graph of Fig.~\ref{fig:graphs}a and the ${\mathcal O}(g^2)$ renormalization correction to the massive single quark loop. Adding these functions together using Eqs.~(\ref{freethy3}), (\ref{lnZ2GI1c}) and (\ref{lnZ2GI2c}), we find
\ba
{\mathcal M}_2&=& \fr{d_A \mu^4}{4\pi^2}\Bigg\{-6z \hat{m}^2 \ln\fr{\bar\Lambda}{m} +2\hat{u}^4 - 4z \hat{m}^2 -3z^2 \Bigg\}. \label{m2}
\ea

To obtain the ${\mathcal O}(g^4)$ contribution to $\Omega^m$, one needs to add together the corresponding parts of the functions $\Omega_\rmi{1L}^m$, $\Omega_\rmi{2GI}^m$ and $\Omega_\rmi{VM}^m$. Due to the length of the resulting expressions, we organize the calculation in terms of the number of cuts and write ${\mathcal M}_3$ in the form
\ba
{\mathcal M}_3&=&{\mathcal M}_3^{1c}+{\mathcal M}_3^{2c}+{\mathcal M}_3^{3c},
\ea
dealing with the three pieces separately.

There are three different types of single cut contributions to ${\mathcal M}_3$: The order $g^4$ renormalization correction to $\Omega_\rmi{1L}^m$ from the last two lines of Eq.~(\ref{freethy3}), the function $\Omega_\rmi{2GI}^{m,1c}$ from Eq.~(\ref{lnZ2GI1c}), as well as the single cut contribution to the massive VM graph $\Omega_\rmi{VM}^m$ from Eq.~(\ref{lnZVMm1c}). Summing up the various pieces, one finds
\ba
{\mathcal M}_3^{1c}&=&\fr{d_A \mu^4}{(2\pi)^2}\Bigg\{\Bigg(
-C_A\Bigg[\(22\,\ln\fr{\bar\Lambda}{m}+\fr{185}{3}\)\ln\fr{\bar\Lambda}{m}+\fr{1111}{24}
-\fr{4\pi^2}{3}+4\pi^2\ln\,2-6\zeta(3)\Bigg]\nn
&-&C_F\Bigg[3\(12\,\ln\fr{\bar\Lambda}{m}+5\)\ln\fr{\bar\Lambda}{m}+\fr{313}{8}+\fr{35\pi^2}{6}-8\pi^2\ln\,2+12\zeta(3)\Bigg]\nn
&+&N_f\Bigg[\fr{2}{3}\(6\,\ln\fr{\bar\Lambda}{m}+13\)\ln\fr{\bar\Lambda}{m}+\fr{71}{12}+\fr{2\pi^2}{3}\Bigg]
+6-2\pi^2\Bigg)\hat{m}^2z\nn
&+&4\hat{m}^2C_F\Bigg[3\(3\,\ln\fr{\bar\Lambda}{m}+4\)\ln\fr{\bar\Lambda}{m}+4\Bigg](\hat{u}-z)\Bigg\},\label{m31c}
\ea
where it is interesting to note that all the $1/\epsilon$ divergences have canceled. This implies that no cancellations need to take place between the single and double cut contributions, which is an important computational simplification.

At the two-cut level, there are two sources of contributions to ${\mathcal M}_3$: The two- and three-loop 2GI diagrams from Eq.~(\ref{lnZ2GI2c}) and the two-cut piece of the function $\Omega_\rmi{VM}^m$ from Eq.~(\ref{lnZVMm2c}). The sum of these parts reads
\ba
\fr{{\mathcal M}_3^{2c}}{(4\pi)^2}&=& d_A\Bigg\{C_A\Bigg(-\frac{16}{9}I_1^2+\frac{62}{9}m^2 I_2 + \frac{5}{3}I_{1c} -\frac{10}{3}m^2  I_{2c}+I_{10}
-\fr{22}{3}\left[I_1^2-2m^2 I_2\right]\ln\frac{\bar{\Lambda}}{m}\Bigg)\nn
&+&C_F\(I_{11}+\left[24(m^2 I_2 -m^2 I_{1b}I_1+2m^4 I_{2b})+48 m^4I_8\right]\ln\frac{\bar{\Lambda}}{m}\)\nn
&+&N_f\Bigg(\frac{10}{9}I_1^2-\frac{20}{9}m^2 I_2-\frac{2}{3}I_{1c}+\frac{4}{3}m^2 I_{2c}
+\left[\frac{4}{3}I_1^2-\frac{8}{3}m^2 I_2\right]\ln\frac{\bar{\Lambda}}{m}\Bigg)
-\frac{2}{3}I_{12}\Bigg\},
\ea
where the definitions and values of the various integrals $I_n$ are given in Appendix \ref{app:idefs}.

Finally, the only triple cut contribution to the grand potential originates from the 2GI graphs of Figs.~\ref{fig:graphs}b, \ref{fig:graphs}c and \ref{fig:graphs}d, implying that it can be read off directly from Eq.~(\ref{lnZ2GI3c}). One finds
\ba
\fr{{\mathcal M}_3^{3c}}{(4\pi)^4}&=&-d_A \Bigg\{C_A\Big[2I_1 I_2-4I_5+8m^4I_6-4I_7\Big]\nn
&+&C_F\Big[2I_1^2I_{1b}-4I_1I_2-8m^2I_1I_{2b}+8m^2I_3+8m^4I_{3b}\nn
&-&2I_4+8I_5-16m^4I_6+8I_7-8m^2I_1I_8+8m^4I_9\Big]\Bigg\}.
\ea

Summing up the single, double and triple cut contributions to ${\mathcal M}_3$ leads to a complicated expression, in which several parts must be evaluated numerically. To this end, we have found it most convenient to express the result in terms of a basis of simple fitting functions, leading to the result
\ba
{\mathcal M}_3&=&
\frac{d_A \mu^4}{2 \pi^2}\Bigg\{-\hat{m}^2\bigg[\(11C_A-2N_f\)z+18C_F\(2z-\hat{u}\)\bigg]\(\ln\fr{\bar\Lambda}{m}\)^2\nonumber\\
&+&\fr{1}{3}\Bigg[C_A\(22\hat{u}^4-\fr{185}{2}z\,\hat{m}^2-33z^2\)
+\fr{9C_F}{2}\(16\hat{m}^2\, \hat{u}(1-\hat{u})-3(7\hat{m}^2-8\hat{u})z-24z^2\)\nn
&-&N_f\(4\hat{u}^4-13z\,\hat{m}^2-6z^2\)\Bigg]\ln\fr{\bar\Lambda}{m}\nn
&+&C_A \(-\fr{11}{3}\ln\fr{\hat{m}}{2}-\fr{71}{9}+{\mathcal G}_1(\hat{m})\) + C_F \(\fr{17}{4}+{\mathcal G}_2(\hat{m})\) \nn
&+&N_f\(\fr{2}{3}\ln\fr{\hat{m}}{2} + \fr{11}{9}+{\mathcal G}_3(\hat{m})\)  + {\mathcal G}_4(\hat{m})\Bigg\}, \label{m3}
\ea
where the functions ${\mathcal G}_n(\hat{m})$ (that are defined so that they vanish in the $\hat{m}\rightarrow 0$ limit) read
\ba
{\mathcal G}_1(\hat{m})&=&32\pi^4 \hat{m}^2\Big[
-0.01863
+0.02038\, \hat{m}^2
-0.03900 \, \hat{m}^2 \log(\hat{m})\nn
&+&0.02581\, \hat{m}^2 \(\log(\hat{m})\)^2
- 0.03153 \, \hat{m}^2 \(\log(\hat{m})\)^3
+0.01151 \, \hat{m}^2 \(\log(\hat{m})\)^4  \Big], \\
{\mathcal G}_2(\hat{m})&=& 32\pi^4 \hat{m}^2\Big[
-0.1998
 - 0.04797  \log(\hat{m})
 + 0.1988\, \hat{m}^2
 -  0.3569\, \hat{m}^2 \log(\hat{m})\nn
  &+& 0.3043\, \hat{m}^2 \(\log(\hat{m})\)^2
   -  0.1611 \, \hat{m}^2 \(\log(\hat{m})\)^3
+ 0.09791 \, \hat{m}^2 \(\log(\hat{m})\)^4 \Big],\\
{\mathcal G}_3(\hat{m})&=&32\pi^4 \hat{m}^2\Big[
-0.05741
-0.02679 \log(\hat{m})
-0.002828 \(\log(\hat{m})\)^2\nn
&+&0.05716\, \hat{m}^2
-0.08777\, \hat{m}^2 \log(\hat{m})
+0.0666\, \hat{m}^2 \(\log(\hat{m})\)^2\nn
&-&0.02381\, \hat{m}^2 \(\log(\hat{m})\)^3
+0.01384\, \hat{m}^2 \(\log(\hat{m})\)^4
\Big],\\
{\mathcal G}_4(\hat{m})&=&32\pi^4 \hat{m}^2\Big[
0.07823
+0.0388 \log(\hat{m})
+0.004873 \(\log(\hat{m})\)^2\nn
&-&0.07822\, \hat{m}^2
+0.1183\, \hat{m}^2 \log(\hat{m})
-0.08755\, \hat{m}^2 \(\log(\hat{m})\)^2\nn
&+&0.03293\, \hat{m}^2 \(\log(\hat{m})\)^3
-0.01644\, \hat{m}^2 \(\log(\hat{m})\)^4
\Big].
\ea

\subsection{$\Omega_\rmi{VM}^x$: The Cross Term VM graph}

Next, we look at the vacuum-matter term $\Omega_\rmi{VM}^{x}$, in which the quarks in the matter part of the diagram are massless, while the vacuum part corresponds to the difference of a massive and a massless quark loop. Performing again one and two cuts and using the results of Appendix \ref{app:VMgraph}, one obtains
\ba
\fr{\Omega_\rmi{VM}^{x}}{V}&=& d_A\fr{g^4}{(4\pi)^2}\frac{m^4}{12 \pi^4}\sum_{i=1}^{N_l}\label{pvmx1} \nn
&\times& \int_0^{\frac{\mu_i^2}{m^2}}{\rm d} w\,\ln\fr{m^2 w}{\mu_i^2} \left[2-2\sqrt{\frac{(1+w)^3}{w}}{\rm arctanh}\bigg(\sqrt{\frac{w}{1+w}}\bigg) +w\log(4w)\right]\nn
&\equiv& d_A\fr{g^4}{(4\pi)^2}\frac{m^4}{12 \pi^4}\sum_{i=1}^{N_l}I_x\(\frac{\mu_i}{m+\mu_i}\),\label{pvmx2}
\ea
where the integral can be approximated by the pocket formula
\ba
I_x(t)&=&-3 t^4\left(1-\ln\,t\right)\Bigg[\frac{0.83}{(1-t)^2}+\frac{0.06}{(1-t)}
-0.056\nn
&+&\frac{\ln(1-t)}{t(1-t)^2}\(1.005-0.272 t (1-t)+0.154 t (1-t)^2\)\Bigg].
\ea

\subsection{$\Omega_\rmi{ring}$: The Plasmon Contribution}
\label{sec:plasmon}

The setup for the plasmon sum calculation can be found in standard textbooks, \textit{e.g.~}Section V of Ref.~\cite{Kapusta:1989tk}, according to which the contribution can be written in the form
\ba
&&\frac{\Omega_\rmi{plas}}{V}=\nn
&&\frac{d_A}{2} \int\frac{d^4K}{(2\pi)^4}
\left[2\ln \left(1-\frac{G(k_0,k)}{K^2}\right)+\ln \left(1-\frac{F(k_0,k)}{K^2}\right)
+\frac{2G(k_0,k)}{K^2}+\frac{F(k_0,k)}{K^2}\right],
\ea
where $k$ denotes the magnitude of the three-vector \textbf{k}. In this equation, we have defined
\ba
G(k_0,k)&=&\frac{1}{2}\left(\Pi^\mu_{\mu}(k_0,k)-\frac{K^2}{{\bf k}^2}\Pi^{00}(k_0,k)\right),\quad
F(k_0,k)=\frac{K^2}{\bf k^2}\Pi^{00}(k_0,k),\label{FGdef}
\ea
with $\Pi^\mu_{\mu}=\Pi^{0}_0-\Pi^{i}_{i}$, where $\Pi^{\mu\nu}$ is the one-loop gluon polarization tensor (\textit{cf.~}Appendix \ref{app:pimunu}). One should note that we have used the notation $F$, $G$, rather than $\Lambda_1$, $\Lambda_2$ sometimes found in the literature \cite{Freedman:1976ub,Toimela:1984xy} to avoid confusion with the renormalization scale $\bar{\Lambda}$.

Taking advantage of the benign IR behavior of the vacuum polarization tensor, we expand the logarithms in powers of the vacuum tensor, keeping only contributions up to and including ${\cal O}(g^4)$. This leads to
\ba
&&\ln \left(1-\frac{G(k_0,k)}{K^2}\right)=\nn
&&\ln\left(1-\frac{G_\rmi{mat}(k_0,k)}{K^2}\right)-\frac{G_\rmi{vac}(k_0,k)}{K^2}-\frac{G_\rmi{vac}(k_0,k) G_\rmi{mat}(k_0,k)}{K^4}- \frac{1}{2}\frac{G_\rmi{vac}^2(k_0,k)}{K^4},
\ea
and a similar expression for the $F(k_0,k)$ part. This separation allows us to split the plasmon sum into the VV, VM and ring sum pieces, the first two of which were already covered in the previous section. The last part, which is considered here, reads
\ba
\frac{\Omega_\rmi{ring}}{V}
&=&\frac{d_A}{2} \int\frac{d^4K}{(2\pi)^4}
\Bigg[2\ln \left(1-\frac{G_\rmi{mat}(k_0,k)}{K^2}\right)\nn
&+&\ln \left(1-\frac{F_\rmi{mat}(k_0,k)}{K^2}\right)
+\frac{2G_\rmi{mat}(k_0,k)}{K^2}+\frac{F_\rmi{mat}(k_0,k)}{K^2}\Bigg]. \label{plasmonring}
\ea

To isolate the $g^4\ln g$ term from the rest of the plasmon sum, we separate from Eq.~(\ref{plasmonring}) the combination (both for $F$ and $G$)
\ba
\int \frac{d^4K}{(2\pi)^4}
\left[\ln \left(1-\frac{F_\rmi{mat}(K=0,\Phi)}{K^2}\right)
+\frac{F_\rmi{mat}(K=0,\Phi)}{K^2}+\frac{F_\rmi{mat}^2(K=0,\Phi)}{2K^2(K^2+\chi^2)}\right],
\ea
where $\chi$ is a fictitious mass scale that will drop out in the final result and we have switched from using the variables $(k_0,\,k)$ to $K\equiv \sqrt{k_0^2+k^2}$ and $\Phi\equiv{\rm arctan}\frac{k}{k_0}$. The reason for introducing this particular function is that it captures the relevant IR physics in a form where the $K$ integration may be performed analytically, giving\footnote{Note that there is an error in the corresponding Eq.~(5.5.2) of Ref.~\cite{Kapusta:1989tk}.}
\ba
\frac{4}{(2\pi)^3} \int_0^{\pi/2} d\Phi\,\sin^2 \Phi\, \frac{F_\rmi{mat}^2(K=0,\Phi)}{4}
\left[-\frac{1}{2}+\ln \left(-F_\rmi{mat}(K=0,\Phi)/\chi^2\right)\right].
\ea
The complete ring sum contribution to the grand potential then becomes
\be
\Omega_\rmi{ring}=\Omega^{(1)}_\rmi{ring}+\Omega^{(2)}_\rmi{ring} + \Omega^{(3)}_\rmi{ring},
\label{totringsum}
\ee
where the ${\mathcal O}(g^4 \ln g)$ contribution has been isolated in
\ba
\frac{\Omega_\rmi{ring}^{(1)}}{V}&=&\frac{d_A\,g^4 \ln g}{(2\pi)^3} \int_0^{\pi/2}d\Phi\,\sin^2 \Phi\,
\left[2 \frac{G_\rmi{mat}^2(K=0,\Phi)}{g^4}+\frac{F_\rmi{mat}^2(K=0,\Phi)}{g^4}\right],
\ea
and the remaining part of the $g^4$ contribution is contained in the functions
\ba
\frac{\Omega_\rmi{ring}^{(2)}}{V}&=&-\frac{d_A}{2}\frac{g^4}{64 \pi^7} \int_0^{\pi/2} d\Phi\,\sin^2\Phi
\left[\left(\frac{-2\pi^2 F_\rmi{mat}(K=0)}{g^2}\right)^2\left(1+\ln \frac{\chi^4 g^4}{F_\rmi{mat}^2(K=0)}\right)
\right.\nonumber\\
&+&\left.\frac{1}{2}\left(\frac{-4\pi^2 G_\rmi{mat}(K=0)}{g^2}\right)^2\left(1+\ln \frac{\chi^4 g^4}{G_\rmi{mat}^2(K=0)}\right)\right],
\nonumber\\
\frac{\Omega_\rmi{ring}^{(3)}}{V}&=&\frac{d_A}{2}\frac{2g^4}{(2\pi)^3}\int_0^\infty dK\,K
\int_0^{\pi/2} d\Phi\,\sin^2\Phi \frac{1}{g^4}\left[
\frac{F_\rmi{mat}^2(K=0)-F_\rmi{mat}^2}{K^2}\right.\nonumber\\
&-&\left.\frac{F_\rmi{mat}^2(K=0)}{K^2+\chi^2}+\frac{2G_\rmi{mat}^2(K=0)-2G_\rmi{mat}^2}{K^2}-\frac{2G_\rmi{mat}^2(K=0)}{K^2+\chi^2}\right].
\ea

Collecting all of the above formulae, one observes that the result for the plasmon ring sum can be written in the form
\ba
\label{plasmonfinal}
\frac{\Omega_\rmi{ring}}{V}&=&\frac{d_A g^4}{512 \pi^6} \Bigg\{
\(\vec{\mu}^2\)^2\Bigg[2\ln \(\frac{g}{4 \pi}\)-\fr{1}{2} \nn
&+&\frac{1}{2}\(-\frac{19}{3}+\frac{2\pi^2}{3}+\frac{I_{15}({\vec{\mu}})}{(\vec{\mu}^2)^2}+\frac{16}{3} (1-\ln 2)\ln\, 2
+I_{16}\(\hat{m},\vec{\hat{\mu}}^2\)\)\Bigg]\nn
&+&2\mu^2 \sum_{i=1}^{N_l} \mu_i^2 \Bigg[I_{14}\(2\ln \(\frac{g}{4 \pi}\)-\fr{1}{2}\)\nn
&+&\frac{1}{2}\(I_{17}\(\hat{m},\hat{\mu}_i\)
+ \frac{16}{3}(1-\ln 2)\ln\, 2  \,I_{18}
+I_{19}\(\hat{m},\vec{\hat{\mu}}^2\)
\)\Bigg]\label{}\\
&+&\mu^4\Bigg[I_{13}\(2\ln \(\frac{g}{4 \pi}\)-\fr{1}{2}\)
+\frac{1}{2}\(I_{20}+ \frac{16}{3}(1-\ln 2)\ln\, 2 \, I_{21}
+I_{22}\(\hat{m},\vec{\hat{\mu}}^2\)\)\Bigg]\Bigg\},\nonumber
\ea
where $\vec{\mu}=\left(\mu_1,\mu_2,\ldots,\mu_{N_l}\right)$ and $\vec{\hat{\mu}}\equiv \vec{\mu}/\mu$, while the definitions of the $I_n$'s --- including approximation formulas to the numerically evaluated functions --- can again be found from Appendix~\ref{app:idefs}. Finally, we note that upon sending $\mu\rightarrow m$ in Eq.~(\ref{plasmonfinal}) and using the result that in this limit $I_{16}\left(\hat{m},\vec{\hat{\mu}}^2\right)\rightarrow -0.85638\ldots$, the result of $N_l$ massless quark flavors presented in Ref.~\cite{Vuorinen:2003fs} is recovered. Similarly, one can show that taking $m\rightarrow 0$, one obtains the correct result for $N_f=N_l+1$ massless flavors.

\subsection{The Result for the QCD Grand Potential \label{sec:result}}

According to our 'master equation' (\ref{lnZparts}), the QCD grand potential, evaluated to the perturbative order $g^4$ in the $\msbar$ scheme, can be written as the sum of four terms, each of which we have evaluated in the previous sections. Of these functions,
\begin{itemize}
\item $\Omega^{m=0}$ corresponds to the contribution of the massless quark flavors, and is available from Eq.~(\ref{masslessp}).
\item $\Omega^{m}$ corresponds to the contribution of the massive quark, and is available from Eq.~(\ref{massivepcollect}), using Eqs.~(\ref{m1}), (\ref{m2}) and (\ref{m3}).
\item $\Omega_\rmi{VM}^x$ corresponds to a cross term vacuum-matter diagram, and is available from Eq.~(\ref{pvmx2}).
\item $\Omega_\rmi{ring}$ corresponds to the plasmon ring sum, and is available from Eq.~(\ref{plasmonfinal}), using integrals listed in Appendix \ref{app:idefs}.
\end{itemize}

Before proceeding to use our result to study various physics questions, we want to stress that it has passed all the consistency checks available in the literature. In particular, the result is renormalization group invariant, \textit{i.e.~}the explicit logarithms of $\bar{\Lambda}$ match with those originating from the running of the coupling and the strange quark mass, and the result correctly approaches the cases of two and three massless quark flavors in the respective limits $\mu\rightarrow m$ and $\mu/m\rightarrow \infty$. Of these, the fact that the massless $N_f=3$ result of Refs.~\cite{Freedman:1976xs} and \cite{Vuorinen:2003fs} is
analytically reproduced is highly non-trivial, and a strong indication that our calculation has been performed correctly.

At this point, a remark regarding the region of applicability of our result is in order. As can be verified from Eqs.~(\ref{m31c}) and (\ref{m3}), the grand potential contains a part that behaves in the $\mu\rightarrow m$ limit as $g^4\mu^2m\sqrt{\mu-m}$, leading to a divergent $1/\sqrt{\mu-m}$ contribution to the strange quark number density. Inspecting the lower order contributions to $\Omega$, one can identify the reason for this behavior: In one part of the result, the expansion parameter is $g^2\mu/(\mu-m)$, which diverges in the limit $m\rightarrow \mu$. It would be interesting to analyze this effect in more detail and investigate what type of a resummation scheme would be required to describe the limit properly, but this is beyond the scope of the present work. For the purposes of this article, we merely note that as long as one confines the analysis to values of $\mu$ satisfying
\ba
\fr{\mu-m}{m}\gtrsim g^2,
\ea
no problems will occur. Due to the rather low value of the strange quark mass, this in practice provides no extra limitation for the applicability of our result.

\section{Analyzing the Result: Quark Matter EoS}
\label{sec:QCDphaset}

In this section, the properties of the cold quark matter EoS are derived from the result presented in the previous section, and issues such as the strange quark mass dependence and the choice of renormalization scale will be discussed. We want to stress that while the calculation of the QCD grand potential is completely unambiguous, the extraction of physical quantities such as the EoS makes it necessary to adopt particular strategies for the use of the result, \textit{e.g.~}regarding to which quantities to truncate at a given perturbative order while still ensuring thermodynamic consistency. With this caveat in mind, it is pleasing that many properties of the resulting EoS are naturally meaningful and intuitive. For example, in this section it will be demonstrated that the result for total quark number density for two massless and one massive quark flavors smoothly interpolates between the cases of two and three massless quarks.

\subsection{Choice of Renormalization Scale}
\label{sec:ren}

With the functional form of $\alpha_s(\bar\Lambda)$ and the strange quark mass $m(\bar\Lambda)$ specified by Eqs.~(\ref{runningalpha}) and (\ref{runningms}), the only undetermined parameter in the perturbative EoS is the choice of the renormalization scale $\bar\Lambda$. At high temperature and low density, the canonical choice for it is the first non-vanishing Matsubara frequency $2 \pi T$, with which resummed perturbation theory has be seen to provide a reasonable description of the lattice results for the EoS and entropy density for temperatures at least a few times the critical temperature of the deconfinement transition \cite{Kajantie:2002wa,Blaizot:2000fc}. For one massless quark with chemical potential $\mu$, phenomenological models suggest the choice $\bar\Lambda= 2 \pi \sqrt{T^2+(\mu/\pi)^2}$ up to an overall factor of two \cite{Schneider:2003uz,Rebhan:2003wn,Cassing:2007nb,Gardim:2009mt}, which also happens to be the scale appearing in the leading order fermionic quasiparticle mass \cite{Blaizot:2000fc}. Finally, a comparison of the perturbative and exact pressure at zero temperature, known in the limit of a large flavor number, also suggests $\bar\Lambda=2 \mu$ \cite{Ipp:2003jy}.

For the above reasons, our canonical choice for the renormalization scale will be $\bar\Lambda=2\sum \mu_i /N_f$, around which we will vary the parameter by a factor of two. We have also investigated more complicated choices of $\bar{\Lambda}$ taking into account the finite value of $m_s$, but have found this to have only minor effects on the results.

\subsection{Warmup: Massless Quarks}
\label{sec:warmup}

To illustrate our approach in a simple setting, let us first consider the case of three massless quarks ($N_f=N_l=3$) at the same chemical potential ($\mu_u=\mu_d=\mu_s=\mu$), corresponding to a system that is locally electrically neutral. Up to ${\cal O}(\alpha_s^2)$, the perturbative grand potential is then given by the sum of Eqs.~(\ref{masslessp}) and (\ref{plasmonfinal}), with the massive quark's chemical potential set to zero in the plasmon integrals. In principle, one could extract the pressure directly from these equations, via $P=-\Omega/V$. The determination of the EoS will, however, also require the total quark number density
\ba
n=-\frac{1}{V}\frac{\partial}{\partial \mu} \left[\Omega^{m=0}+\Omega^{m=0}_{\rm ring}\right],
\label{totalqn}
\ea
which, when evaluated using this formula, will receive corrections at higher order in $\alpha_s$ for instance from $\frac{\partial}{\partial \mu} \alpha_s^2$. These higher order terms are beyond the accuracy of our calculation, and hence typically ill-behaved, but cannot be simply dropped because this would imply $n\, d\mu\neq V dP$ and therefore ruin thermodynamic consistency. However, it is also possible to choose the quark number density as the fundamental quantity, keeping only terms up to (and including) ${\cal O}(\alpha_s^2)$, and use it to determine the other parts of the EoS by requiring thermodynamic consistency. Using Eq.~(\ref{totalqn}), we thus obtain (for general $N_f$) the three-loop result
\ba
n^{(2)}(\mu,\bar\Lambda)&=&n^{(0)}(\mu)\left[1-2\,\frac{\alpha_s}{\pi}-\left(\frac{\alpha_s}{\pi}\right)^2
\left(\fr{61}{4}-11\,\ln 2 -0.369165 N_f\right.\right.\nonumber\\
&+&\left.\left.N_f\ln\,\frac{N_f\alpha_s}{\pi}+\beta_0\ln \frac{\bar\Lambda}\mu\right)\right].
\label{masslessn2}
\ea
Here, $\beta_0$ is given by Eq.~(\ref{betafunc}), and the non-interacting quark number density reads $n^{(0)}(\mu)=N_f \mu^3/\pi^2$.

\begin{figure}[t]
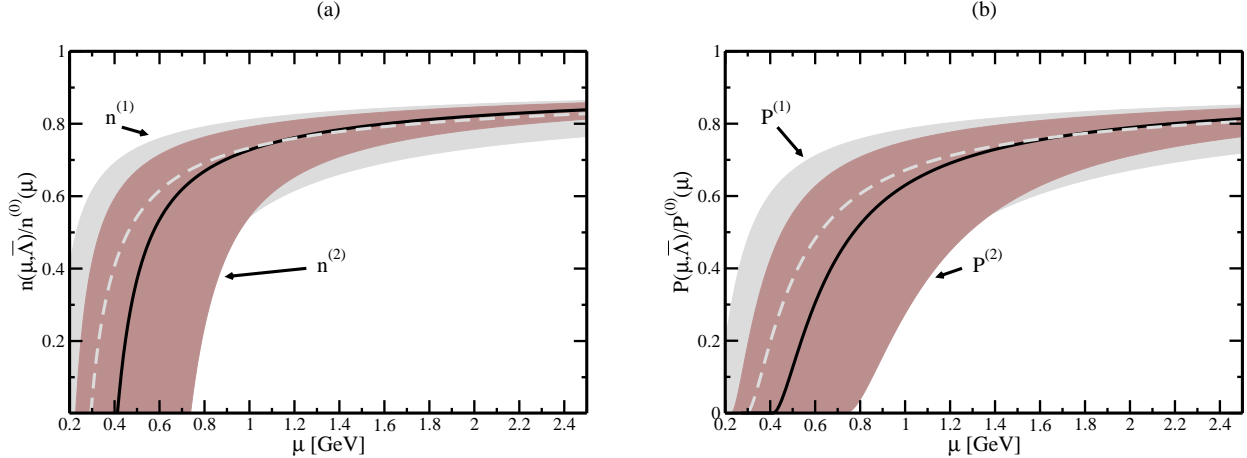

\center
\includegraphics[width=.47\linewidth]{fig4a.eps}
\hfill
\includegraphics[width=.47\linewidth]{fig4b.eps}
\caption{The renormalization scale dependence of the massless quark number density and pressure for perturbation theory to order ${\cal O}(\alpha_s^1)$ and ${\cal O}(\alpha_s^2)$ (light and dark shaded regions, respectively), with $\Lambda_{\tiny \msbar}=0.378$ GeV and $\bar\Lambda$ ranging from $\mu$ (right boundary) to $4\mu$ (left boundary). The dashed (full) lines correspond to the first (second) order results
with $\bar\Lambda=2\mu$.}
\label{fig:masslessn}
\end{figure}

In Fig.~\ref{fig:masslessn}a, we display the behavior of the function $n^{(2)}(\mu,\bar\Lambda)$, evaluated numerically using Eq.~(\ref{runningalpha}) with the value of the renormalization scale varied between $\bar\Lambda=\mu$ and $4\mu$. To assess the convergence properties of the perturbative expansion, we show in the same figure also the two-loop result $n^{(1)}$, obtained by truncating Eq.~(\ref{masslessn2}) at order $\alpha_s$ and --- to be consistent --- also truncating Eq.~(\ref{runningalpha}) by setting $\beta_1=0$. Inspecting the result, we find that the total quark number density $n(\mu,\bar\Lambda)$ is (even when normalized to the free quark density) a monotonically increasing function of $\mu$, and vanishes for a particular chemical potential $\mu_0(\bar\Lambda)$. This critical chemical potential is below the regime where our computation is reliable, and the fact that the quark number densities become negative for $\mu<\mu_0$ can be seen to signal the breakdown of our calculation. We will therefore set $n(\mu,\bar\Lambda)=0$ for $\mu<\mu_0(\bar\Lambda)$.

As one can see from Fig.~\ref{fig:masslessn}a, our results for $n^{(2)}$ are within the uncertainty band of the two-loop result, given by the renormalization scale dependence of $n^{(1)}$, for $\mu\gtrsim1$ GeV, indicating that the perturbative expansion for the quark number density converges reasonably well. More surprisingly, even at lower $\mu$, where the uncertainty band generated by $\bar\Lambda/\mu \in [1,4]$ is large, our values for $\mu_0(\bar\Lambda)$ do not differ much between $n^{(1)}$ and $n^{(2)}$. This should be contrasted with the case of high temperatures and low densities, where the convergence properties of weak coupling expansions are in most cases substantially worse.

Finally, from the quark number density one also obtains the pressure through the relation
\be
P(\mu,\bar\Lambda)=-B+\int_{\mu_0(\bar\Lambda)}^\mu {\rm d}\mu  \,\,  n(\mu,\bar\Lambda)\, ,
\label{pressuredef}
\ee
where $B$ is an integration constant, equal to minus the pressure at $\mu=\mu_0(\bar\Lambda)$. While in a purely perturbative calculation $B$ would usually be set to zero, in a realistic description it should be taken to be non-vanishing as the pressure can in any case only be determined up to an additive constant, representing the pressure difference between the physical and perturbative vacua. In our work, we will consider $B$ a free parameter, which allows us to take into account non-perturbative effects not captured by the weak coupling expansion. In fact, using the free quark number density $n^{(0)}$ with $\mu_0=0$ instead of $n^{(2)}$ in Eq.~(\ref{pressuredef}), one recovers the expression for the pressure in the original MIT bag model, with $B$ taking the role of the bag constant.

When performed in the way described above, our originally purely perturbative calculation can be seen to offer the possibility of adding non-perturbative effects to a result that is guaranteed to have the correct behavior at high energy densities. Due to physics criteria (\textit{e.g.~}requiring the energy density to be positive), the possible values for $B$ are, however, typically rather restricted, allowing us to make quantitative statements that are not possible in the original MIT bag model. In particular, if the equation of state is to be used down to $\mu=\mu_0$, one must require $B\geq 0$, because otherwise the energy density
\be
\varepsilon(\mu,\bar\Lambda)=-P(\mu,\bar\Lambda)+\mu\, n(\mu,\bar\Lambda)
\label{epsilondef}
\ee
will be negative at $\mu=\mu_0$. It is only if one ceases to use the perturbative EoS at some $\mu>\mu_0$ --- for instance when matching to a hadronic EoS at smaller $\mu$ --- that $B$ can take on negative values.

\subsection{Massive Quarks}
\label{sec:mquarks}

In a realistic attempt to describe cold deconfined quark matter, the non-vanishing mass of the strange quark must be taken into account. Up to and including the perturbative order $\alpha_s^2$, the grand potential is then given by the main result of our calculation, collected in Section \ref{sec:result}. To extract physically meaningful information from this expression, we follow the same strategy as in Section \ref{sec:warmup} up to some minor modifications necessary to maintain thermodynamic consistency.

We begin by evaluating the up ($u$) quark number density to ${\mathcal O}(\alpha_s^2)$, truncating terms of higher order, and use this to obtain expressions for the down ($d$) and strange ($s$) quark number densities that are
thermodynamically consistent, \textit{i.e.~}satisfy relations such as $\partial n_s/\partial \mu_u = \partial n_u/\partial \mu_s$. This procedure leads to the results
\ba
n_u(\mu_u,\mu_d,\mu_s)&=&n_u^{(2)}(\mu_u,\mu_d,\mu_s),\,\nonumber\\
n_d(\mu_u,\mu_d,\mu_s)&=&\int_{u_0(\mu_d,\mu_s)}^{\mu_u} d\mu_u^\prime
\partial_{\mu_d} n_u^{(2)}(\mu_u^\prime,\mu_d,\mu_s)+
n_d^{(2)}(\mu_u=u_0,\mu_d,\mu_s),\,\nonumber\\
n_s(\mu_u,\mu_d,\mu_s)&=&\int_{u_0(\mu_d,\mu_s)}^{\mu_u} d\mu_u^\prime
\partial_{\mu_s} n_u^{(2)}(\mu_u^\prime,\mu_d,\mu_s)+
\int_{d_0(\mu_s)}^{\mu_d} d\mu_d^\prime
\partial_{\mu_s} n_d^{(2)}(u_0(\mu_d^\prime,\mu_s),\mu_d^\prime,\mu_s)\nonumber\\
&&+n_s^{(2)}(\mu_u=u_0,\mu_d=d_0,\mu_s)\, , \label{nirels}
\ea
where the functions $n_i^{(2)}$ are defined so that they contain no terms beyond ${\mathcal O}(\alpha_s^2)$, and $u_0$ and $d_0$ are free integration functions. For reasons discussed in Section \ref{sec:ren} --- and to better facilitate comparison with the massless case --- the renormalization scale is taken
to be of the form
\be
\bar\Lambda\propto \frac{\mu_s+\mu_u+\mu_d}{3}\, ,
\ee
where the canonical choice of two for the proportionality constant will be varied by a factor of two to test how strongly our result depends on this choice.

In this work, we consider deconfined quark matter that is locally charge neutral and in beta equilibrium.\footnote{Relaxing the assumption of local charge neutrality to a global one in systems with a mixed phase is typically only a minor effect in comparison with the renormalization scale dependence \cite{Glendenning:1997wn}.} Chemical equilibrium is reached via the weak processes
\ba
d\rightarrow u+e+\bar{\nu}_e,& & u+e\rightarrow d+\nu_e\, , \nn
s\rightarrow u+e+\bar{\nu}_e,& & u+e\rightarrow s+\nu_e\, , \nn
s+u &\leftrightarrow& d+u\, ,\nonumber
\ea
which imply the conditions
\ba
\mu_s=\mu_d\equiv\mu\,,\qquad \mu_u=\mu-\mu_e\, ,
\label{CEconditons}
\ea
with $\mu_e$ being the electron chemical potential. The contribution of the neutrinos has been neglected here because for the quasi-static compact star systems we have in mind, neutrinos escape quickly, as their mean free path is considerably larger than the physical size of the system. Local charge neutrality on the other hand relies on the presence of electrons and leads to the relation
\ba
\frac{2}{3}n_u-\frac{1}{3}n_d-\frac{1}{3}n_s-n_e=0\, ,
\label{chargeeq}
\ea
where $n_e=\mu_e^3/(3\pi^2)$ is the electron density. Solving Eq.~(\ref{chargeeq}) with Eq.~(\ref{CEconditons}) fixes the electron chemical potential $\mu_e$ as a function of $\mu$.

The above constraints of beta equilibrium and charge neutrality enable us to simplify the expressions of Eq.~(\ref{nirels}) significantly. Namely, for a very particular choice of the integration functions $d_0$ and $u_0$, $d_0(\mu_s)=\mu_s$ and $u_0(\mu_d,\mu_s) = \mu_s-\mu_e(\mu_d,\mu_s)$, we see that each of the integral terms in Eq.~(\ref{nirels}) vanishes on the one-dimensional curve in $\mu_i$ space where both beta equilibrium and charge neutrality are maintained. In this physical subspace --- and in particular when calculating $\mu_e(\mu_d,\mu_s)$ from Eq.~(\ref{chargeeq}) --- we may use the truncated expressions $n_i^{(2)}$ for all quark flavors,
\ba
n_u(\mu,\bar{\Lambda})&=&n_u^{(2)}(\mu-\mu_e(\mu),\mu,\mu) + {\mathcal O}(\alpha_s^3)\, ,\nn
n_d(\mu,\bar{\Lambda})&=&n_d^{(2)}(\mu-\mu_e(\mu),\mu,\mu) + {\mathcal O}(\alpha_s^3)\, ,\nn
n_s(\mu,\bar{\Lambda})&=&n_s^{(2)}(\mu-\mu_e(\mu),\mu,\mu) + {\mathcal O}(\alpha_s^3)\, ,
\label{truncns}
\ea
where we have emphasized the dependence of the result on $\bar\Lambda$ by explicitly reinstating it as an argument of the functions. In these results, we will again set $n_{i}=0$ whenever the number density in question becomes negative. We also note that while in our starting point Eq.~(\ref{nirels}) we chose $n_u$ rather than $n_d$ or $n_s$ as the fundamental quantity, our end result Eq.~(\ref{truncns}) treats all flavors in a symmetric fashion.

In terms of the above results, the total quark number density is finally given by
\ba
n(\mu,\bar\Lambda)&=&n_u(\mu,\bar{\Lambda})+n_d(\mu,\bar{\Lambda})+n_s(\mu,\bar{\Lambda})\,,
\ea
while the pressure and energy density are evaluated through
\ba
P(\mu,\bar\Lambda)&=&-B + \int_{\mu_0(\bar\Lambda)}^\mu d\mu \left[
n_u\left(1-\frac{d\mu_e(\mu)}{d\mu}\right)+n_d + n_s + n_e \fr{d\mu_e(\mu)}{d\mu}\right]\,,\nonumber\\
\varepsilon(\mu,\bar\Lambda)&=&-P(\mu,\bar\Lambda)+\mu\left(n_u+n_d+n_s\right)-\mu_e(\mu)\(n_u-n_e(\mu)\)\,,
\ea
in analogy with the massless quark case, \textit{cf.~}Eqs.~(\ref{pressuredef}) and (\ref{epsilondef}). Note that the contribution coming from the electron number density in the above equations is rather small.

\begin{figure}[t]
\center
\includegraphics[width=.7\linewidth]{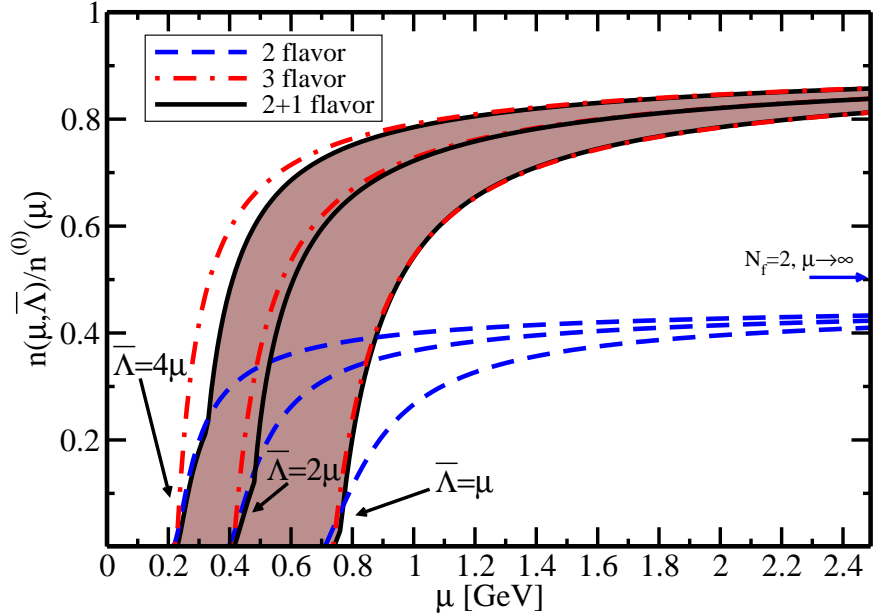}
\caption{The total quark number density evaluated to ${\cal O}(\alpha_s^2)$ for locally charge neutral systems of 2 and 3 massless quark flavors, as well as for the two light and one massive flavor case ('2+1'). All results are normalized to the density of three free massless flavors $3\mu^3/\pi^2$, and assume the values $\Lambda_{\tiny \msbar}=0.378$ GeV, $m(2{\rm GeV})=0.1$ GeV, while the renormalization scale takes the values $3\bar{\Lambda}/(\mu_u+\mu_d+\mu_s)=1,\,2,\,4$ (for $N_f=2$, $2 \bar\Lambda/(\mu_u+\mu_d)=1,\,2,\,4$). As expected, the 2+1 flavor result matches the three flavor result at large $\mu$ and approaches the two flavor result at small $\mu$.}
\label{fig:mnd2}
\end{figure}

In Fig.~\ref{fig:mnd2}, we display our result for the total quark number density in cold deconfined quark matter as a function of the $d$ quark chemical potential $\mu$, and compare it to the cases of two and three massless quark flavors (for which beta equilibrium and charge neutrality require that $\mu_e=0$ for $N_f=3$, but $\mu_e\neq 0$ for $N_f=2$). Upon comparison with the two-loop ${\cal O}(\alpha_s)$ result (not shown in the plot), we find that perturbation theory exhibits convergence for $\mu>1$ GeV, in analogy with the massless case. Somewhat visible in Fig.~\ref{fig:mnd2} are kinks at the critical chemical potentials at which the strange quark density drops to zero, and below which the quark matter is net strange quark free. We suspect that this is simply a consequence of not having enough energy to produce strange quarks with a non-vanishing in-medium mass: The chemical potential is required to satisfy the condition $\mu > m_{\rm medium}(\mu_s)$, where the parameter $m_{\rm medium}(\mu)$ can be evaluated by studying the effects of the finite chemical potential on the poles of a massive quark propagator. A study of the one-loop quark self energy at finite temperature was recently performed in Ref.~\cite{Chesler:2009yg} (cf. \cite{Schertler:1996tq}), and a simple generalization of these results to finite $\mu$ shows that for $g\mu\ll M \ll \mu$, the mass can be approximated by the formula
\ba
m_{\rm medium}(\bar\Lambda,\mu)&\sim&\sqrt{m^2+\frac{8 \alpha_s}{3 \pi} \mu^2}+{\cal O}(\alpha_s^2) \;>\;m\, .
\ea
This leads us to argue that the chemical potential at which the strange quark density vanishes does not need to receive large non-perturbative corrections, as instead of confinement physics only energy conservation is involved in the mechanism. As a consequence, we can expect perturbative results to give quantitatively reasonable estimates for this critical chemical potential, at least if $\bar\Lambda>1$ GeV at this point (\textit{cf.~}the discussion in Section \ref{sec:renorm}).

Moving up on the chemical potential axis, we note that for $\mu\gtrsim 1$ GeV, the strange quark mass becomes unimportant and one recovers the result of three massless flavors, discussed in Section \ref{sec:warmup}. Interestingly, studying the ${\mathcal O}(\alpha_s^0)$ and ${\mathcal O}(\alpha_s)$ quark number densities, one observes the trend that the effects of the strange quark mass become less important with increasing perturbative order. This indicates that accounting for the interactions accurately in fact makes the QCD EoS less sensitive to the quark masses, which we suspect may well generalize to further perturbative orders as well.

\section{Strange Quark Matter and the Phase Transition}
\label{sec:sqmpt}

In this section, we will use the results derived above to investigate two mutually exclusive scenarios: The properties of absolutely stable strange quark matter and the onset of the confinement transition from deconfined quark matter to the hadronic phase. In both of these cases, we will evaluate the EoS for bulk matter (in contrast to systems of finite size), which is the relevant setting for the astrophysical applications considered in Section \ref{sec:astro}. In the strange quark matter calculation, our main goal will be to sweep the parameter space of the theory to find out if there is a region that allows for the existence of stable strange quark matter, while for the confinement phase transition case we will investigate if there are density windows that allow for a smooth matching of the quark matter and hadronic EoSs.

For completeness --- and to better describe real world finite density quark matter in beta equilibrium --- we will include in our EoS a contribution modeling the effects of color superconductivity (CSC). This is accomplished by adding to the pressure a term accounting for the condensation energy of Cooper pairs in the Color-Flavor-Locked (CFL) phase (see \textit{e.g.~}Refs.~\cite{Alford:1998mk,Alford:2007xm,Orsaria:2007zza}),
\ba
P_\rmi{CSC}&\equiv& \fr{\Delta^2\mu_B^2}{3\pi^2}, \label{pcfc}
\ea
where the baryon chemical potential is $\mu_B\equiv\mu_u+\mu_d+\mu_s$
and the gap parameter $\Delta$ approaches at asymptotically high densities the form \cite{Son:1998uk}
\ba
\Delta &=&\fr{b\mu}{(4 \pi \alpha_s)^{5/2}}{\rm e}^{-3\pi^2/\sqrt{8\pi \alpha_s}},
\ea
with $b$ a constant. In this work, $\Delta$ itself will for simplicity be assumed to be a constant, which suffices to at least estimate the magnitude of the CSC effects. We will in each case study the values $\Delta=0$ (corresponding to normal, unpaired quark matter) and $\Delta=$100 MeV, of which the latter can be viewed as an upper limit for the size of the gap. It should also be noted that while Eq.~(\ref{pcfc}) in principle obtains corrections due to the finiteness of $m_s$, the condensation term has its largest value in the $m_s=0$ limit. Thus, Eq.~(\ref{pcfc}) serves the purpose of indicating the maximal effect that quark pairing may have on the EoS of the system.

\subsection{Strange Quark Matter Hypothesis}
\label{sec:sqm}

Stable strange quark matter configurations in vacuum can exist if there is a (strange) quark chemical potential $\mu$ for which the strange quark density in the system is non-zero, while the pressure is vanishing and the energy per baryon
\ba
E/A\equiv\,\varepsilon(\mu,\bar\Lambda)/n_B(\mu,\bar\Lambda)\,,
\ea
is lower than for the most stable nucleus \cite{Weber:2004kj} ($^{56}{\rm Fe}$ and $^{62}{\rm Ni}$),
\ba
E/A\leq 0.93\,{\rm GeV}.
\ea
Here, the baryon density is related to the total quark number density $n$ via $n_B=n/3$. If stable strange quark matter were to exist, ordinary nuclear matter (made up of up and down quarks) would be only metastable, with a lifetime determined by the probability to generate strange quarks via several simultaneous weak interactions \cite{Page:2006ud}. In particular, this would imply that nuclei with baryon number $A\gtrsim 6$ would have a lifetime of $\sim\!10^{60}$ years \cite{Weber:2004kj}, consistent with current observations. In contrast, stable two flavor quark matter (zero strange quark density) is clearly ruled out by experiment, as it would imply an extremely small lifetime for ordinary nuclei because no strangeness needs to be created.

In the following, we will consider the stability criteria for bulk strange quark matter ($A\gg1$), which is both the relevant case for astrophysical applications as well as the most stable configuration (finite size effects for small $A$ only make strange quark matter more unstable \cite{Weber:2004kj}). We will also only consider quasi-static configurations, where beta equilibrium is maintained and the system is locally electrically neutral, of which the latter requirement follows from the fact that for strange quark matter no hadronic admixture can be present. Hence, we will require $P^{(2)}(\mu,\bar\Lambda)+P_\rmi{CSC}(\mu)=0$, where the three-loop pressure for 2+1 flavor quark matter, $P^{(2)}$, is taken from Section \ref{sec:mquarks} and $P_\rmi{CSC}$ from Eq.~(\ref{pcfc}). In addition, we will naturally enforce the condition that the density of strange quarks is non-zero.

\begin{figure}[t]
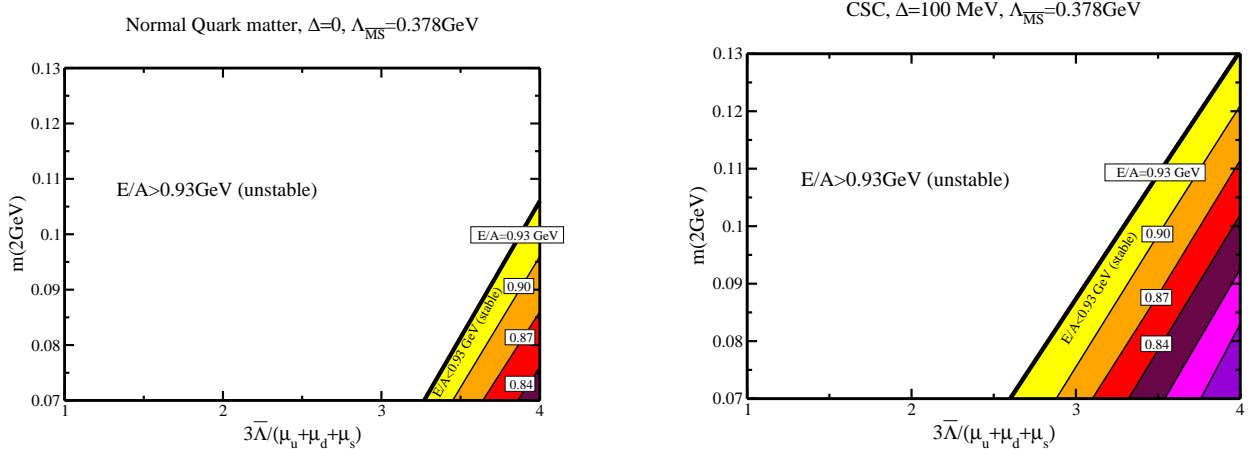

\center
\includegraphics[width=.45\linewidth]{fig6a.eps}
\hfill
\includegraphics[width=.45\linewidth]{fig6b.eps}
\caption{
Exclusion plots of stable strange quark matter, where the smallest possible value of $E/A$ (required to be lower than 0.93 GeV for stability) is plotted as a function of the parameters of the theory. Left: Normal (unpaired) stable strange quark matter can only occur for comparatively large values of the renormalization scale. Right: The incorporation of the effects of the CSC gap significantly increases the allowed parameter space. Reducing the value of $\Lambda_{\tiny \msbar}$ has an effect similar to that of increasing $\bar\Lambda/(\mu_u+\mu_d+\mu_s)$.}
\label{fig:sqm}
\end{figure}

In practice, we sweep through the parameter space of the theory as follows. For both $\Delta=0$ and $\Delta = 100$ MeV, we vary the renormalization scale $\bar{\Lambda}$, the $\msbar$ scale $\Lambda_{\tiny \msbar}$ as well as the strange quark mass $m$ in the ranges $3\bar{\Lambda}/(\mu_u+\mu_d+\mu_s)=1...\,4$, $\Lambda_{\tiny \msbar}=0.378+0.034-0.032$ GeV and $m(2\, {\rm GeV})\simeq 0.100\pm0.030$ GeV. At each point of this parameter space, we first search for the smallest value of $\mu$ for which the non-zero strangeness condition $n_s(\mu,\bar\Lambda)>0$ is fulfilled, which also corresponds to the minimal value of $E/A$. After determining the value of the latter quantity, we then choose $B>0$ so that the stability criterion $P=0$ is satisfied. It turns out that these constraints imply that our results do not probe physics below $\bar\Lambda\simeq0.95$ GeV, which gives us confidence in our method. We find that that for $\Delta=0$ strange quark matter is either unstable ($E/A>0.93$ GeV) or violates experimental evidence ($n_s=0$) for most parts of the above uncertainty ranges, while a finite value for the gap parameter has the effect of relaxing the constraints significantly (\textit{cf.~}Refs.~\cite{Lugones:2002va,Alford:2002rj} for similar conclusions). These effects are summarized in the exclusion plots of Fig.~\ref{fig:sqm}.

The conclusions one can draw from the above results are clearly strongly dependent on whether one considers $\Delta=0$ or $\Delta=100$ MeV. While the former case suggests a rather hostile parameter space for stable strange quark matter, perhaps offering an explanation for the absence of experimental evidence for it in direct searches \cite{Sandweiss:2004bu,Abelev:2007zz,Cecchini:2008su,Han:2009sj}, in the CSC case our results are inconclusive. Thus, within the current uncertainties of the different parameter values, we can neither confirm nor rule out the existence of strange quark matter. Alternatively, if stable strange quark matter is indeed realized, then the restricted parameter space allows us to set limits on its properties,\footnote{It has been noted in the literature (see \textit{e.g.~}Ref.~\cite{Fraga:2001id}) that for massless strange quark matter at high density, the EoS is extremely well (to the per cent level) approximated by the simple formula
\ba
\varepsilon&=&4 B_{\rm eff}+ a P\,,
\label{strangeeos}
\ea
where $B_{\rm eff}$ is an effective bag constant. In our case, one sees that a similar ansatz works only in the limit of large energy densities, but breaks down where the effects of the strange quark mass (as well the running of $\alpha_s$) become important.} as we will observe in the next section where we consider the masses and radii of dense stars made up of strange quark matter.

\subsection{Phase Transition from Quark Matter to Hadronic Matter}
\label{sec:matching}

\begin{figure}[t]
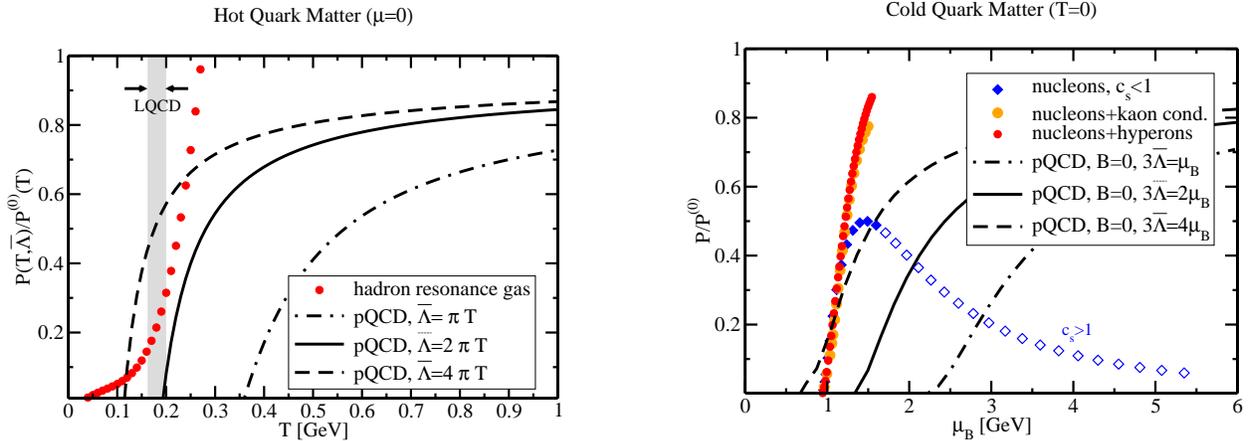

\center
\includegraphics[width=.45\linewidth]{fig7a.eps}
\hfill
\includegraphics[width=.45\linewidth]{fig7b.eps}
\caption{The matching of perturbative results for the quark matter EoS to hadronic EoSs at high temperature and density. Left: The pressure at $\mu=0$, obtained from resummed ${\cal O}(\alpha_s^{5/2})$ pQCD \cite{Blaizot:2003iq} (using Eq.~(\ref{runningalpha}) for $\alpha_s$ with $\Lambda_{\tiny \msbar}=0.378$ GeV and $N_f=3$) and compared with the result of summing up the effects of all hadron resonances with masses smaller than $2$ GeV \cite{Karsch:2003vd,Amsler:2008zzb}. For a cross-over transition, the pQCD results should match smoothly onto the hadronic EoS. The range in $\bar\Lambda$ for which this is possible corresponds quite well to the transition region determined by lattice QCD \cite{Aoki:2009sc,Bazavov:2009zn} (the grey band labeled 'LQCD' in the figure). Right: The $T=0$ quark matter pressure, taken from our ${\cal O}(\alpha_s^2)$ result with $\Lambda_{\tiny \msbar}=0.378$ GeV, $m(2{\rm GeV})=0.1$ GeV and $B=0$, compared to three different hadronic EoSs \cite{Akmal:1998cf,Schulze:2006vw}. The matching in general involves $B\neq 0$ and does not have to be smooth, as the phase transition at zero temperature could be of first order.}
\label{fig:matching}
\end{figure}

If strange quark matter is not the true ground state of nuclear matter, then at some critical chemical potential (or range of chemical potentials) quarks must become confined into hadrons. In this section, we try to estimate the location of this confinement/deconfinement transition by matching our perturbative EoS for quark matter to existing results for the hadronic EoS. While it is clear that our framework can never capture the details of the confinement process itself, it is not excluded that our results could be used to get a reasonable estimate for the critical chemical potential. One should contrast the situation to that of high temperatures and small $\mu$, where the matching between the hadronic and quark gluon plasma EoSs --- the latter obtained through resummed perturbation theory --- suggests a location for the deconfinement transition that agrees reasonably well with lattice data (see Fig.~\ref{fig:matching}). In the case of small $T$ and high $\mu$ considered here, a quantitative test of our predictions will have to await further advances in non-perturbative solutions to QCD.

To describe zero temperature hadronic matter in beta equilibrium above nuclear saturation densities, we consider three different EoSs, representing three classes of physical pictures. These are a nucleonic 'baseline' EoS by Akmal, Pandharipande and Ravenhall (\cite{Akmal:1998cf}, denoted '$A_{18}+\delta v+{\rm UIX}^*$' in this reference), an EoS including kaon condensation by Glendenning and Schaffner-Bielich (\cite{Glendenning:1998zx}, denoted '$U_K=-140$ MeV' in this reference), as well as one including the effects of hyperons by Schulze \textit{et al.} (\cite{Schulze:2006vw}, denoted '$V_{18}+{\rm UIX}+{\rm NSC89}$' in this reference). We have selected these three results because they constitute the most realistic and accurate calculations for the different physical scenarios,\footnote{In the case of kaon condensation, one may argue that $U_K=-100$ MeV to $-120$ MeV would be more realistic values for the potential \cite{Koch:1994mj,Waas:1997pe}, but we were unable to obtain a tabulated EoS for these cases.} in contrast to being 'maximally different' as in the selection criteria of Ref.~\cite{Lattimer:2000nx}.

The matching of hadronic and quark matter EoSs relies on imposing the following conditions:
\begin{enumerate}
\item
At the matching point, the pressure of the hadronic phase is equal to that of the quark matter phase.
\item
Both the hadronic and quark matter phases are locally charge neutral.
\item
The speed of sound $c_s\equiv \sqrt{dP/d\varepsilon}$ has to be less than the speed of light in both phases.
\item
The energy density has to increase monotonically with baryon chemical potential
\end{enumerate}
The second of these criteria can be easily relaxed by considering a two component system that is only globally charge neutral \cite{Glendenning:1997wn}. This is, however, only a minor effect in comparison with the other uncertainties in the calculation, and we consider it misplaced accuracy to perform a detailed analysis of the two-component phase transition here. The last two criteria are on the other hand meant to impose naturalness on the resulting EoS: $c_s<1$ is required to maintain causality, and it would be quite bizarre if for any given $\mu$ matter composed of nuclei could have an energy density higher than that of quark matter (except if nuclei did not correspond to the true ground state of hadronic matter, \textit{cf.~}Section \ref{sec:sqm}). Together with the first condition, the monotonicity of the energy density also implies that above the critical chemical potential, the physical phase is always the one with larger pressure. This is an important consistency check of our procedure.

In practice, we implement the matching as follows: For both $\Delta=0$ and $\Delta=100$ MeV, and any of the three hadronic EoSs, we first fix $\Lambda_{\tiny \msbar}$ and $m(2 {\rm GeV})$ within the ranges given in Section \ref{sec:renorm}, and then vary $\bar\Lambda$ in its usual range to determine whether the criterion 1 above can be fulfilled for any (positive or negative) integration constant $B$ and quark chemical potential. The successful cases are then subjected to the criteria 2--4, and the parameter values and hadronic EoSs that violate any of these are subsequently ruled out. As a first result, one observes that matching is only possible if $3\bar\Lambda/(\mu_u+\mu_d+\mu_s)\gtrsim 3.3$.

\begin{figure}[t]
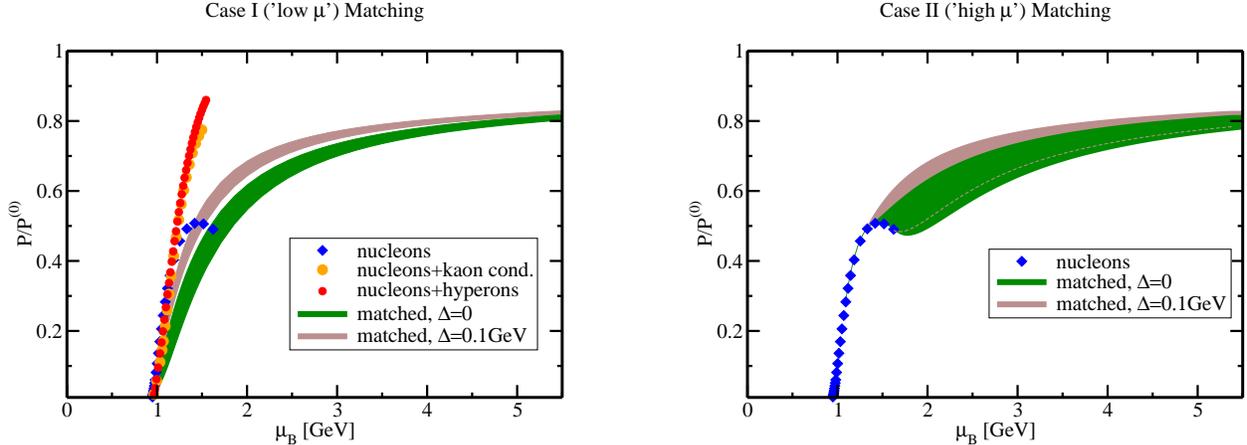

\center
\includegraphics[width=.45\linewidth]{fig8a.eps}
\hfill
\includegraphics[width=.45\linewidth]{fig8b.eps}
\caption{The equation of state of hybrid quark matter, obtained by matching the quark matter and nucleonic EoSs at low ($n_B\lesssim 0.39\ {\rm fm}^{-3}$, Case I) and high ($n_B\gtrsim 0.64\ {\rm fm}^{-3}$, Case II) densities, respectively. The shaded regions are obtained by varying $\bar{\Lambda}$, the integration constant $B$, and thus also the exact matching point, while keeping $\Lambda_{\tiny \msbar}$ and $m(2\, {\rm GeV})$ at their central values 0.378 GeV and 0.1 GeV, respectively.}
\label{fig:highmumatch}
\end{figure}

Summarizing the results of the matching calculation, we find that for both normal quark matter ($\Delta=0$) as well as for the CSC case ($\Delta=100$ MeV), matching to a hadronic EoS is only possible in two disjoint density windows: At low baryon densities, $n_B\lesssim 0.39\ {\rm fm}^{-3}$ ('Case I' in the following), where all three hadronic EoSs are degenerate and above which the criterion 4 is violated, as well as at high baryon densities, $n_B\gtrsim 0.64\ {\rm fm}^{-3}$ ('Case II'), where matching is only possible to the purely nucleonic EoS and above which the criterion 3 is violated in the hadronic sector. The matched EoSs of both cases are displayed in Figure \ref{fig:highmumatch}, from where we notice a significant decrease in the uncertainty of our results in comparison with the unmatched pure quark matter case of Figure \ref{fig:matching}.

For quark chemical potentials above 0.4 GeV, the nucleonic EoS differs from the hyperon and kaon EoSs significantly. If one is to trust (one of) the latter two --- and strange quark matter is not the true ground state of nuclear matter --- our results suggest a confinement transition from quark to hadronic matter at or around the density of atomic nuclei (\hbox{'case I'}, $0.15\ {\rm fm}^{-3}\lesssim n_B \lesssim 0.39\ {\rm fm}^{-3}$).\footnote{One should, however, note that the Case I matching can only be carried out in a rather small region of our parameter space, in particular requiring a large value for the renormalization scale $\bar{\Lambda}$, and is thus in some sense less robust than Case II. We thank David Blaschke and Thomas Kl\"ahn for drawing our attention to this issue.} This is predominantly a consequence of our matching criterion 4 regarding the monotonic increase of the energy density, as well as the very high energies (corresponding to half of the Fermi pressure of a gas of free quarks) predicted by all of the hadronic EoSs at around $n_B\sim 0.16\ {\rm fm}^{-3}$. While keeping in mind that the matching process carries sizable quantitative uncertainties due to the perturbative nature of our calculation, we note that if the deconfinement transition at high temperature and low density  --- where $\varepsilon(T)/\varepsilon^{(0)}(T)\lesssim 0.5$ in the transition region \cite{Bazavov:2009zn} --- is any guide for the physics at zero temperature, our findings may after all not be so unreasonable.

\section{Astrophysics Applications}
\label{sec:astro}

Having obtained EoSs for both stable strange quark matter as well as hybrid quark/hadronic matter undergoing a phase transition, we are now finally able to compare our results to nature. At present, the only available observational window into the properties of cold and dense nuclear matter are compact stars, which makes them the natural application of our results.

\begin{figure}[t]
\center
\includegraphics[width=.5\linewidth]{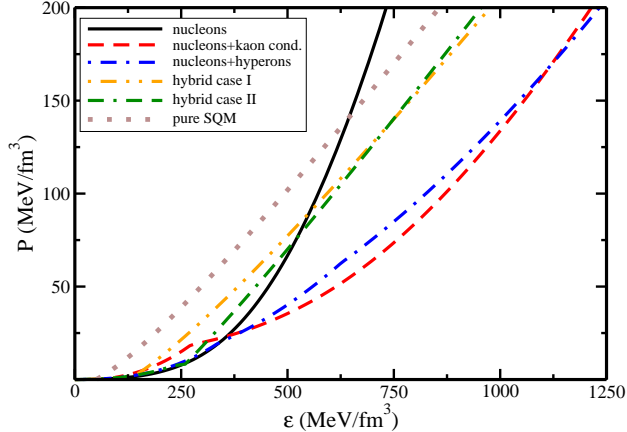}
\caption{The functional dependence between the pressure and energy density, given for a selection of different EoSs considered in this section.}
\label{fig:EP}
\end{figure}

Of special interest for us is the structure of a non-rotating compact star, in particular the relation between its mass and radius, because it is highly sensitive to the details of the underlying EoS of high density nuclear and/or quark matter. In addition, the mass-radius relation for a hydrostatic compact star is straightforward to determine in general relativity by solving the Tolman-Oppenheimer-Volkoff (TOV) equations \cite{TOV},
\ba
dM(r)&=&4 \pi r^2 \varepsilon(r) dr,\nn
dP(r)&=&-\frac{G \left(P(r)+\varepsilon(r)\right)\left(M(r)+4 \pi r^3 P(r)\right)}
{r \left(r-2 G M(r)\right)} dr\, .
\label{TOV}
\ea
Here, $G=\left(1.22\times 10^{19}\right)^{-2}\, {\rm GeV}^{-2}$ is Newton's constant in natural units and $r$ the radial coordinate of the star, while the EoS enters through the function $\varepsilon(P)$ (\textit{cf.~}Fig.~\ref{fig:EP}). In practice, the TOV equations are solved by choosing a value for $P$ at the center of the star ($r=0$) and then integrating outwards to the surface, where the pressure vanishes. The resulting mass and radius can be calculated for any central pressure allowed by the EoS under consideration, and differ markedly between the different classes of EoSs.

\begin{figure}[t]
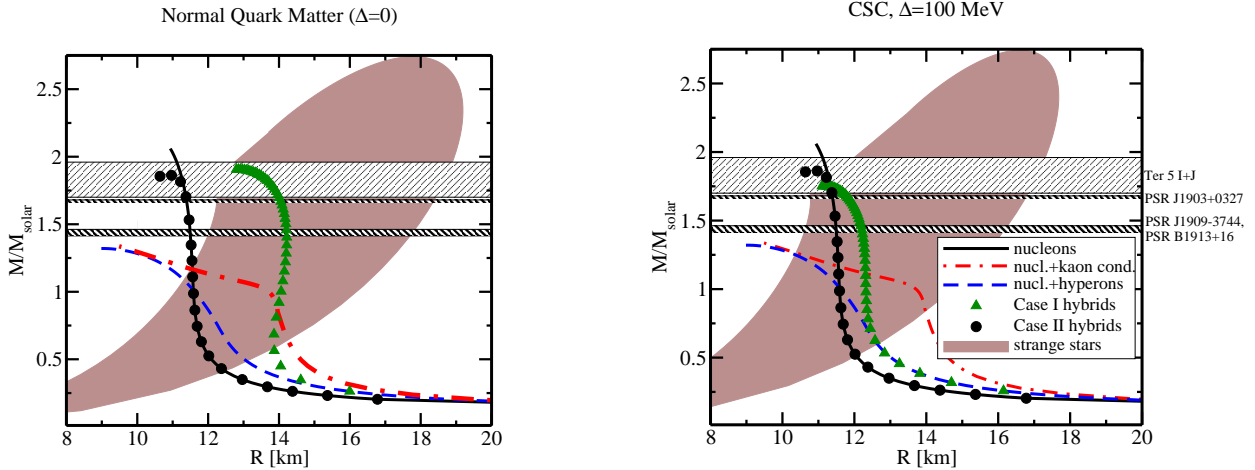

\center
\includegraphics[width=.4\linewidth]{fig10a.eps}
\hfill
\includegraphics[width=.48\linewidth]{fig10b.eps}
\caption{The mass-radius relation for compact stars, obtained using $\Delta=0$ (left) and $\Delta=100$ MeV (right) in the quark matter EoS. We display the results for purely hadronic stars (containing only nucleons \cite{Akmal:1998cf}, nucleons with kaon condensation \cite{Glendenning:1998zx}, or nucleons and hyperons \cite{Schulze:2006vw}), pure quark matter stars ('strange stars', \textit{cf.~}Section \ref{sec:sqm}) and hybrid stars including both hadronic and quark matter (see text for details). Also shown in the plots are compact star mass observations from Refs.~\cite{Freire:2009dr,Weisberg:2002nv,Jacoby:2005qg,Ransom:2005,Champion:2008ge}.}
\label{fig:MR}
\end{figure}

In Fig.~\ref{fig:MR}, we show the mass-radius relations for several classes of compact stars, including
\begin{itemize}
\item Purely nucleonic stars, obeying the EoS of Ref.~\cite{Akmal:1998cf}.
\item Stars composed of hadronic matter including the effect from kaon
condensation, \textit{cf.~}Ref.~\cite{Glendenning:1998zx}.
\item Stars composed of hadronic matter including nucleons and hyperons, \textit{cf.~}Ref.~\cite{Schulze:2006vw}.
\item Hybrid stars composed mostly of nucleonic matter and a small quark core, corresponding to Case II of the previous section.
\item Hybrid stars composed mostly of quark matter and a small nucleonic crust, corresponding to Case I of the previous section.
\item Strange stars composed of (stable) strange quark matter.
\end{itemize}
These cases cover all the different realistic EoSs of hadronic and/or quark matter at small and moderate densities, and in particular include both the scenarios of stable strange quark matter and a confinement transition to the hadronic phase. Note that the branches for different EoSs in Fig.~\ref{fig:MR} could in principle be extended by allowing stars with higher central densities. These stars are, however, unstable with respect to radial oscillations \cite{Kokkotas:2000up,Glass}, so we do not display them in the figure\footnote{
The exception to this statement would be the so-called
'third family' of compact stars \cite{Gerlach68,Schertler:2000xq}, which could occur
if the matching between nuclear and quark matter takes place
at densities above those reached in the center of the maximum
mass neutron star. In our calculation, this possibility is not realized.}.

\begin{figure}[t]
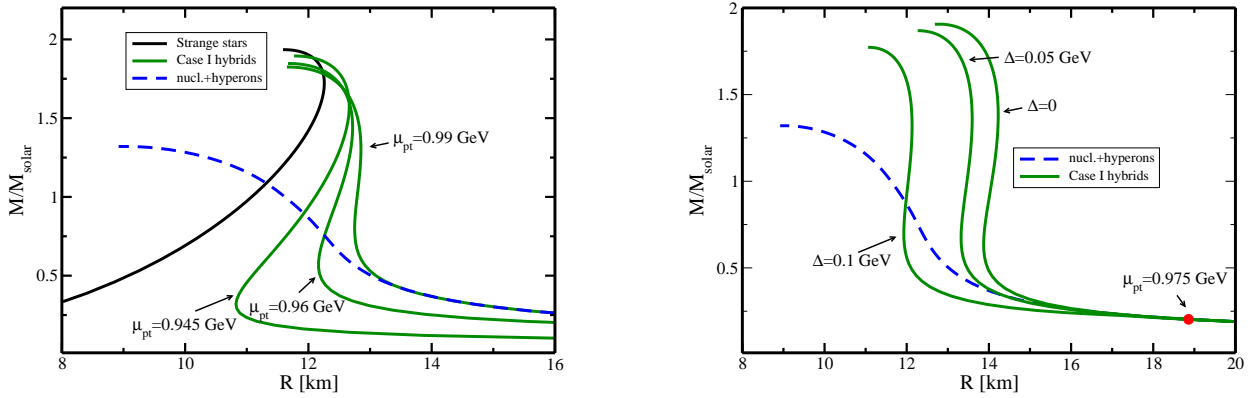

\center
\includegraphics[width=.45\linewidth]{fig11a.eps}
\hfill
\includegraphics[width=.45\linewidth]{fig11b.eps}
\caption{
Left: The effect of the variation of the baryon chemical potential $\mu_\rmi{pt}$, at which the 'case I' matching between the hadronic and quark matter EoSs is performed (with $3\bar\Lambda/(\mu_u+\mu_d+\mu_s)=4$, $\Lambda_{\tiny \msbar}=0.346$ GeV, $m(2 {\rm GeV})=0.07$ GeV). The value $\mu_\rmi{pt}=0.99$ GeV corresponds to the maximal value where the matching is possible. Lowering $\mu_\rmi{pt}$ towards $0.93$ GeV, the result smoothly approaches that of the strange stars, which are marginally stable with respect to neutron stars. Right: The effect of CSC obtained by varying the value of the gap parameter $\Delta$. Relative to unpaired quark matter ($\Delta=0$), CSC generally decreases the radii of stable stars while the masses are hardly affected.}
\label{fig:effectofmuc}
\end{figure}

For the Case I of hybrid stars, the curves shown in Fig.~\ref{fig:MR} corresponds to the maximal value of $\mu$, for which matching is possible using the parameter values $3\bar\Lambda/(\mu_u+\mu_d+\mu_s)=4$, $\Lambda_{\tiny \msbar}=0.378$ GeV and $m(2 {\rm GeV})=0.10$ GeV. Varying these numbers in the usual ranges has a large effect on the radius of the star, while the mass seems to be quite stable. The effect of varying the chemical potential at which the matching is performed is depicted in Fig.~\ref{fig:effectofmuc}.

For the Case II stars, the effects of the above parameter variations are considerably smaller, and our predictions thus more robust. The reason for this is that the structure of the star is most sensitive to the behavior of the EoS at relatively low densities, which in Case II is uniquely determined by the nucleonic EoS. One should, however, note that because we did not perform a detailed matching of the two-component mixture of quark matter and hadrons, including \textit{e.g.~}the effects of quark matter droplets in nuclear matter, our predictions for the radii of the resulting hybrid stars may contain sizable uncertainties. As discussed \textit{e.g.~}in Ref.~\cite{Alford:2004pf}, the masses of hybrid stars we obtain should on the other hand be fairly independent on the details of the matching process, and can thus be considered accurate.

Corresponding to stars containing only strange quark matter ('strange stars'), we show in Figure \ref{fig:MR} the grey shaded regions obtained by varying $\bar{\Lambda}$, $\Lambda_{\tiny \msbar}$ and $m(2 {\rm GeV})$ in the usual ranges.\footnote{Note that stars of this type have been considered in the literature before, using different lower order or resummed perturbative EoSs. See \textit{e.g.~}Refs.~\cite{Fraga:2004gz,Andersen:2002jz} and references therein.} The large size of these areas reflects the sizable uncertainty related to the behavior of the quark matter EoS at low densities. One should note that the maximal strange star masses are of the order of $2.75\ M_\odot$, with radii in excess of 17 km, in principle offering a clear signature for the detection of stable strange quark matter. The general effect of the color superconducting gap $\Delta$ is to somewhat decrease the largest possible radii of the stars, leaving the maximal mass basically unchanged.

In addition to our predictions, we show in Fig.~\ref{fig:MR} observational data corresponding to compact star masses \cite{Freire:2009dr,Weisberg:2002nv,Jacoby:2005qg,Ransom:2005,Champion:2008ge}, which allow one to exclude any EoS that does not pass through all the measurements. In principle, data on compact star radii could provide even more stringent constraints, but these are also much more difficult to obtain, and at present no radius measurement exists that is both precise and unambiguous. Within the next decade, there is nevertheless hope for the emergence of new data in particular for quiescent and isolated neutron stars \cite{Paerels:2009pz} that could improve the situation significantly.

From neutron star mass measurements alone, one can exclude two classes of hadronic EoSs, namely those of condensed kaons and hyperons without phase transitions to quark matter even at very high densities \cite{Glendenning:1998zx,Schulze:2006vw} (see Fig.~\ref{fig:MR}). Interestingly, our calculations for all types of quark matter EoSs (corresponding to both hybrid stars and strange stars) as well as for the purely nucleonic EoS lead to masses that are compatible with current observations. On one hand, this is unfortunate because it does not allow us to discriminate between the different physical scenarios. On the other hand, the fact that our treatment of the quark matter EoS naturally (and without any fine tuning) leads to compact star masses that are in agreement with observations is also rather pleasing.

In light of the marked difference between our results (and hence conclusions) and those of Refs.~\cite{Alford:2004pf,Ozel:2006bv}, some remarks are in order. The presence of a finite strange quark mass leads to a significant stiffening in the EoSs containing quark matter, which is not captured by either the usual MIT bag model or the ansatz of Ref.~\cite{Alford:2004pf}, which can be seen to describe our EoS well only at $\mu\gtrsim 1$ GeV. For this reason, our calculation results in pure strange stars that can be more massive than pure hadronic stars, turning the argument in Ref.~\cite{Ozel:2006bv} on its head (\textit{cf.~}also the discussion in Ref.~\cite{Alford:2006vz}). Also, it seems to us that the largest neutron star masses displayed in Ref.~\cite{Ozel:2006bv} result from nuclear EoSs that can no longer be considered realistic. Therefore, we argue that a confirmed determination of a compact star with a mass in excess of 2 $M_{\odot}$, as suggested in Ref.~\cite{Ozel:2006bv}, would seem to strongly favor --- rather than rule out --- the existence of quark matter in astrophysical objects.

Clearly, our calculations can (and should) be improved in several ways, and also further astrophysical applications of our results are possible. Notably, a detailed calculation of the properties of the two-component system of quark and hadronic matter in hybrid stars can be performed with our quark matter EoS, which would allow us to quantitatively calculate also hybrid star radii. Furthermore, in this work the effects of quark pairing were modeled in a rather crude way, and one could certainly study them in a more detailed and realistic setting. Finally, our results can also be used as a basis for studying other astrophysical observables, such as a compact star's moment of inertia, as well as dynamical processes like compact star oscillations and cooling rates \cite{Lattimer:2006xb}. We hope that our result for the quark matter EoS will lay the foundations for bringing the same rigor to compact star core physics than is today routinely applied to hadronic matter in their crust and surface.

\section{Conclusions and Outlook}

Investigating the properties of cold and dense QCD matter with first principles calculations is a notoriously difficult task. For densities relevant to real world physical systems, the strong coupling constant is not small, thus making perturbation theory converge at best slowly, while lattice QCD simulations are inapplicable due to the famous sign problem. Up to now, most studies of such systems have been performed using phenomenological models, most importantly the MIT bag model, where the effects of the QCD interactions are only visible in a single constant that is added to the grand potential of a system of free quarks and gluons. If one aims at a quantitative description of the physics of cold quark matter, such a crude approach is clearly inadequate.

In the paper at hand, we have tackled the challenge of performing first principles calculations for zero temperature QCD matter using the machinery of perturbation theory. To gain better understanding of the thermodynamics of the system, and in particular of the effects of a finite strange quark mass, we have performed an ${\mathcal O}(\alpha_s^2)$ evaluation of the equation of state of $T=0$ quark matter, keeping both the quark chemical potentials and the strange quark mass nonzero. We feel that such a calculation was long overdue for several reasons. One one hand, the existing ${\mathcal O}(\alpha_s)$ results available for the massive EoS exhibit a sizable dependence on the renormalization scale and offer very few possibilities for analyzing the convergence of the expansion --- issues that can only be addressed by computing the next perturbative order for the quantity. In addition, in most of the practical applications of our results, the density of the matter is not large enough to warrant the approximation $m_s\ll \mu$, thus leaving the door open for significant quark mass effects \cite{Fraga:2004gz}.

Our main result for the EoS of cold, deconfined quark matter with an arbitrary number of massless and one massive quark flavor can be found from Section \ref{sec:result}, and has been analyzed in Section \ref{sec:QCDphaset}. As expected, our result for the 2+1 flavor case smoothly interpolates between the results of two and three massless flavors in the limits of low and high $\mu$, making the quark mass effects clearly visible. The ambiguity related to the choice of the renormalization scale is still sizable, but already at $\mu\sim$ 1 GeV clearly smaller than in the two-loop result. Investigating the effects of the strange quark mass, we find an interesting trend indicating that they become smaller with increasing perturbative order, suggesting that the EoS becomes less sensitive to $m_s$ when interactions are correctly accounted for.

As physical applications of our result, we studied the (mutually exclusive) scenarios of stable strange quark matter as well as a confinement transition between quark matter and hadronic matter. Normal unpaired strange quark matter turned out to be stable only for large values of the renormalization scale $\bar{\Lambda}$, in addition to which one needs to have relatively low values for both the strange quark mass and the $\msbar$ scale parameter $\Lambda_{\tiny \msbar}$. Incorporating the effects of quark pairing by adding to the pressure a term accounting for the condensation energy of the Cooper pairs, we on the other hand found that the constraints on stable strange quark matter were considerably relaxed.

To study the hadronic-to-quark-matter deconfinement transition, we matched our quark matter EoS to various hadronic EoSs, including purely nucleonic matter as well as nuclear matter in the presence of hyperons or kaon condensation. We found the physical requirements for a successful matching to restrict the behavior of the EoS considerably, discovering two distinct density windows where matching was possible: The 'low $\mu$' region $n_B\lesssim 0.39\ {\rm fm}^{-3}$ and the 'high $\mu$' region $n_B\gtrsim 0.64\ {\rm fm}^{-3}$. Interestingly, we were unable to match the quark matter EoS to either the hyperon or kaon condensation EoS in the 'high $\mu$' region. Thus, if we assume these hadronic EoSs to correctly describe nature, we must conclude that the deconfinement transition takes place at or around the density of nuclear matter $n_B\sim 0.16\ {\rm fm}^{-3}$.

Finally, we applied the various EoSs to describe non-rotating compact stars, investigating the mass-radius relations following from the different cases. Using reliable observations for the masses of the stars, we found that it is possible to rule out the cases of hyperon stars as well as stars with kaon condensation, in accordance with earlier findings~\cite{Schulze:2006vw}. Conversely, EoSs with a transition to deconfined quark matter can without any fine-tuning accommodate compact star masses close to 2$M_{\odot}$ --- close to the upper limit of confirmed observations --- and up to 2.75$M_{\odot}$ for pure strange stars. Similar results for hybrid stars were also reported in Ref.~\cite{Alford:2004pf}.

While our evaluation of the ${\mathcal O}(\alpha_s^2)$ EoS for zero temperature quark matter has evidently opened the door for a plethora of physics applications, it is clear in particular from the sizable renormalization scale dependence of our result --- often overwhelming the uncertainties in the experimental values of $m_s$ and $\Lambda_{\tiny \msbar}$ --- that an extension of the calculation to include four loop contributions would be desirable. Although the full $\alpha_s^3$ computation seems quite demanding, a meaningful and considerably more straightforward challenge would be to determine the coefficient of the ${\mathcal O}(\alpha_s^3\ln\,\alpha_s)$ term in the expansion, which originates from the IR sensitive ring diagrams. This term would in fact only require a numerical evaluation of the two-loop gluon polarization tensor, a task of difficulty comparable to the calculation we have performed in this work. In addition to simply extending the region of validity of the current calculation, this result would allow one to apply new optimization schemes for the renormalization scale, thus hopefully significantly decreasing the uncertainty involved in its choice.

Finally, the physics of cold quark matter can naturally also be tackled in various other ways apart from purely perturbative calculations. As we have tried to highlight in Section~\ref{sec:QCDphaset}, the calculation performed here can be used as an ingredient in several models, with an obvious example being the replacement of the simple MIT bag model by our three-loop EoS. To this end, we have attempted to make our results as accessible as possible, presenting them in terms of simple fitting functions above, and making all of them electronically available \cite{URL}.

\section*{Acknowledgments}

We are indebted to Mikko Laine and York Schr\"oder for discussions and advice on the technical aspects of our calculation, and to James Lattimer, Sanjay Reddy and Hans-Josef Schulze for discussions and for providing various hadronic EoSs in a tabulated form. In addition, we would like to acknowledge useful discussions with Mark Alford, David Blaschke, Philippe de Forcrand, Andreas Ipp, Keijo Kajantie, Thomas Kl\"ahn, Pierpaolo Mastrolia, Krishna Rajagopal, Anton Rebhan, J\"urgen Schaffner-Bielich, Andreas Schmitt and Fridolin Weber. AK was supported by the SNF grant 20-122117, PR by the US Department of Energy, grant number DE-FG02-00ER41132, and AV in part by the Austrian Science Foundation, FWF, project No. M1006, as well as the Sofja Kovalevskaja Award of the Humboldt foundation.

\appendix

\section{The One-Loop Gluon Polarization Tensor}
\label{app:pimunu}

Several parts of the derivation of the grand potential, presented in Section \ref{sec:computation}, relied on various properties of the one-loop gluon polarization tensor at zero temperature and finite density. To this end, we will in this first Appendix review what is known about this quantity.

The one-loop gluon polarization tensor is defined by the graphs of Fig.~\ref{fig:ring}a. We divide the function into its vacuum ($T=\mu=0$) and matter (vacuum subtracted) parts through
\ba
\Pi^{\mu\nu}(K)&=&\Pi_{\rmi{vac}}^{\mu\nu}(K)+\Pi_{\rmi{mat}}^{\mu\nu}(K),
\ea
where we have suppressed the trivial color indices, noting that both parts of the tensor are proportional to $\delta_{ab}$. A result that proves quite helpful in evaluating the VM graphs is the simple form, in which one can write the vacuum tensor, evaluated with $N_l$ massless and one massive quark flavor. A straightforward computation leads to the result
\ba
\Pi^{\mu\nu}_{\rmi{vac}}(K)&=&\Pi_{\rmi{vac}}\(\fr{m^2}{K^2}\)\fr{g^2}{(4\pi)^2}\(\fr{\Lambda^2}{K^2}\)^{\!\!\e}\(P_{\mu}P_{\nu}-
K^2\delta_{\mu\nu}\), \label{polarvac}
\ea
where the function $\Pi_{\rmi{vac}}$ can be written in the form
\ba
\Pi_{\rmi{vac}}\(\fr{m^2}{K^2}\)&=&-2^{5-2d}\pi^{(7-d)/2}\bigg\{(3d-2)C_A-2(d-2)N_l\bigg\}\fr{{\rm csc}(\pi d/2)}{\Gamma((d+1)/2)}\nn
&-&2^{6-d}\pi^{(4-d)/2}\Gamma(2-d/2)\int_0^1{\rm d}x (x(1-x))^{d/2-1}\(1+\fr{m^2/K^2}{x(1-x)}\)^{d/2-2}\nn
&\equiv& f_0+f_1\(\fr{m^2}{K^2}\).\label{f0f1}
\ea
Here, $f_0$ and $f_1$ refer to the constant and one-dimensional function appearing on the first and second lines of Eq.~(\ref{f0f1}), respectively.

For the matter part of the tensor, we on the other hand obtain \textit{e.g.~}from Ref.~\cite{Toimela:1984xy}
\begin{eqnarray}
\Pi^{00}_{\rmi{mat}}(k_0,k)&=&-\frac{g^2}{2\pi^2} \sum_{i=1}^{N_f} \int_0^{u_i}dp\,
\frac{ p^2}{\epsilon_p}
\Bigg[1+\frac{4\epsilon_p^2-K^2}{8 p\, k}\,
\ln{\frac{(K^2+2 p k)^2+4 \epsilon_p^2 k_0^2}{(K^2-2 p k)^2+4 \epsilon_p^2 k_0^2}}
\nn
&-&\frac{\epsilon_p k_0}{p\, k}\left(\frac{\pi}{2}-\arctan{
\frac{K^4+4 \epsilon_p^2 k_0^2-4 p^2 k^2}{8\, \epsilon_p \,p\, k\, k_0}}\right)
\Bigg],\\
\(\Pi_\rmi{mat}\)^{\mu}_{\mu}(k_0,k)&=&-2 \frac{g^2}{2\pi^2}\sum_{i=1}^{N_f} \int_0^{u_i}
dp\,\frac{ p^2}{\epsilon_p}
\left[1+\frac{2 m^2-K^2}
{8 p k}
\ln{\frac{(K^2+2 p k)^2+4 \epsilon_p^2 k_0^2}{(K^2-2 p k)^2+4 \epsilon_p^2 k_0^2}}
\right],\nonumber
\end{eqnarray}
where the sum is over $N_l$ massless flavors with $u_i=\mu_i$ and one massive flavor with $u_{N_f}=u$. After a considerable amount of algebra, the remaining integrals in $\Pi^{\mu \nu}_{\rmi{mat}}$ may be evaluated analytically, giving \cite{Toimela:1984xy}
\begin{eqnarray}
\Pi^{00}_{\rmi{mat}}(K,\Phi)&=&-\frac{g^2}{2 \pi^2}\sum_{i=1}^{N_f} \left[\frac{2 \mu u}{3}-\frac{K^2}{6} \sin^2 \Phi\, \ln \frac{\mu +u}{m}
+\frac{\sqrt{4 m^2+K^2}(2 m^2-K^2) \sin^2 \Phi}{24 K}\times\right.\nonumber\\
&&\left.\ln \frac{2 m^4 \cos 2\Phi+(2m^2+K^2)(2 \mu^2-m^2)-2 u \mu K \sqrt{4 m^2 +K^2}}
     {2 m^4 \cos 2\Phi+(2m^2+K^2)(2 \mu^2-m^2)+2 u \mu K \sqrt{4 m^2 +K^2}}\right.\nonumber\\
&&\left.+\frac{\mu}{K \sin\Phi}\left(\frac{\mu^2}{6}-\frac{K^2}{8}\right)
\ln
\frac{4 \mu^2 \cos^2\Phi+\left(K+2 u \sin \Phi\right)^2}
     {4 \mu^2 \cos^2\Phi+\left(K-2 u \sin \Phi\right)^2}\right.\\
&&\left.-\frac{-2 K^2+12 \mu^2+K^2 \cos 2 \Phi}{24 \tan \Phi} \left(\frac{\pi}{2}
-\arctan \frac{2 (2\mu^2-m^2) \cos 2 \Phi+K^2+2 m^2}{4 u \mu \sin 2 \Phi}\right)
\right],\nonumber\\
\(\Pi_\rmi{mat}\)^{\mu}_{\mu}(K,\Phi)&=&-\frac{g^2}{2\pi^2}\sum_{i=1}^{N_f} \left[\mu u-\frac{K^2}{2} \ln \frac{\mu +u}{m}
+\frac{\sqrt{4 m^2+K^2}(2 m^2-K^2)}{8 K}\times\right.\nonumber\\
&&\left.\ln \frac{2 m^4 \cos 2\Phi+(2m^2+K^2)(2 \mu^2-m^2)-2 u \mu K \sqrt{4 m^2 +K^2}}
     {2 m^4 \cos 2\Phi+(2m^2+K^2)(2 \mu^2-m^2)+2 u \mu K \sqrt{4 m^2 +K^2}}\right.\nonumber\\
&&\left.+\frac{\mu(2m^2-K^2)}{4K \sin\Phi}
\ln
\frac{4 \mu^2 \cos^2\Phi+\left(K+2 u \sin \Phi\right)^2}
     {4 \mu^2 \cos^2\Phi+\left(K-2 u \sin \Phi\right)^2}\right.\nn
&&\left.+\frac{K^2-2m^2}{4 \tan\Phi}
\left(\frac{\pi}{2}-\arctan \frac{2 (2\mu^2-m^2) \cos 2 \Phi+K^2+2 m^2}{4 u \mu \sin 2 \Phi}\right)
\right],
\end{eqnarray}
where $\Phi\equiv{\rm arctan}\frac{k}{k_0}$, and in the sum it should be understood that $m=0$ for the $N_l$ light flavors. It is also useful to consider the limits
\begin{eqnarray}
\lim_{K\rightarrow \infty}\Pi^{00}_{\rmi{mat}}(K,\Phi)&=&-\frac{g^2}{2\pi^2}\frac{\sin^2\Phi}{3 K^2}\sum_{i=1}^{N_f}
\left(\mu u (m^2+2 \mu^2)-3 m^4 \ln\frac{\mu+u}{m}\right),\\
\lim_{K\rightarrow 0}\Pi^{00}_{\rmi{mat}}(K,\Phi)&=&-\frac{g^2}{2\pi^2}\sum_{i=1}^{N_f}\left(\mu u-\mu^2 \cot \Phi \arctan \frac{u}{\mu \cot \Phi}\right),\\
\lim_{K\rightarrow \infty}\(\Pi_\rmi{mat}\)^{\mu}_{\mu}(K,\Phi)&=&-\frac{g^2}{2\pi^2}\sum_{i=1}^{N_f}\left(\frac{\mu u (7m^2+2 \mu^2)}{12 K^2}
+\frac{4}{3}\frac{\mu u^3 \cos 2\Phi}{K^2}-\frac{3}{4}\frac{m^4}{K^2} \ln \frac{\mu +u}{m}\right),\\
\lim_{K\rightarrow 0}\(\Pi_\rmi{mat}\)^{\mu}_{\mu}(K,\Phi)&=&-\frac{g^2}{2\pi^2}\sum_{i=1}^{N_f}\left(\mu u-m^2 \cot \Phi \arctan \frac{u}{\mu \cot \Phi}\right),
\end{eqnarray}
which can be used show that the plasmon ring sum of Eq.~(\ref{totringsum}) is both UV and IR finite.

Finally, we note that upon separation of the contributions of the massless and massive quark flavors to $\Pi^{\mu \nu}_{\rmi{mat}}$, one can decompose the combinations $F$ and $G$, defined in Eq.~(\ref{FGdef}), in the forms
\ba
-4 \pi^2 g^{-2} G_\rmi{mat}(K,\Phi)&=&\mu^2 G_{h}(\hat{K},\Phi,\hat{m})+\sum_{i=1}^{N_l}
\mu_i^2 G_{l}\(\frac{K}{\mu_i},\Phi\),\nonumber\\
-2 \pi^2 g^{-2} F_\rmi{mat}(K,\Phi)&=&\mu^2 F_{h}(\hat{K},\Phi,\hat{m})+\sum_{i=1}^{N_l} \mu_i^2 F_{l}\(\frac{K}{\mu_i},\Phi\),
\ea
where $\hat K \equiv K/\mu$, $G_l(0,\Phi)=\frac{\Phi\cot \Phi-\cos^2\Phi}{\sin^2\Phi}$ and $F_l(0,\Phi)=\frac{1-\Phi \cot \Phi}{\sin^2\Phi}$ (\textit{cf.~}the Appendices of Ref.~\cite{Freedman:1976ub}).

\section{The Massive 2GI Graphs}
\label{app:m2GI}

In this Appendix, we will go through the evaluation of the massive 2GI graphs of Section~\ref{sec:2gi} in detail, filling in the pieces missing from our earlier discussion.

\subsection{Scalarization}

\begin{figure}[t]

\centerline{\epsfxsize=13cm \epsfbox{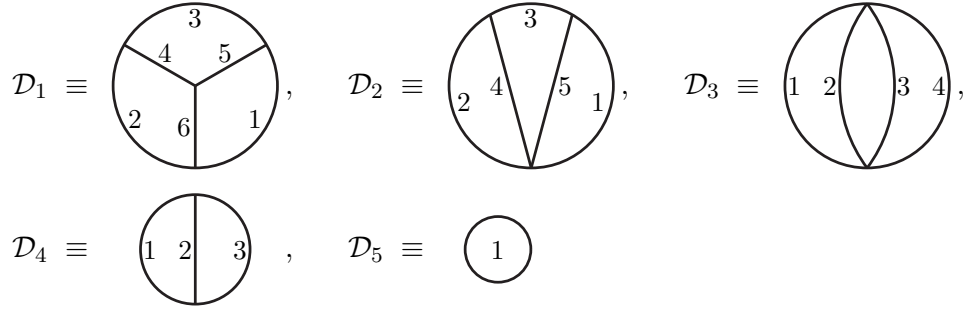}}

\caption[a]{The topologies resulting from the scalarization procedure. Our notation for the functions ${\mathcal D}_i$ from Eqs.~(\ref{A1}) onwards is such that a lower index at position $k$ indicates the power of the propagator $k$ in the figure, while the corresponding upper index gives the mass of the line in question, with $c$ indicating that the line has been cut. For the uncut graphs, the chemical potential is present in each massive propagator, but not in the massless ones.
    \label{fig:topos}
    }
\end{figure}

The first step in evaluating the diagrams a--d of Fig.~\ref{fig:graphs} is to perform all the Lorentz and color algebra associated with them and write them in terms of the scalar topologies $\mathcal{D}_1$--$\mathcal{D}_5$ of Fig.~\ref{fig:topos}. The procedure can be automatized with the program FORM \cite{Vermaseren:2000nd}, and leads to the results
\ba
D_a&=&d_A g_\rmi{B}^2\Bigg\{2m^2
\mathcal{D}_4
\left(
\begin{array}{ccc}
m & m & 0  \\
1 & 1 & 1
\end{array}
\right)
-(1-\e)
\Bigg(\mathcal{D}_5
\left(
\begin{array}{c}
m  \\
1
\end{array}
\right)
\Bigg)^2\Bigg\},  \label{A1}\\
D_b&=&d_AC_Ag^4\Bigg\{2m^2
\mathcal{D}_2
\left(
\begin{array}{ccccc}
m & m & m & 0 & 0  \\
1 & 1 & 1 & 1 & 1
\end{array}
\right)
+(1-\e)
\mathcal{D}_3
\left(
\begin{array}{cccc}
m & m & 0 & 0  \\
1 & 1 & 1 & 1
\end{array}
\right)\nn
&-&2(1-\e)
\mathcal{D}_4
\left(
\begin{array}{ccc}
m & m & 0  \\
1 & 1 & 1
\end{array}
\right)
\mathcal{D}_5
\left(
\begin{array}{c}
m  \\
1
\end{array}
\right)\Bigg\},\\
D_c&=&d_AC_F g^4\Bigg\{
-2(1-\e)^2
\mathcal{D}_1
\left(
\begin{array}{cccccc}
m & m & m & 0 & 0 & 0 \\
1 & 1 & 1 & 1 & 1 & -1
\end{array}
\right)
+4m^2
\mathcal{D}_2
\left(
\begin{array}{ccccc}
m & m & m & 0 & 0  \\
1 & 1 & 1 & 1 & 1
\end{array}
\right)\nn
&-&8m^4
\mathcal{D}_2
\left(
\begin{array}{ccccc}
m & m & m & 0 & 0  \\
1 & 1 & 2 & 1 & 1
\end{array}
\right)
-2(1-\e)^2
\mathcal{D}_3
\left(
\begin{array}{cccc}
m & m & 0 & 0   \\
1 & 1 & 1 & 1
\end{array}
\right)\nn
&+&4(1-\e)^2
\mathcal{D}_4
\left(
\begin{array}{ccc}
m & m & 0  \\
1 & 1 & 1
\end{array}
\right)
\mathcal{D}_5
\left(
\begin{array}{c}
m  \\
1
\end{array}
\right)
+8(1-\e)m^2
\mathcal{D}_4
\left(
\begin{array}{ccc}
m & m & 0  \\
2 & 1 & 1
\end{array}
\right)
\mathcal{D}_5
\left(
\begin{array}{c}
m  \\
1
\end{array}
\right)\nn
&-&2(1-\e)^2
\Bigg(\mathcal{D}_5
\left(
\begin{array}{c}
m  \\
1
\end{array}
\right)
\Bigg)^2
\mathcal{D}_5\left(
\begin{array}{c}
m  \\
2
\end{array}
\right)\Bigg\},\\
D_d&=& d_A(C_F-\frac{1}{2}C_A)g^4 \Bigg\{-4m^4 \mathcal{D}_1
\left(\begin{array}{cccccc}
0 & m & m & 0 & m & m \\
1 & 1 & 1 & 1 & 1 & 1
\end{array}
\right)\nn
&+&2(1-\e)\mathcal{D}_1
\left(
\begin{array}{cccccc}
0 & m & m & 0 & m & m \\
-1& 1 & 1 & 1 & 1 & 1
\end{array}
\right)
+4m^2 \mathcal{D}_2
\left(
\begin{array}{ccccc}
m & m & m & 0 & 0 \\
1 & 1 & 1 & 1 & 1
\end{array}
\right)\nn
&-&4m^2\e\mathcal{D}_2
\left(
\begin{array}{ccccc}
m & m & 0 & m & m\\
1& 1 & 1 & 1 & 1
\end{array}
\right)
-2\e(1-\e)\mathcal{D}_3
\left(
\begin{array}{cccc}
m & m & 0 & 0 \\
1 & 1 & 1 & 1
\end{array}
\right)\nn
&+&(2-\e-\e^2)\mathcal{D}_3
\left(
\begin{array}{cccc}
m & m & m & m \\
1 & 1 & 1 & 1
\end{array}
\right)
-8(1-\e) \mathcal{D}_4
\left(
\begin{array}{ccc}
m & m & 0  \\
1 & 1 & 1
\end{array}
\right)
\mathcal{D}_5
\left(
\begin{array}{c}
m  \\
1
\end{array}
\right)\Bigg\}, \label{A4}
\ea
where the appearance of the $g_\rmi{B}$ in $D_a$ reminds us that it should eventually be replaced by the renormalized coupling constant.

In the above expressions, we have used a compact notation, in which the functions $\mathcal{D}_1,\,\mathcal{D}_2,\, \mathcal{D}_3,\, \mathcal{D}_4$ and $\mathcal{D}_5$ refer to the scalar topologies of Fig.~\ref{fig:topos}. In each case, a lower index at position $k$ indicates the power of the propagator $k$ in the figure, and the corresponding upper index gives the mass of the line in question, so that \textit{e.g.~}the two-loop graph of Fig.~\ref{fig:topos} with two massive and one massless lines, with the massless one raised to power two, would read
\ba
\mathcal{D}_4
\left(
\begin{array}{ccc}
m & m & 0  \\
1 & 1 & 2
\end{array}
\right)&=& \int_{-\infty+i\mu}^{-\infty+i\mu}{\rm d}p_0\int\fr{{\rm d}^{d-1} p}{(2\pi)^{d-1}}\int_{-\infty+i\mu}^{-\infty+i\mu}{\rm d}q_0\int\fr{{\rm d}^{d-1} q}{(2\pi)^{d-1}}\nn
&\times& \fr{1}{(p_0^2+p^2+m^2)(q_0^2+q^2+m^2)((p_0-q_0)^2+(\mbox{\textbf{p}}-\mbox{\textbf{q}})^2)^2}.
\ea

The chemical potential is naturally present in each massive propagator, but not in the massless ones, and therefore the massive scalar lines carrying $\mu$ will be referred to as 'fermionic' below. Each scalar graph we encounter can be viewed as containing exactly one solid fermion loop, the orientation of which we are free to choose. We make the choice of following the numbering of the propagators in Fig.~\ref{fig:topos}, so that $\mu$ is always flowing in the direction of increasing propagator number.

\subsection{Cutting of the Graphs}

Next, we proceed to apply the cuts, introduced in Sec.~\ref{sec:2gi}, to the above 2GI diagrams, and list the results obtained after performing one, two, and three cuts on the graphs. The cutting should be thought of just a bookkeeping tool, helping us to keep track of the various contributions to the grand potential. The correctness of the procedure can be verified by explicitly performing all the energy integrations in the graphs.

\subsubsection{Single-cut Diagrams}

Cutting always exactly one of the fermionic lines in Eqs.~(\ref{A1})--(\ref{A4}), we obtain the following results, where the symbol '$c$' indicates that the corresponding massive line in the diagram has been cut:
\ba
D^{1c}_a &=& -d_A g_\rmi{B}^2\frac{2\(3-\e(5-2\e)\)}{1-2\e}\,\mathcal{D}_5
\!\left(
\begin{array}{c}
m  \\
1
\end{array}
\right), \label{D1ca} \\
D^{1c}_b &=& d_AC_A g^4\Bigg\{\frac{2(3-12\e+14\e^2-4\e^3)}{(1-2\e)^2}\mathcal{D}_3\!\left( \begin{array}{cccc}m & m & 0 & 0  \\ c & 1 & 1 & 1 \end{array} \right) \nn
&+&\frac{12(1-\e)^3}{(1-2\e)^2m^2}\Bigg(\,\mathcal{D}_5 \!\left( \begin{array}{c} m  \\ 1 \end{array} \right)\Bigg)^2\Bigg\},\label{D1cb}\\
D^{1c}_c&=&d_AC_F g^4\Bigg\{-\frac{2(3-2\e)(1+\e-4\e^2)}{1-4\e}\mathcal{D}_3\!\left( \begin{array}{cccc}m & m & 0 & 0  \\ c & 1 & 1 & 1 \end{array} \right)\nn
&+&\frac{(1-\e)^2(3-2\e)(7-14\e+8\e^2)}{(1-2\e)^2m^2}\Bigg(\,\mathcal{D}_5 \!\left( \begin{array}{c} m  \\ 1 \end{array} \right)\Bigg)^2 \nn
&+&8m^4 \(\fr{\partial}{\partial \widetilde{m}^2}
\mathcal{D}_2
\left(
\begin{array}{ccccc}
m & m & \widetilde{m} & 0 & 0  \\
1 & 1 & c & 1 & 1
\end{array}
\right)\)\nn
&-&8(1-\e)m^2\(\fr{\partial}{\partial \widetilde{m}^2}
\mathcal{D}_4
\left(
\begin{array}{ccc}
\widetilde{m} & m & 0  \\
c & 1 & 1
\end{array}
\right)\)
\mathcal{D}_5
\left(
\begin{array}{c}
m  \\
1
\end{array}
\right)\nn
&+&\frac{2(3-5\e+2\e^2)^2}{(1-2\e)^2}\Bigg(\,\mathcal{D}_5 \!\left( \begin{array}{c} m  \\ 1 \end{array} \right)\Bigg)^2\fr{\partial}{\partial \widetilde{m}^2}\Bigg\},\label{D1cc}\\
D^{1c}_d &=& d_A(C_F-\fr{1}{2}C_A)g^4\Bigg\{-\frac{(1+\e)(1-2\e)(2+\e-2\e^2)}{\e^2} \mathcal{D}_3\!\left( \begin{array}{cccc}m & m & m & m  \\ c & 1 & 1 & 1 \end{array} \right)\nn
&+&\fr{4 (2 - 11 \e + 15 \e^2 + \e^3 - 14 \e^4 + 8 \e^5)}{\e (1 - 2 \e) (1 - 4 \e) }\mathcal{D}_3\!\left( \begin{array}{cccc}m & m & 0 & 0  \\ c & 1 & 1 & 1 \end{array} \right)\nn
&-&\fr{3(1-\e)^3(1+2\e)(1-4\e)}{\e^2(1-2\e)^2m^2}\Bigg(\,\mathcal{D}_5 \!\left( \begin{array}{c} m  \\ 1 \end{array} \right)\Bigg)^2
   \Bigg\}.\label{D1cd}
\ea
The terms with a mass derivative originate from integrals with squared lines, and in them it is assumed that $\widetilde{m}$ is set equal to $m$ after performing the differentiation. In the last term of Eq.~(\ref{D1cc}), the derivative is acting only on the integration measure of the external \textbf{p} integration. It should be recalled that in all the integrals appearing on the right hand side of these equations (including $\mathcal{D}_5$), the chemical potential has been set to zero.

In deriving the above expressions, we have used the symmetry of the graphs in equating terms where a distinct, but topologically equivalent propagator has been cut. We have also used the FIRE algorithm \cite{Smirnov:2008iw} to simplify the final result by finding relations between the various two-point functions originating from the cuts.

\subsubsection{Double-cut Diagrams}

Continuing to use the same notation as above, we obtain for the double-cut diagrams
\ba
D^{2c}_a &=& d_Ag_\rmi{B}^2\Bigg\{\fr{2m^2}{(P-Q)^2} -1+\e\Bigg\},\label{D2ca}\\
D^{2c}_b &=& d_AC_Ag^4\Bigg\{2m^2\mathcal{D}_2
\left(
\begin{array}{ccccc}
m & m & m & 0 & 0 \\
c & c & 1 & 1 & 1
\end{array}
\right)
+(1-\e)\,\mathcal{D}_3\!\left( \begin{array}{cccc}m & m & 0 & 0  \\ c & c & 1 & 1 \end{array} \right) \nn &-&\frac{2(1-\e)}{1-2\e}\(\fr{3-2\e}{(P-Q)^2}-\fr{2(1-\e)}{m^2}\)\,\mathcal{D}_5 \!\left( \begin{array}{c} m  \\ 1 \end{array} \right)\Bigg\},\label{D2cb}\\
D^{2c}_c&=&d_AC_Fg^4\Bigg\{2\(2(1-2\e)m^2-(1-\e)^2(P-Q)^2\)\,\mathcal{D}_2
\left(
\begin{array}{ccccc}
m & m & m & 0 & 0 \\
c & c & 1 & 1 & 1
\end{array}
\right)\nn
&-&2\((1-\e)^2-\fr{4(1-2\e)m^2}{(P-Q)^2}\)\mathcal{D}_3\!\left( \begin{array}{cccc}m & m & 0 & 0  \\ c & c & 1 & 1 \end{array} \right)\nn
&-&\fr{2(1-\e)}{1-2\e}\Bigg(\fr{(1-\e)(3-2\e)\e}{m^2}+\fr{4}{(P-Q)^2}\nn
&-&2(3-2\e)\bigg(1-\e-\fr{2m^2}{(P-Q)^2}\bigg)\fr{\partial}{\partial \widetilde{m}^2}\Bigg) \mathcal{D}_5 \!\left( \begin{array}{c} m  \\ 1 \end{array} \right)\nn
&+&16m^4 \(\fr{\partial}{\partial \widetilde{m}^2}
\mathcal{D}_2
\left(
\begin{array}{ccccc}
m & m & \widetilde{m} & 0 & 0  \\
c & 1 & c & 1 & 1
\end{array}
\right)\)\nn \label{D2cc}
&-&8(1-\e)m^2\Bigg[\(\fr{\partial}{\partial \widetilde{m}^2}
\mathcal{D}_4
\left(
\begin{array}{ccc}
\widetilde{m} & m & 0  \\
c & c & 1
\end{array}
\right)\)
\mathcal{D}_5
\left(
\begin{array}{c}
m  \\
1
\end{array}
\right)+\(\fr{\partial}{\partial \widetilde{m}^2}
\mathcal{D}_4
\left(
\begin{array}{ccc}
\widetilde{m} & m & 0  \\
c & 1 & 1
\end{array}
\right)\)\Bigg]\Bigg\},\\
D^{2c}_d &=& d_A(C_F-\fr{1}{2}C_A)g^4\Bigg\{\nn
&-&4m^2\Bigg[m^2\mathcal{D}_1
\left(
\begin{array}{cccccc}
m & m & 0 & m & m & 0 \\
c & 1 & 1 & c & 1 & 1
\end{array}
\right)
+m^2\mathcal{D}_1
\left(
\begin{array}{cccccc}
m & m & 0 & m & m & 0 \\
1 & c & 1 & 1 & c & 1
\end{array}
\right)\nn
&-&
\mathcal{D}_2
\left(
\begin{array}{ccccc}
m & m & m & 0 & 0 \\
c & c & 1 & 1 & 1
\end{array}
\right)\Bigg]\nn
&-&\Bigg(\fr{2-6\e+3\e^2+\e^3}{\e}-\fr{4(1-2\e)m^2}{(P-Q)^2}\Bigg)\Bigg[
\mathcal{D}_3\!\left( \begin{array}{cccc}m & m & m & m  \\ c & 1 & c & 1 \end{array} \right)+\mathcal{D}_3\!\left( \begin{array}{cccc}m & m & m & m  \\ 1 & c & 1 & c \end{array} \right)\Bigg]\nn
&+&\fr{1}{2\e}\Bigg(\fr{8m^4}{(P-Q)^2(P+Q)^2}+\fr{8(1-\e)^2m^2}{(P+Q)^2}
+(2-3\e+3\e^2-2\e^3)\Bigg)\nn
&\times&\Bigg[\mathcal{D}_3\!\left( \begin{array}{cccc}m & m & m & m  \\ c & c & 1 & 1 \end{array} \right)+\mathcal{D}_3\!\left( \begin{array}{cccc}m & m & m & m  \\ c & 1 & 1 & c \end{array} \right)\nn
&+&\mathcal{D}_3\!\left( \begin{array}{cccc}m & m & m & m  \\ 1 & c & c & 1 \end{array} \right)+\mathcal{D}_3\!\left( \begin{array}{cccc}m & m & m & m  \\ 1 & 1 & c & c \end{array} \right)\Bigg]\nn
&-&2(1-\e)\e\,\mathcal{D}_3\!\left( \begin{array}{cccc}m & m & 0 & 0  \\ c & c & 1 & 1 \end{array} \right)  \nn
&+&   \fr{4(1-\e)}{(1-2\e)\e m^2}\Bigg(\fr{4(1+\e)(1-2\e)m^4}{(P-Q)^2(P+Q)^2}-\fr{4\e^2 m^2}{(P-Q)^2}
+\fr{(4-\e)(1-2\e)m^2}{(P+Q)^2}\nn
&+&2\e(1-\e)\Bigg) \mathcal{D}_5 \!\left( \begin{array}{c} m  \\ 1 \end{array} \right)\Bigg\}.\label{D2cd}
\ea
In terms where a squared propagator has been cut, it is understood that the momentum flowing along that line is $P$, so the external mass derivative on the fourth line of Eq.~(\ref{D2cc}) only acts on the external \textbf{p} integration measure.

\subsubsection{Triple-cut Diagrams}

For three cuts, the two-loop graph in Fig.~\ref{fig:graphs}a naturally vanishes, so we are only left with the diagrams of Fig.~\ref{fig:graphs}b--d. For these, we obtain
\ba
D_b^{3c}&=&d_AC_Ag^4\Bigg\{
\fr{2m^2}{(P-Q)^2(P-R)^2}
-\fr{2(1-\e)}{(P-Q)^2}\Bigg\},\\
D_c^{3c}&=&d_AC_Fg^4\Bigg\{
-\fr{2(1-\e)^2(Q-R)^2}{(P-Q)^2(P-R)^2}
+\fr{4m^2+8m^4\fr{\partial}{\partial \widetilde{m}^2}}{(P-Q)^2(P-R)^2}\nn
&+&\fr{4(1-\e)^2-8(1-\e)m^2\fr{\partial}{\partial \widetilde{m}^2}}{(P-Q)^2}
+2(1-\e)^2 \fr{\partial}{\partial \widetilde{m}^2}\nn
&+&8m^4 \(\fr{\partial}{\partial \widetilde{m}^2}
\mathcal{D}_2
\left(
\begin{array}{ccccc}
m & m & \widetilde{m} & 0 & 0  \\
c & c & c & 1 & 1
\end{array}
\right)\)\nn
&-&8(1-\e)m^2\(\fr{\partial}{\partial \widetilde{m}^2}
\mathcal{D}_4
\left(
\begin{array}{ccc}
\widetilde{m} & m & 0  \\
c & c & 1
\end{array}
\right)\)\Bigg\},\\
D_d^{3c}&=& d_A(C_F-\frac{1}{2}C_A) g^4 \Bigg\{
\fr{-16m^4}{(P-Q)^2(P-R)^2((P-Q-R)^2+m^2)}\nn
&+&\fr{8(1-\e)(P-Q)^2}{(P-R)^2((P-Q-R)^2+m^2)}
-\fr{16\e m^2}{(P-Q)^2((P-Q-R)^2+m^2)}\nn
&+&\fr{4m^2}{(P-Q)^2(P-R)^2}
+\fr{4(2-\e-\e^2)}{(P-Q-R)^2+m^2}
-\fr{8(1-\e)}{(P-Q)^2}\Bigg\},
\ea
where it is again understood that whenever a squared propagator has been cut, the momentum flowing along that line is chosen as $P$.

\subsection{Collecting Results}

Simplifying the above formulae with the help of Eqs.~(\ref{2GIfirstdef}) and (\ref{2GIsecdefinition}) as well as Appendices \ref{app:idefs} and \ref{app:4dints}, we finally arrive at the following expressions for the one-, two-, and three-cut parts of the 2GI graphs:
\ba
\fr{\Omega_\rmi{2GI}^{m,1c}}{V}&=&d_A m^2 \(\fr{6}{\e}+8\)\(\fr{\bar\Lambda^2}{m^2}\)^{\e}I_1\,\fr{g^2(\bar{\Lambda})}{(4\pi)^2}\nn
&-&d_A m^2\Bigg\{C_A\Bigg[\fr{16}{\e^2}+\fr{10}{3}\(6\ln\fr{\bar\Lambda}{m}+1\)\fr{1}{\e}\nn
&-&\(4\ln\fr{\bar\Lambda}{m}+\fr{136}{3}\)\ln\fr{\bar\Lambda}{m}-\fr{82}{3}+\pi^2\(\fr{41}{6}-8\ln\,2\)+12\zeta(3)\Bigg]I_1\nn
&+&C_F\Bigg[\bigg\{\fr{18}{\e^2}-\fr{3}{2\e}
-6\(12\ln\fr{\bar\Lambda}{m}+5\)\ln\fr{\bar\Lambda}{m}-\fr{313}{4}\nn
&-&\pi^2\(\fr{35}{3}-16\ln\,2\)-24\zeta(3)\bigg\}I_1\nn
&+&m^2 \bigg\{\fr{18}{\e^2}-36\(2\ln\fr{\bar\Lambda}{m}+3\)\ln\fr{\bar\Lambda}{m}-32\bigg\}I_{1b}\Bigg] \label{lnZ2GI1c}\\
&-&N_f\Bigg[\fr{4}{\e^2}+8\bigg(\ln\fr{\bar\Lambda}{m}+\fr{2}{3}\bigg)\fr{1}{\e}
+8\(\ln\fr{\bar\Lambda}{m}+\fr{4}{3}\)\ln\fr{\bar\Lambda}{m}+\fr{32}{3}+\fr{\pi^2}{3}\Bigg]I_1\Bigg\}\,\fr{g^4(\bar{\Lambda})}{(4\pi)^4},\nn
\fr{\Omega_\rmi{2GI}^{m,2c}}{V}&=& -d_A\Big\{2m^2 I_2-I_1^2\Big\}g^2(\bar{\Lambda})\nn
&-&d_A \Bigg\{ C_A\left[
\left(\frac{5}{3}I_1^2-\frac{10}{3}m^2 I_2\right)/\epsilon + I_{10}
+ (-4I_1^2+8 m^2 I_2)\ln\fr{\bar\Lambda}{m}\right]\nonumber\nn
&+&C_F\left[I_{11}+\left[24(m^2 I_2 -m^2 I_{1b}I_1+2m^4 I_{2b})+48 m^4I_8\right]\ln\fr{\bar\Lambda}{m}\right]\nn
&+&N_f \Bigg[\left(-\frac{2}{3}I_1^2+\frac{4}{3}m^2 I_2\right)/\epsilon + \frac{2}{3}I_1^2\Bigg]\Bigg\}\,\fr{g^4(\bar{\Lambda})}{(4\pi)^2}, \label{lnZ2GI2c}\\
\fr{\Omega_\rmi{2GI}^{m,3c}}{V}&=& d_A \Bigg\{C_A\bigg[2I_1 I_2-4I_5+8m^4I_6-4I_7\bigg]\nn
&+&C_F\Bigg[2I_1^2I_{1b}-4I_1I_2-8m^2I_1I_{2b}+8m^2I_3+8m^4I_{3b}-2I_4+8I_5-16m^4I_6\nn
&+&8I_7-8m^2I_1I_8+8m^4I_9\Bigg]\Bigg\}g^4(\bar{\Lambda}).\label{lnZ2GI3c}
\ea
In these expressions, the coupling $g$ and the quark mass $m$ now correspond to the physical, renormalized quantities, and the $I_n$'s again denote integrals defined in Appendix \ref{app:idefs}.

\section{The VM Graph}
\label{app:VMgraph}

Here, we go through the evaluation of the vacuum-matter diagrams $\Omega_\rmi{VM}^m$ and $\Omega_\rmi{VM}^x$ in detail. Starting from the massive VM graph $\Omega_\rmi{VM}^m$, we first perform the Lorenz algebra and obtain
\ba
\fr{\Omega_\rmi{VM}^m}{V}&=&-2d_A\fr{g^4}{(4\pi)^2}\Lambda^{2\e}\Bigg\{2m^2
\(f_0\,\mathcal{D}_4
\left(
\begin{array}{ccc}
m & m & 0  \\
1 & 1 & 1+\e
\end{array}
\right)+\widetilde{\mathcal{D}_4}
\left(
\begin{array}{ccc}
m & m & 0  \\
1 & 1 & 1+\e
\end{array}
\right)\)\nn
&-&(1-\e)
\(f_0\,\mathcal{D}_4
\left(
\begin{array}{ccc}
m & m & 0  \\
1 & 1 & \e
\end{array}
\right)+\widetilde{\mathcal{D}_4}
\left(
\begin{array}{ccc}
m & m & 0  \\
1 & 1 & \e
\end{array}
\right)\) \nn
&+&2(1-\e)
\widetilde{\mathcal{D}_5}
\left(
\begin{array}{c}
0  \\
1+\e
\end{array}
\right)\mathcal{D}_5
\left(
\begin{array}{c}
m  \\
1
\end{array}
\right)
\Bigg\},
\ea
where the tilde implies that the integrand in the scalar graph has been multiplied by the function $f_1$ prior to integration over $P$ and $Q$. The four-momentum appearing in the argument of $f_1$ is naturally that of the bosonic propagator in the graph.

The evaluation of the above integrals proceeds the same fashion as before, \textit{i.e.~}by applying zero to two cuts in the massive fermion lines. Dropping the $\mu$ independent zero-cut term, we obtain at the one-cut level
\ba
\fr{\Omega_\rmi{VM}^{m,1c}}{V}&=&-4d_A\fr{g^4}{(4\pi)^2}\Lambda^{2\e}\Bigg\{2m^2
\(f_0\,\mathcal{D}_4
\left(
\begin{array}{ccc}
m & m & 0  \\
c & 1 & 1+\e
\end{array}
\right)+\widetilde{\mathcal{D}_4}
\left(
\begin{array}{ccc}
m & m & 0  \\
c & 1 & 1+\e
\end{array}
\right)\)\nn
&-&(1-\e)
\(f_0\,\mathcal{D}_4
\left(
\begin{array}{ccc}
m & m & 0  \\
c & 1 & \e
\end{array}
\right)+\widetilde{\mathcal{D}_4}
\left(
\begin{array}{ccc}
m & m & 0  \\
c & 1 & \e
\end{array}
\right)\) \nn
&+&(1-\e)
\widetilde{\mathcal{D}_5}
\left(
\begin{array}{c}
0  \\
1+\e
\end{array}
\right)\Bigg\},
\ea
and for two cuts
\ba
\fr{\Omega_\rmi{VM}^{m,2c}}{V}&=&-2d_A\fr{g^4}{(4\pi)^2}\Lambda^{2\e}\Bigg\{2m^2
\(\fr{f_0}{((P-Q)^2)^{1+\e}}+\fr{1}{(P-Q)^2}\widetilde{\mathcal{D}_4}
\left(
\begin{array}{ccc}
m & m & 0  \\
c & c & \e
\end{array}
\right)\)\nn
&-&(1-\e)
\(\fr{f_0}{((P-Q)^2)^{\e}}+\widetilde{\mathcal{D}_4}
\left(
\begin{array}{ccc}
m & m & 0  \\
c & c & \e
\end{array}
\right)\)\Bigg\}. \label{pvmmres}
\ea
To complete the evaluation of Eq.~(\ref{VMmfirstdef}), we now finally plug in the values of the various integrals from Appendix \ref{app:4dints}, arriving at
\ba
\fr{\Omega_\rmi{VM}^{m,1c}}{V}&=&d_A m^2\Bigg\{C_A\Bigg[\fr{5}{\e^2}+\bigg(20\,\ln\fr{\bar\Lambda}{m}
+\fr{39}{2}\bigg)\fr{1}{\e}\nn
&+&2\(20\,\ln\fr{\bar\Lambda}{m}+39\)\ln\fr{\bar\Lambda}{m}+\fr{261}{4}+\fr{25\pi^2}{6}\Bigg]I_1\nn
&-&N_f\Bigg[\fr{2}{\e^2}+\bigg(8\,\ln\fr{\bar\Lambda}{m}+7\bigg)\fr{1}{\e}
+4\(4\,\ln\fr{\bar\Lambda}{m}+7\)\ln\fr{\bar\Lambda}{m}+\fr{45}{2}+\fr{5\pi^2}{3}\Bigg]I_1\nn
&-&4\bigg[3-\pi^2\bigg]I_1\Bigg\}\, \fr{g^4(\bar \Lambda )}{(4\pi)^4}, \label{lnZVMm1c} \\
\fr{\Omega_\rmi{VM}^{m,2c}}{V}&=&d_A\Bigg\{C_A\Bigg[\left(\frac{5}{3}I_1^2-\frac{10}{3}m^2 I_2\right)\fr{1}{\e}+\left[\frac{10}{3}I^2_1-\frac{20}{3}m^2 I_2\right]\ln\frac{\bar{\Lambda}}{m}\nn
&+&\frac{16}{9}I_1^2-\frac{62}{9}m^2 I_2 - \frac{5}{3}I_{1c} +\frac{10}{3}m^2  I_{2c}\Bigg]\nn
&-&N_f\Bigg[\left(\frac{2}{3}I_1^2-\frac{4}{3}m^2 I_2\right)\fr{1}{\e}+\left[\frac{4}{3}I_1^2-\frac{8}{3}m^2 I_2\right]\ln\frac{\bar{\Lambda}}{m}\nn
&+&\frac{4}{9}I_1^2-\frac{20}{9}m^2 I_2-\frac{2}{3}I_{1c}+\frac{4}{3}m^2 I_{2c}-\frac{2}{3}I_{12}
\Bigg\}\, \fr{g^4(\bar \Lambda )}{(4\pi)^2}. \label{lnZVMm2c}
\ea

Next, we look at the cross term $\Omega_\rmi{VM}^x$. Proceeding along the same lines as with $\Omega_\rmi{VM}^m$, we write
\ba
\Omega_\rmi{VM}^x &=&\Omega_\rmi{VM}^{x,1c}+\Omega_\rmi{VM}^{x,2c},
\ea
in which the first function is seen to vanish, while the second is given by
\ba
\fr{\Omega_\rmi{VM}^{x,2c}}{V}&=& 2(1-\e)d_A\fr{g^4}{(4\pi)^2}\Lambda^{2\e}\,\sum_{i=1}^{N_l}\, \bar{\mathcal{D}_4}
\left(
\begin{array}{ccc}
0 & 0 & 0  \\
c & c & \e
\end{array}\right).\label{D2cVMm0}
\ea
Here, the bar indicates multiplication of the integrand of $\mathcal{D}_4$ by the function $f_1(m^2/P^2)-f_1(0)$, and just as before, the index $c$ means that the line in question has been cut, the only difference being that now $m=0$.

In the end, one finds
\ba
\fr{\Omega_\rmi{VM}^{x}}{V} &=& \fr{4d_A}{3}\fr{g^4}{(4\pi)^2}\sum_{i=1}^{N_l}\int\fr{{\rm d}^3p}{(2\pi)^3}\fr{\theta(\mu_i-p))}{2p}\int\fr{{\rm d}^3q}{(2\pi)^3}\fr{\theta(\mu_i-q)}{2q}\Delta(a), \label{pvmxres}
\ea
where we have defined
\ba
a &\equiv& \fr{(P-Q)^2}{m^2}, \;\;\;\;\;\;\;\;\; \Delta(a)=\fr{4}{a}-\ln\,a+\fr{2(a-2)}{a}\sqrt{\fr{4+a}{a}}{\rm arctanh}\bigg[\sqrt{\fr{a}{4+a}}\bigg]. \label{adef}
\ea
An analytic evaluation of two of the remaining three integrals finally leads to the result displayed in Eq.~(\ref{pvmx1}).

\section{Integral Definitions and Results}
\label{app:idefs}

In this Appendix, we give the definitions of all of the integrals $I_n$ that have appeared in our discussion so far, as well as provide results for the ones needed in constructing the result in Section \ref{sec:computation}, \textit{i.e.~}the ones not included in the functions ${\mathcal G}_n(\hat{m})$. When available, the results will be in the form of analytic expressions, while for the numerically evaluated ones we provide approximating pocket formulae.

Starting from the integrals needed in various parts of the 2GI computation, we have
\ba
I_1(\hat{m})&=&\int\! \fr{{\rm d}^{3} p}{(2\pi)^{3}} \fr{\theta(\mu-E(\mbox{\boldmath$p$}))}{2E(\mbox{\boldmath$p$})}\;\equiv\; \int_p=\fr{\mu^2 z}{8\pi^2},\\
I_{1b}(\hat{m})&=&\fr{\partial}{\partial \widetilde{m}^2}\int_p=\fr{1}{8\pi^2}\fr{z-\hat{u}}{\hat{m}^2},\\
I_{1c}(\hat{m})&=& \int_p\int_q \ln\(\frac{(P-Q)^2}{m^2}\)\\
I_2(\hat{m})&=&\int_p\int_q \fr{1}{(P-Q)^2}=\fr{\mu^2}{64\pi^4}\fr{\hat{u}^4-z^2}{\hat{m}^2},\\
I_{2b}(\hat{m})&=&\fr{\partial}{\partial \widetilde{m}^2}\int_p\int_q \fr{1}{(P-Q)^2}\nn
&=&\fr{1}{128\pi^4}\Bigg\{\({\rm arctan}\(\fr{\hat{u}}{\hat m}\)\)^2-\fr{2u}{m}{\rm arctan}\(\fr{\hat u}{\hat m}\) -\(\ln\(\fr{1+\hat u}{\hat m}\)\)^2\Bigg\},\\
I_{2c}(\hat{m})&=& \int_p \int_q \frac{1}{(P-Q)^2 }\ln\(-\frac{(P-Q)^2}{m^2}\)\\
I_3(\hat{m})&=&\int_p\int_q\int_r \fr{1}{(P-Q)^2(P-R)^2},\\
I_{3b}(\hat{m})&=&\fr{\partial}{\partial \widetilde{m}^2}\int_p\int_q\int_r \fr{1}{(P-Q)^2(P-R)^2},\\
I_4(\hat{m})&=&\int_p\int_q\int_r \fr{(Q-R)^2}{(P-Q)^2(P-R)^2},\\
I_5(\hat{m})&=&\int_p\int_q\int_r \fr{1}{(P-Q-R)^2+m^2},\\
I_6(\hat{m})&=&\int_p\int_q\int_r \fr{1}{(P-Q)^2(P-R)^2\left((P-Q-R)^2+m^2\right)},\\
I_7(\hat{m})&=&\int_p\int_q\int_r \fr{(P-Q)^2}{(P-R)^2\left((P-Q-R)^2+m^2\right)},\\
I_8(\hat{m})&=&\int_p\int_q \fr{\partial}{\partial \widetilde{m}^2}
\mathcal{D}_4
\left(
\begin{array}{ccc}
\widetilde{m} & m & 0  \\
c & c & 1
\end{array}
\right)=-\frac{1}{128 \pi ^4}\left(\frac{\hat u}{\hat m}-{\rm arctan}\left(\frac{\hat u}{\hat m}\right)\right)^2,\\
I_9(\hat{m})&=&\int_p\int_q\int_r \fr{\partial}{\partial \widetilde{m}^2}
\mathcal{D}_2
\left(
\begin{array}{ccccc}
m & m & \widetilde{m} & 0 & 0  \\
c & c & c & 1 & 1
\end{array}
\right),\\
I_{10}(\hat{m})&=&\int_p\int_q\Bigg\{ -\frac{23}{3}+\frac{ 8 }{a}- \ln a\nn
&&+\sqrt{\frac{a}{4+a}}\Bigg[-\pi^2-\frac{4\pi^2}{3a^2}+\left(\ln a\right)^2-2\left(\ln\left(\frac{1}{2}\left[\sqrt{a(4+a)}-a \right]\)\)^2\nn
&&-2\(\frac{4}{a^2}-1\)\(\ln\(\frac{1}{2}\left[4+a-\sqrt{a(4+a)}\right]\)\)^2  \nn
&&+\frac{8}{a^2}\ln \left( \frac{1}{2}\left[2+a-\sqrt{a(4+a)}\right]\right)\ln a  \nn
&&
+\(\frac{4}{a^2}-1\)\left[\ln(4+a)\right]^2
+4\text{Li}_2\(\frac{1}{2} \left[1-\sqrt{\frac{4+a}{a}}\right]\)\nn
&&+4\(\frac{4}{a^2}-1\)\text{Li}_2\(\frac{1}{2}\left[1-\sqrt{\frac{a}{4+a}}\right]\)\Bigg]\nn
&&+\sqrt{\frac{4+a}{a}}\left(\frac{4}{a}-3 \right)\ln\left( \frac{1}{2}\left[2+a-\sqrt{a(4+a)}\right]\right)\Bigg\},\\
I_{11}(\hat{m})&=&-16m^2 I_1 I'_1+32m^4 I'_2 + \int_p\int_q\Bigg\{  12+32\frac{ (E_p-E_q)}{E_p}\frac{1}{ a^2}
+\left(2-\frac{8}{\pmq}\right) \ln~a\nn
&& +\sqrt{\frac{a}{4+a}}\Bigg[\left( 2+\frac{16}{a}+\frac{8}{a^2}\right)\frac{\pi^2}{3}-2\left( \ln a\right)^2\nn
&&+4\(\ln\(\frac{1}{2}\left[\sqrt{a(4+a)}-a\right]\)\)^2\nn
&&+4\left(\frac{4}{a^2}-1\right) \(\ln\(\frac{1}{2} \left[4+a-\sqrt{a (4+a)}\right]\)\)^2-2\left(\frac{4}{a^2}-1\right) \left[\ln\left(4+a\right)\right]^2\nn
&&+4\left(\frac{4}{a}-1\right)\text{Li}_2\( \frac{1}{2}\left[2+a-\sqrt{a(4+a)}\right]\)\nn
&&-8\text{Li}_2\(\frac{1}{2} \left[1-\sqrt{\frac{4+a}{a}}\right]\)-8\left(\frac{4}{a^2}-1\right)\text{Li}_2\(\frac{1}{2} \left[1-\sqrt{\frac{a}{4+a}}\right]\)\nn
&&+\(\frac{4}{a}-1\)\(\ln\(\frac{1}{2}\left[2+a-\sqrt{a(4+a)}\right]\)\)^2-\frac{16}{a^2}\ln a\ln\(\frac{1}{2}\left[2+a-\sqrt{a(4+a)}\right]\)\Bigg]\nn
&&+\sqrt{\frac{4+a}{a}}\left(6-\frac{8}{\pmq}\right)\ln\(\frac{1}{2} \left[2+a-\sqrt{a (4+a)}\right]\)
\Bigg\},\\
I_{12}(\hat{m})&=&\int_p\int_q \left(\frac{a-2}{a}\right)\Delta(a),
\ea
where again $a\equiv (P-Q)^2/m^2$ and $\Delta(a)$ is as defined in Eq.~(\ref{adef}). In these integrals, all four-momenta are massive and taken to be on shell ($P^2=-m^2$), while $\mu$ only appears in the integration measures. When outside of the integral, the mass derivative again acts only on the mass appearing in the $p$ integration measure (\textit{i.e.~}not on the ones in the $q$ or $r$ integration measures or on the masses inside the integrand), after which $\widetilde{m}$ is set equal to $m$.

Finally, the integrals introduced in Section \ref{sec:plasmon} are defined by
\ba
I_{13}(\hat{m})&=&\frac{8}{\pi}\int_0^{\pi/2} d\Phi\,\sin^2\Phi
\left[F_{h}^2(0,\Phi,\hat{m})+\frac{1}{2}G_{h}^2(0,\Phi,\hat{m})\right]\nn
&\simeq&\frac{8}{3}(1-\ln 2)\hat{u}^3+0.318 \hat{u}^6-0.137 \hat{u}^7,\\
I_{14}(\hat{m})&=&\frac{8}{\pi}\int_0^{\pi/2} d\Phi\,\sin^2\Phi
\left[F_{l}(0,\Phi)F_{h}(0,\Phi,\hat{m})+\frac{1}{2}G_{l}(0,\Phi)G_{h}(0,\Phi,\hat{m})\right]\nn
&\simeq&
1.99\hat{v}-0.99\hat{v}^2+\ln\hat{v}\left(
-0.27 \hat{v}+0.26\hat{v}^2-0.119 \hat{v}^2\ln\hat{v}\right),\\
I_{15}({\vec{\mu}})&=&\frac{16}{3} \ln 2 \sum_i \mu_i^4
-2 \vec{\mu}^2 \sum_i \mu_i^2 \ln \frac{\mu_i^2}{\vec{\mu}^2}
+\frac{2}{3}\sum_{i=1}^{N_l}\sum_{j>i}^{N_l}\left[(\mu_i-\mu_j)^4 \ln \frac{|\mu_i^2-\mu_j^2|}{\mu_i\mu_j}
\right.\nn
&&\hspace*{3cm}\left.
+4 \mu_i\mu_j (\mu_i^2+\mu_j^2) \ln \frac{(\mu_i+\mu_j)^2}{\mu_i\mu_j}
-(\mu_i^4-\mu_j^4)\ln\frac{\mu_i}{\mu_j}\right],\\
I_{16}(\hat{m},\vec{\hat{\mu}}^2)&=&\frac{16}{\pi}\int_0^{\pi/2}\,\sin^2\Phi\,
\left\{
F_{l}^2(0,\Phi)
\ln \left[F_{l}(0,\Phi)+F_{h}(0,\Phi,\hat{m})/\vec{\hat{\mu}}^2\right]
\right.\nonumber\\
&&\hspace*{3cm}\left.
+\frac{1}{2} G_{l}^2(0,\Phi) \ln \left[G_{l}(0,\Phi)+G_{h}(0,\Phi,\hat{m})/\vec{\hat{\mu}}^2\right]
\right\}\nn
&\simeq&-0.8564+2 \ln\,\left( \frac{\vec{\mu}^4+2 \vec{\mu}^2 I_{14}
+I_{14} I_{16h1}}{\vec{\mu}^4+ \vec{\mu}^2 I_{14}}\right)\nn
I_{16h1}(\hat{m})&\simeq&
4.629\hat{v}-3.629\hat{v}^2+\ln\hat{v}\left(0.6064 \hat{v}+
2.021\hat{v}^2-0.566\hat{v}^2\ln \hat{v}\right),\\
I_{17}(\hat{m},\hat{\mu}_i)&=&\frac{32}{\pi}\int_0^\infty \frac{dq}{q}
\int_0^{\pi/2} d\Phi\,\sin^2\Phi\left\{
-\left[F_{l}(q,\Phi)F_{h}(q\, \hat{\mu}_i,\Phi,\hat{m})+\frac{1}{2}G_{l}(q,\Phi)
G_{h}(q\, \hat{\mu}_i,\Phi,\hat{m})
\right]
\right.\nonumber\\
&&\left.
+\left[F_{l}(0,\Phi)F_{h}(0,\Phi,\hat{m})+\frac{1}{2} G_{l}(0,\Phi)
G_{h}(0,\Phi,\hat{m})\right]\frac{4 q^2}{q^2(q^2 \hat{\mu}_i^2+4)}
\right\}\nonumber\\
&=&\frac{I_{14}}{6 \hat{\mu}_i^2}\left[
\left(1-\hat{\mu}_i\right)^4\ln\,\left(1-\frac{1}{\hat{\mu}_i}\right)^2+
\left(1+\hat{\mu}_i\right)^4\ln\,\left(1+\frac{1}{\hat{\mu}_i}\right)^2
-2 \ln\,\frac{1}{\hat{\mu}_i^{2}}\right]+I_{17h1}\nn
&&+I_{17h2}\frac{\hat{\mu}_i}{\hat{\mu}_i+1}+I_{17h3}\frac{1}{\hat{\mu}_i^2+1}
+I_{17h4} \frac{\hat{\mu}_i}{3.38 \sqrt{1-\hat{m}}+7.96 (1-\hat{m})+\hat{\mu}_i^3},\\
I_{17h1}(\hat{m})&\simeq&-7.5439\hat{v}+1.1422 \hat{u}+6.639 \hat{u}^5-8.8809 \hat{u}^3\ln(1+\hat{m})\nn
&&+\ln\hat{v}\left(
-2.005 \hat{v}+5.865 \hat{v}^2 \ln \hat{v}-10.622 \hat{v}^3 \ln^2 \hat{v}\right)
,\\
I_{17h2}(\hat{m})&\simeq&-8.57 \hat{v}+3.09 \hat{u}+9.654 \hat{u}^2+10.86 \hat{u}^3-12.445 \hat{u}^4\nn
&&+\hat{u}\ln \hat{v}\left(0.374 +8.73 \hat{u}^2\right)-\frac{7}{3}I_{14}-I_{17h1}
,\\
I_{17h3}(\hat{m})&\simeq&4.804 \hat{v}-2.224 \hat{v}^2+\ln \hat{v}\left(
-0.411 \hat{v}+0.056 \hat{v}^2 -0.0718 \hat{v}^2 \ln \hat{v}\right)\nn
&-&\frac{7}{3}I_{14}-I_{17h1},\\
I_{17h4}(\hat{m})&\simeq&-8.085 \hat{v}^3+6.646 \hat{v}^5-0.5846 \hat{v}^{12}
+0.3465 \hat{m}\hat{u}\nn
&&+1.359 \hat{u}^2-3.77 \hat{u}^3+5.395 \hat{u}^4-0.9525 \hat{u}^5,\\
I_{18}(\hat{m})&=&\frac{3}{\pi(1-\ln2)}\int_0^{\pi/2} d\Phi\,\sin^2\Phi\, F_{l}(0,\Phi) F_{h}(0,\Phi,\hat{m})
\nonumber\\
&\simeq&2.3 \hat{v}-1.3 \hat{v}^2+\ln \hat{v}\left(-0.3477\hat{v}+0.6484 \hat{v}^2-0.241
\hat{v}^2 \ln \hat{v}\right),\\
I_{19}(\hat{m},\vec{\hat{\mu}}^2)&=&\frac{16}{\pi}\int_0^{\pi/2}\,\sin^2\Phi\,
\left\{F_{l}(0,\Phi)F_{h}(0,\Phi,\hat{m})\ln \left[\vec{\hat{\mu}}^2 F_{l}(0,\Phi)+F_{h}(0,\Phi,\hat{m})
\right]\right.\nonumber\\
&&\hspace*{2cm}\left.
+\frac{1}{2} G_{l}(0,\Phi)G_{h}(0,\Phi,\hat{m}) \ln \left[\vec{\hat{\mu}}^2 G_{l}(0,\Phi)
+ G_{h}(0,\Phi,\hat{m})\right]\right\}\nn
&=&I_{19h1}+I_{14}\left[2 \ln \vec{\hat{\mu}}^2+2 \ln \left(
\frac{I_{14}\,\vec{\hat{\mu}}^4+2 I_{13}\, \vec{\hat{\mu}}^2 +I_{14} I_{19h2}}
{I_{14}\,\vec{\hat{\mu}}^4+ I_{13}\,\vec{\hat{\mu}}^2}\right)\right],\\
I_{19h1}(\hat{m})&\simeq&-3.6643\hat{v}+2.808 \hat{v}^2+\ln \hat{v}
\left(-0.7305 \hat{v}-1.22 \hat{v}^2\right)\nn
&&+\ln^2\hat{v}\left(-0.0437 \hat{v}+0.312 \hat{v}^2-0.0237\hat{v}^2\ln \hat{v}\right),\\
I_{19h2}(\hat{m})&\simeq&0.104 \hat{v}+0.464 \hat{v}^2+0.432 \hat{u}^5+\ln \hat{v}
\left(-0.04 \hat{v}-0.99 \hat{v}^2+0.455 \hat{v}^2 \ln \hat{v}\right),\\
I_{20}(\hat{m})&=&\frac{32}{\pi}\int_0^\infty \frac{dq}{q}
\int_0^{\pi/2} d\Phi\,\sin^2\Phi\left\{
-\left[F_{h}^2(q,\Phi,\hat{m})+\frac{1}{2}G_{h}^2(q,\Phi,\hat{m})
\right]
\right.\nonumber\\
&&\left.
+\left[F_{h}^2(0,\Phi,\hat{m})+\frac{1}{2} G_{h}^2(0,\Phi,\hat{m})\right]
\frac{4 q^2}{q^2(q^2+4)}
\right\}\nonumber\\
&\simeq&3.84\hat{v}-7.94\hat{v}^2+8.06 \hat{u}^2+\ln \hat{v}\left(
3.81 \hat{v}+9.7 \hat{v}^2-7.42 \hat{v}^2 \ln \hat{v}\right),\\
I_{21}(\hat{m})&=&\frac{3}{\pi(1-\ln2)}\int_0^{\pi/2} d\Phi\,\sin^2\Phi\, F_{h}^2(0,\Phi,\hat{m})\nonumber\\
&\simeq&-0.11 \hat{v}+1.11 \hat{v}^2+\ln \hat{v}\left(-0.059 \hat{v}-2.07 \hat{v}^2+0.4
\hat{v}^2 \ln \hat{v}\right),\\
I_{22}(\hat{m},\vec{\hat{\mu}}^2)&=&\frac{16}{\pi}\int_0^{\pi/2}\,\sin^2\Phi\,
\left\{F_{h}^2(0,\Phi,\hat{m})\ln \left[\vec{\hat{\mu}}^2 F_{l}(0,\Phi)+F_{h}(0,\Phi,\hat{m})
\right]\right.\nonumber\\
&&\hspace*{2cm}\left.
+\frac{1}{2} G_{h}^2(0,\Phi,\hat{m}) \ln \left[\vec{\hat{\mu}}^2 G_{l}(0,\Phi)+G_{h}(0,\Phi,\hat{m})\right]\right\},\\
&=&I_{22h1}+I_{13}\left[2 \ln \vec{\hat{\mu}}^2+2 \ln \left(
\frac{\vec{\hat{\mu}}^4+ \vec{\hat{\mu}}^2 I_{22h2}+I_{22h3}}{\vec{\hat{\mu}}^4+0.5
\vec{\hat{\mu}}^2 I_{22h2}}\right)\right],\\
I_{22h1}(\hat{m})&\simeq&-0.8255\hat{u}^2-0.03084 \hat{u}^7+0.00161 \hat{v} \ln \hat{v},\\
I_{22h2}(\hat{m})&\simeq&1.313 \hat{u}+0.434 \hat{u}^2+0.253 \hat{u}^5,\\
I_{22h3}(\hat{m})&\simeq&1.844 \hat{v}-0.844 \hat{v}^2+\ln \hat{v}\left(
0.194\hat{v}-0.392 \hat{v}^2-0.704 \hat{v}^2 \ln \hat{v}\right),
\ea
where $F_{l,h}$ and $G_{l,h}$ are defined in Appendix \ref{app:pimunu}, and we have used the notation $\hat{v}\equiv 1-\hat{m}$.

\section{$4d$ Integrals}
\label{app:4dints}

In this last Appendix, we complete our presentation by providing results for the various vacuum ($T=\mu=0$) amplitudes that are needed as input in the single, double and triple cut integrals. They are given as power series expansions in $\e=(4-d)/2$ up to the order required, using the integration measure defined in the first section.

According to the notation introduced earlier, and using results from Refs.~\cite{Laporta:2004rb,Bonciani:2003cj,Bonciani:2003hc,Schroder:2005va} we obtain
\ba
\mathcal{D}_5 \!\left( \begin{array}{c} m  \\ 1 \end{array} \right)&=&-\fr{m^2}{(4\pi)^2}\(\fr{\bar\Lambda^2}{m^2}\)^{\e}\Bigg\{\fr{1}{\e}+1+\bigg[1+\fr{\pi^2}{12}\bigg]\e
+\bigg[1+\fr{\pi^2}{12}-\fr{\zeta(3)}{3}\bigg]\e^2\nn
&+&\bigg[1+\fr{\pi^2}{12}-\fr{\zeta(3)}{3}+\fr{\pi^4}{160}\bigg]\e^3\nn
&+&\bigg[\(1+\fr{\pi^2}{12}\)\(1-\fr{\zeta(3)}{3}\)+\fr{\pi^4}{160}-\fr{\zeta(5)}{5}\bigg]\e^4\Bigg\},\\
\Lambda^{2\e}\,\widetilde{\mathcal{D}}_5\!
\left(
\begin{array}{c}
0  \\
1+\e
\end{array}
\right)&=&\fr{2m^2}{(4\pi)^2}\(\fr{\bar\Lambda^2}{m^2}\)^{2\e}
\Bigg\{\fr{1}{\e}+1+\bigg[5+\fr{\pi^2}{6}\bigg]\e\Bigg\},\\
\mathcal{D}_4 \!\left( \begin{array}{ccc} m & m & 0  \\ c & 1 & 1 \end{array} \right)&=&\fr{1}{(4\pi)^2}\(\fr{\bar\Lambda^2}{m^2}\)^{\e}\Bigg\{\fr{1}{\e}+2\Bigg\},\\
\mathcal{D}_3\!\left( \begin{array}{cccc}m & m & 0 & 0  \\ c & 1 & 1 & 1 \end{array} \right)&=&-\fr{m^2}{(4\pi)^4}\(\fr{\bar\Lambda^2}{m^2}\)^{2\e}\Bigg\{\fr{1}{2\e^2}+\fr{5}{4\e}+\fr{11}{8}+\fr{5\pi^2}{12}\nn
&+&\bigg[-\fr{55}{16}+\fr{25\pi^2}{24}+\fr{11\zeta(3)}{3}\bigg]\e\Bigg\},\\
\mathcal{D}_3\!\left( \begin{array}{cccc}m & m & m & m  \\ c & 1 & 1 & 1 \end{array} \right)&=&-\fr{m^2}{(4\pi)^4}\(\fr{\bar\Lambda^2}{m^2}\)^{2\e}\Bigg\{\fr{3}{2\e^2}+\fr{17}{4\e}+\fr{59}{8}+\fr{\pi^2}{4}\nn
&+&\bigg[\fr{65}{16}+\fr{49\pi^2}{24}-\zeta(3)\bigg]\e\\
&+&\bigg[-\fr{1117}{32}+\fr{475\pi^2}{48}+\fr{151\zeta(3)}{6}+\fr{7\pi^4}{240}-8\pi^2\,\ln\,2\bigg]\e^2\Bigg\}\nn
\mathcal{D}_4\!
\left(
\begin{array}{ccc}
m & m & 0  \\
c & 1 & \e
\end{array}
\right)&=&-\fr{m^2}{2(4\pi)^2}\(\fr{\bar\Lambda^2}{m^4}\)^{\e}\Bigg\{\fr{1}{\e}+\fr{1}{2}-\fr{1}{4}\bigg[9-\fr{11\pi^2}{3}\bigg]\e\Bigg\},\\
\mathcal{D}_4\!
\left(
\begin{array}{ccc}
m & m & 0  \\
c & 1 & 1+\e
\end{array}
\right)&=&\fr{1}{2(4\pi)^2}\(\fr{\bar\Lambda^2}{m^4}\)^{\e}\Bigg\{\fr{1}{\e}+3+\bigg[9+\fr{11\pi^2}{12}\bigg]\e\Bigg\},\\
\Lambda^{2\e}\,\widetilde{\mathcal{D}}_4\!
\left(
\begin{array}{ccc}
m & m & 0  \\
c & 1 & \e
\end{array}
\right)&=&\fr{m^2}{3(4\pi)^2}\(\fr{\bar\Lambda^2}{m^2}\)^{2\e}\Bigg\{\fr{1}{\e^2}+\fr{49}{6\e}+\fr{1213}{36}-\fr{5\pi^2}{2}\Bigg\},\\
\Lambda^{2\e}\,\widetilde{\mathcal{D}}_4\!
\left(
\begin{array}{ccc}
0 & 0 & 0  \\
c & 1 & \e
\end{array}
\right)&=&\fr{2m^2}{(4\pi)^2}\(\fr{\bar\Lambda^2}{m^2}\)^{2\e}\Bigg\{\fr{1}{\e}+1\Bigg\},\\
\Lambda^{2\e}\,\widetilde{\mathcal{D}}_4\!
\left(
\begin{array}{ccc}
m & m & 0  \\
c & 1 & 1+\e
\end{array}
\right)&=&-\fr{1}{3(4\pi)^2}\(\fr{\bar\Lambda^2}{m^2}\)^{2\e}\Bigg\{\fr{1}{\e^2}+\fr{14}{3\e}+\fr{118}{9}-\fr{\pi^2}{2}\Bigg\},\\
\fr{\partial}{\partial \widetilde{m}^2}
\mathcal{D}_4
\left(
\begin{array}{ccc}
\widetilde{m} & m & 0  \\
c & 1 & 1
\end{array}
\right)&=&-\fr{1}{2m^2}\mathcal{D}_4 \!\left( \begin{array}{ccc} m & m & 0  \\ c & 1 & 1 \end{array} \right),\\
\fr{\partial}{\partial \widetilde{m}^2}
\mathcal{D}_2
\left(
\begin{array}{ccccc}
m & m & \widetilde{m} & 0 & 0  \\
1 & 1 & c & 1 & 1
\end{array}
\right)&=&-\fr{1}{m^2}\(\mathcal{D}_4 \!\left( \begin{array}{ccc} m & m & 0  \\ c & 1 & 1 \end{array} \right)\)^2,\\
\fr{\partial}{\partial \widetilde{m}^2}
\mathcal{D}_4
\left(
\begin{array}{ccc}
\widetilde{m} & m & 0  \\
c & c & 1
\end{array}
\right)&=&\(1-\fr{E(q)}{E(p)}\)\fr{1}{\((P-Q)^2\)^2},\nn
\fr{\partial}{\partial \widetilde{m}^2}
\mathcal{D}_2
\left(
\begin{array}{ccccc}
m & m & \widetilde{m} & 0 & 0  \\
c & 1 & c & 1 & 1
\end{array}
\right)&=&\(\fr{1-\fr{E(q)}{E(p)}}{\((P-Q)^2\)^2}-\fr{1}{2m^2(P-Q)^2}\)\mathcal{D}_4
\left(
\begin{array}{ccc}
m & m & 0  \\
c & 1 & 1
\end{array}
\right),\nn
\fr{\partial}{\partial \widetilde{m}^2}
\mathcal{D}_2
\left(
\begin{array}{ccccc}
m & m & \widetilde{m} & 0 & 0  \\
c & c & c & 1 & 1
\end{array}
\right)&=&\(1-\fr{E(q)}{E(p)}\)\fr{2}{\((P-Q)^2\)^2(P-R)^2},\\
\mathcal{D}_3\!\left( \begin{array}{cccc}m & m & 0 & 0  \\ c & c & 1 & 1 \end{array} \right)&=& \fr{1}{(4\pi)^2}\(\fr{\bar\Lambda^2}{m^2}\)^{\e}\Bigg\{\fr{1}{\e}+2-\ln\fr{(P-Q)^2}{m^2}\Bigg\},\\
\mathcal{D}_3\!\left( \begin{array}{cccc}m & m & m & m  \\ c & c & 1 & 1 \end{array} \right)&=&\fr{1}{(4\pi)^2}\(\fr{\bar\Lambda^2}{m^2}\)^{\e}\Bigg\{\fr{1}{\e}+2+\fr{1+x}{1-x}\ln\,x\nn
&+&\Bigg[4+\fr{\pi^2}{12}
-\fr{1+x}{2(1-x)}\Bigg(\fr{\pi^2}{3}-4\,\ln\,x+2\(\ln\(1+x\)\)^2\nn
&-&\(\ln\fr{(1+x)^2}{x}\)^2
-4\,{\rm Li}_2\(\fr{x}{1+x}\)\Bigg)\Bigg]\e\Bigg\}\nn
&=&\mathcal{D}_3\!\left( \begin{array}{cccc}m & m & m & m  \\ 1 & 1 & c & c \end{array} \right),\\
\mathcal{D}_3\!\left( \begin{array}{cccc}m & m & m & m  \\ c & 1 & c & 1 \end{array} \right)&=&\fr{1}{(4\pi)^2}\(\fr{\bar\Lambda^2}{m^2}\)^{\e}\Bigg\{\fr{1}{\e}+2+\fr{1+y}{1-y}\ln\,y\nn
&+&\Bigg[4+\fr{\pi^2}{12}
-\fr{1+y}{2(1-y)}\Bigg(\fr{\pi^2}{3}-4\,\ln\,y+2\(\ln\(1+y\)\)^2\nn
&-&\(\ln\fr{(1+y)^2}{y}\)^2
-4\,{\rm Li}_2\(\fr{y}{1+y}\)\Bigg)\Bigg]\e\Bigg\}\nn
&=&\Bigg\{\mathcal{D}_3\!\left( \begin{array}{cccc}m & m & m & m  \\ 1 & c & 1 & c \end{array} \right)\Bigg\}^*,\\
\mathcal{D}_2
\left(
\begin{array}{ccccc}
m & m & m & 0 & 0 \\
c & c & 1 & 1 & 1
\end{array}
\right) &=& \fr{1}{(4\pi)^2m^2}\fr{x}{1-x^2}\Bigg\{4\zeta(2)+\fr{1}{2}\(\ln\,x\)^2+2\,{\rm Li}_2(x)\Bigg\}, \\
\mathcal{D}_1
\left(
\begin{array}{cccccc}
m & m & 0 & m & m & 0 \\
c & 1 & 1 & c & 1 & 1
\end{array}
\right) &=& \fr{2}{(4\pi)^2m^4}\fr{x}{(1-x)^2}\fr{y\,\ln\,y}{1-y^2}\(\fr{\bar\Lambda^2}{m^2}\)^{\e}\nn
&\times&\Bigg\{\fr{1}{\e}+\ln\,x-2\,\ln(1-x)\Bigg\}\nn
&=&\Bigg\{\mathcal{D}_1
\left(
\begin{array}{cccccc}
m & m & 0 & m & m & 0 \\
1 & c & 1 & 1 & c & 1
\end{array}
\right)\Bigg\}^*,\\
\Lambda^{2\e}\,\widetilde{\mathcal{D}}_4\!
\left(
\begin{array}{ccc}
m & m & 0  \\
c & c & \e
\end{array}
\right)&=&\(\fr{\bar\Lambda^2}{m^2}\)^{\e}\Bigg\{-\fr{2}{3\e}-\fr{10}{9}+\fr{8}{3a}\nn
&+&\fr{4(a-2)}{3a}\sqrt{\fr{4+a}{a}}
{\rm arctanh}\bigg[\sqrt{\fr{a}{4+a}}\bigg]\Bigg\}.
\ea
In accordance with the notation of Refs.~\cite{Bonciani:2003cj,Bonciani:2003hc}, we have denoted here
\ba
x &\equiv& \fr{\sqrt{-(P+Q)^2}-\sqrt{(P-Q)^2}}{\sqrt{-(P+Q)^2}+\sqrt{(P-Q)^2}},\\
y &\equiv& \fr{\sqrt{(P-Q)^2}-\sqrt{-(P+Q)^2}}{\sqrt{(P-Q)^2}+\sqrt{-(P+Q)^2}}\;=\;-x.
\ea
In simplifying the results, we have found the Mathematica package HypExp very useful \cite{Huber:2005yg}.



\begin{thebibliography}{99}

\bibitem{Aoki:2006we}
  Y.~Aoki, G.~Endrodi, Z.~Fodor, S.~D.~Katz and K.~K.~Szabo,
  Nature {\bf 443} (2006) 675
  [arXiv:hep-lat/0611014].


\bibitem{Bazavov:2009zn}
  A.~Bazavov {\it et al.},
  Phys.\ Rev.\  D {\bf 80} (2009) 014504
  [arXiv:0903.4379 [hep-lat]].

\bibitem{Kajantie:2002wa}
 K.~Kajantie, M.~Laine, K.~Rummukainen and Y.~Schroder,
 Phys.\ Rev.\ ÊD {\bf 67} (2003) 105008
 [arXiv:hep-ph/0211321].

\bibitem{Blaizot:2003iq}
  J.~P.~Blaizot, E.~Iancu and A.~Rebhan,
  Phys.\ Rev.\  D {\bf 68} (2003) 025011
  [arXiv:hep-ph/0303045].



\bibitem{Vuorinen:2003fs}
  A.~Vuorinen,
  Phys.\ Rev.\  D {\bf 68} (2003) 054017
  [arXiv:hep-ph/0305183].


\bibitem{Andersen:2004fp}
  J.~O.~Andersen and M.~Strickland,
  Annals Phys.\  {\bf 317}, 281 (2005)
  [arXiv:hep-ph/0404164].


\bibitem{Adcox:2004mh}
  K.~Adcox {\it et al.}  [PHENIX Collaboration],
  Nucl.\ Phys.\  A {\bf 757} (2005) 184.
  [arXiv:nucl-ex/0410003].

\bibitem{Back:2004je}
  B.~B.~Back {\it et al.},
  Nucl.\ Phys.\  A {\bf 757} (2005) 28.
  [arXiv:nucl-ex/0410022].

\bibitem{Arsene:2004fa}
  I.~Arsene {\it et al.}  [BRAHMS Collaboration],
  Nucl.\ Phys.\  A {\bf 757} (2005) 1.
  [arXiv:nucl-ex/0410020].

\bibitem{Adams:2005dq}
  J.~Adams {\it et al.}  [STAR Collaboration],
  Nucl.\ Phys.\  A {\bf 757} (2005) 102.
  [arXiv:nucl-ex/0501009].

\bibitem{Carminati:2004fp}
  F.~Carminati {\it et al.}  [ALICE Collaboration],
  J.\ Phys.\ G {\bf 30} (2004) 1517.Vermaseren:1997fq



\bibitem{Barrette:1994xr}
  J.~Barrette {\it et al.}  [E877 Collaboration],
  Phys.\ Rev.\ Lett.\  {\bf 73} (1994) 2532
  [arXiv:hep-ex/9405003].

\bibitem{Abreu:2000ni}
  M.~C.~Abreu {\it et al.}  [NA50 Collaboration],
  Phys.\ Lett.\  B {\bf 477} (2000) 28.


\bibitem{Stephans:2006tg}
  G.~S.~F.~Stephans,
  J.\ Phys.\ G {\bf 32} (2006) S447
  [arXiv:nucl-ex/0607030].

\bibitem{Senger:2008zz}
  P.~Senger,
  Phys.\ Part.\ Nucl.\  {\bf 39} (2008) 1055.

\bibitem{Heiselberg:2000dn}
  H.~Heiselberg and V.~Pandharipande,
  Ann.\ Rev.\ Nucl.\ Part.\ Sci.\  {\bf 50} (2000) 481
  [arXiv:astro-ph/0003276].

\bibitem{Bender:2003jk}
  M.~Bender, P.~H.~Heenen and P.~G.~Reinhard,
  Rev.\ Mod.\ Phys.\  {\bf 75} (2003) 121.

\bibitem{Weber:2004kj}
  F.~Weber,
  Prog.\ Part.\ Nucl.\ Phys.\  {\bf 54} (2005) 193
  [arXiv:astro-ph/0407155].


\bibitem{Lattimer:2000nx}
  J.~M.~Lattimer and M.~Prakash,
  Astrophys.\ J.\  {\bf 550} (2001) 426
  [arXiv:astro-ph/0002232].

\bibitem{pionco}
A.B.~Migdal, Zh.\ Eksp.\ Teor.\ Fiz. {\bf 61} (1971), 2210
[Sov.\ Phys.\ JETP {\bf 36} (1973), 1052];
R.F.~ Sawyer, Phys.\ Rev.\ Lett.\ {\bf 29} (1972), 382;
D.J.~Scalapino, Phys.\ Rev.\ Lett.\ {\bf 29} (1972), 386.


\bibitem{Kaplan:1986yq}
D.~B.~Kaplan and A.~E.~Nelson,
Phys.\ Lett.\  B {\bf 175} (1986) 57.

\bibitem{Alford:2007xm}
  M.~G.~Alford, A.~Schmitt, K.~Rajagopal and T.~Schafer,
  Rev.\ Mod.\ Phys.\  {\bf 80} (2008) 1455
  [arXiv:0709.4635 [hep-ph]].

\bibitem{Bodmer:1971we}
  A.~R.~Bodmer,
  Phys.\ Rev.\  D {\bf 4} (1971) 1601.


\bibitem{Witten:1984rs}
  E.~Witten,
  Phys.\ Rev.\  D {\bf 30} (1984) 272.

\bibitem{Farhi:1984qu}
  E.~Farhi and R.~L.~Jaffe,
  Phys.\ Rev.\  D {\bf 30} (1984) 2379.

\bibitem{Alcock:1986hz}
  C.~Alcock, E.~Farhi and A.~Olinto,
  Astrophys.\ J.\  {\bf 310} (1986) 261.



\bibitem{Chodos:1974pn}
  A.~Chodos, R.~L.~Jaffe, K.~Johnson and C.~B.~Thorn,
  Phys.\ Rev.\  D {\bf 10} (1974) 2599.

\bibitem{Fraga:2001id}
  E.~S.~Fraga, R.~D.~Pisarski and J.~Schaffner-Bielich,
  Phys.\ Rev.\  D {\bf 63} (2001) 121702
  [arXiv:hep-ph/0101143].


\bibitem{Freedman:1976xs}
  B.~A.~Freedman and L.~D.~McLerran,
  Phys.\ Rev.\  D {\bf 16} (1977) 1130.


\bibitem{Freedman:1976ub}
  B.~A.~Freedman and L.~D.~McLerran,
  Phys.\ Rev.\  D {\bf 16} (1977) 1169.


\bibitem{Baluni:1977ms}
  V.~Baluni,
  Phys.\ Rev.\  D {\bf 17} (1978) 2092.

\bibitem{Amsler:2008zzb}
  C.~Amsler {\it et al.}  [Particle Data Group],
  Phys.\ Lett.\  B {\bf 667} (2008) 1.

\bibitem{Fraga:2004gz}
  E.~S.~Fraga and P.~Romatschke,
  Phys.\ Rev.\  D {\bf 71} (2005) 105014
  [arXiv:hep-ph/0412298].

\bibitem{Fleischer:1998dw}
  J.~Fleischer, F.~Jegerlehner, O.~V.~Tarasov and O.~L.~Veretin,
  Nucl.\ Phys.\  B {\bf 539} (1999) 671
  [Erratum-ibid.\  B {\bf 571} (2000) 511]
  [arXiv:hep-ph/9803493].

\bibitem{vanRitbergen:1997va}
  T.~van Ritbergen, J.~A.~M.~Vermaseren and S.~A.~Larin,
  Phys.\ Lett.\  B {\bf 400} (1997) 379
  [arXiv:hep-ph/9701390].

\bibitem{Czakon:2004bu}
  M.~Czakon,
  Nucl.\ Phys.\  B {\bf 710} (2005) 485
  [arXiv:hep-ph/0411261].


\bibitem{Chetyrkin:1997dh}
  K.~G.~Chetyrkin,
  Phys.\ Lett.\  B {\bf 404} (1997) 161
  [arXiv:hep-ph/9703278].




\bibitem{Vermaseren:1997fq}
  J.~A.~M.~Vermaseren, S.~A.~Larin and T.~van Ritbergen,
  Phys.\ Lett.\  B {\bf 405} (1997) 327
  [arXiv:hep-ph/9703284].

\bibitem{Gupta:2000hr}
  S.~Gupta,
  Phys.\ Rev.\  D {\bf 64} (2001) 034507
  [arXiv:hep-lat/0010011].

\bibitem{Cheng:2006qk}
  M.~Cheng {\it et al.},
  Phys.\ Rev.\  D {\bf 74} (2006) 054507
  [arXiv:hep-lat/0608013].

\bibitem{Aoki:2006br}
  Y.~Aoki, Z.~Fodor, S.~D.~Katz and K.~K.~Szabo,
  Phys.\ Lett.\  B {\bf 643} (2006) 46
  [arXiv:hep-lat/0609068].


\bibitem{Rodrigo:1993hc}
  G.~Rodrigo and A.~Santamaria,
  Phys.\ Lett.\  B {\bf 313} (1993) 441
  [arXiv:hep-ph/9305305].


\bibitem{Smirnov:2008iw}
  A.~V.~Smirnov,
  JHEP {\bf 0810} (2008) 107
  [arXiv:0807.3243 [hep-ph]].


\bibitem{Bonciani:2003hc}
  R.~Bonciani, P.~Mastrolia and E.~Remiddi,
  Nucl.\ Phys.\  B {\bf 690} (2004) 138
  [arXiv:hep-ph/0311145].

\bibitem{Bonciani:2003cj}
  R.~Bonciani, A.~Ferroglia, P.~Mastrolia, E.~Remiddi and J.~J.~van der Bij,
  Nucl.\ Phys.\  B {\bf 681} (2004) 261
  [Erratum-ibid.\  B {\bf 702} (2004) 364]
  [arXiv:hep-ph/0310333].

\bibitem{Schroder:2005va}
  Y.~Schroder and A.~Vuorinen,
  JHEP {\bf 0506} (2005) 051
  [arXiv:hep-ph/0503209].

\bibitem{Laporta:2004rb}
  S.~Laporta and E.~Remiddi,
  Nucl.\ Phys.\  B {\bf 704} (2005) 349
  [arXiv:hep-ph/0406160].


\bibitem{Kapusta:1989tk}
  J.~I.~Kapusta,
  ''Finite-Temperature Field Theory'',
  Cambridge University Press (1989).

\bibitem{Toimela:1984xy}
  T.~Toimela,
  Int.\ J.\ Theor.\ Phys.\  {\bf 24} (1985) 901
  [Erratum-ibid.\  {\bf 26} (1987) 1021].


\bibitem{Blaizot:2000fc}
  J.~P.~Blaizot, E.~Iancu and A.~Rebhan,
  Phys.\ Rev.\  D {\bf 63} (2001) 065003
  [arXiv:hep-ph/0005003].

\bibitem{Schneider:2003uz}
  R.~A.~Schneider,
  arXiv:hep-ph/0303104.


\bibitem{Rebhan:2003wn}
  A.~Rebhan and P.~Romatschke,
  Phys.\ Rev.\  D {\bf 68} (2003) 025022
  [arXiv:hep-ph/0304294].

\bibitem{Cassing:2007nb}
  W.~Cassing,
  Nucl.\ Phys.\  A {\bf 795} (2007) 70
  [arXiv:0707.3033 [nucl-th]].


\bibitem{Gardim:2009mt}
  F.~G.~Gardim and F.~M.~Steffens,
  Nucl.\ Phys.\  A {\bf 825} (2009) 222
  [arXiv:0905.0667 [nucl-th]].

\bibitem{Ipp:2003jy}
  A.~Ipp and A.~Rebhan,
  JHEP {\bf 0306} (2003) 032
  [arXiv:hep-ph/0305030].


\bibitem{Glendenning:1997wn}
  N.~K.~Glendenning,
  ''Compact stars: Nuclear physics, particle physics, and general relativity,''
2nd edition, {\it  New York, USA: Springer (2000)}


\bibitem{Chesler:2009yg}
  P.~M.~Chesler, A.~Gynther and A.~Vuorinen,
  JHEP {\bf 0909} (2009) 003
  [arXiv:0906.3052 [hep-ph]].

\bibitem{Schertler:1996tq}
  K.~Schertler, C.~Greiner and M.~H.~Thoma,
  Nucl.\ Phys.\  A {\bf 616} (1997) 659
  [arXiv:hep-ph/9611305].



\bibitem{Alford:1998mk}
  M.~G.~Alford, K.~Rajagopal and F.~Wilczek,
  Nucl.\ Phys.\  B {\bf 537} (1999) 443
  [arXiv:hep-ph/9804403].

\bibitem{Orsaria:2007zza}
  M.~Orsaria, S.~B.~Duarte and H.~Rodrigues,
  Braz.\ J.\ Phys.\  {\bf 37} (2007) 20.

\bibitem{Son:1998uk}
  D.~T.~Son,
  Phys.\ Rev.\  D {\bf 59} (1999) 094019
  [arXiv:hep-ph/9812287].

\bibitem{Page:2006ud}
  D.~Page and S.~Reddy,
  Ann.\ Rev.\ Nucl.\ Part.\ Sci.\  {\bf 56} (2006) 327
  [arXiv:astro-ph/0608360].


\bibitem{Lugones:2002va}
  G.~Lugones and J.~E.~Horvath,
  Phys.\ Rev.\  D {\bf 66}, 074017 (2002)
  [arXiv:hep-ph/0211070].

\bibitem{Alford:2002rj}
  M.~Alford and S.~Reddy,
  Phys.\ Rev.\  D {\bf 67} (2003) 074024
  [arXiv:nucl-th/0211046].





\bibitem{Sandweiss:2004bu}
  J.~Sandweiss,
  J.\ Phys.\ G {\bf 30} (2004) S51.

\bibitem{Abelev:2007zz}
  B.~I.~Abelev {\it et al.}  [STAR Collaboration],
  Phys.\ Rev.\  C {\bf 76} (2007) 011901
  [arXiv:nucl-ex/0511047].

\bibitem{Cecchini:2008su}
  S.~Cecchini {\it et al.}  [SLIM Collaboration],
  Eur.\ Phys.\ J.\  C {\bf 57} (2008) 525
  [arXiv:0805.1797 [hep-ex]].


\bibitem{Han:2009sj}
  K.~Han {\it et al.},
  Phys.\ Rev.\ Lett.\  {\bf 103} (2009) 092302
  [arXiv:0903.5055 [nucl-ex]].





\bibitem{Karsch:2003vd}
  F.~Karsch, K.~Redlich and A.~Tawfik,
  Eur.\ Phys.\ J.\  C {\bf 29} (2003) 549
  [arXiv:hep-ph/0303108].

\bibitem{Aoki:2009sc}
  Y.~Aoki, S.~Borsanyi, S.~Durr, Z.~Fodor, S.~D.~Katz, S.~Krieg and K.~K.~Szabo,
  JHEP {\bf 0906} (2009) 088
  [arXiv:0903.4155 [hep-lat]].



\bibitem{Akmal:1998cf}
  A.~Akmal, V.~R.~Pandharipande and D.~G.~Ravenhall,
  Phys.\ Rev.\  C {\bf 58} (1998) 1804
  [arXiv:nucl-th/9804027].



\bibitem{Schulze:2006vw}
  H.~J.~Schulze, A.~Polls, A.~Ramos and I.~Vidana,
  Phys.\ Rev.\  C {\bf 73} (2006) 058801.


\bibitem{Glendenning:1998zx}
  N.~K.~Glendenning and J.~Schaffner-Bielich,
  Phys.\ Rev.\ Lett.\  {\bf 81} (1998) 4564
  [arXiv:astro-ph/9810284].

\bibitem{Koch:1994mj}
  V.~Koch,
  Phys.\ Lett.\  B {\bf 337} (1994) 7
  [arXiv:nucl-th/9406030].

\bibitem{Waas:1997pe}
  T.~Waas and W.~Weise,
  Nucl.\ Phys.\  A {\bf 625} (1997) 287.




\bibitem{TOV}
R.C.~Tolman, Phys.\ Rev.\ {\bf 55} (1939) 364;
J.R.~Oppenheimer and G.M.~Volkoff, Phys.\ Rev.\ {\bf 55} (1939) 374.


\bibitem{Freire:2009dr}
  P.~C.~Freire, D.~Nice, J.~Lattimer, I.~Stairs, Z.~Arzoumanian, J.~Cordes and J.~Deneva,
  arXiv:0902.2891 [astro-ph.HE].



\bibitem{Weisberg:2002nv}
  J.~M.~Weisberg and J.~H.~Taylor,
  arXiv:astro-ph/0211217.

\bibitem{Jacoby:2005qg}
  B.~A.~Jacoby, A.~Hotan, M.~Bailes, S.~Ord and S.~R.~Kulkarni,
  Astrophys.\ J.\  {\bf 629} (2005) L113
  [arXiv:astro-ph/0507420].


\bibitem{Ransom:2005}
S.~Ransom {\it et al.},
Science {\bf 307} (2005) 892.

\bibitem{Champion:2008ge}
  D.~J.~Champion {\it et al.},
  arXiv:0805.2396 [astro-ph].
  Science {\bf 320} (2008) 1309.


\bibitem{Kokkotas:2000up}
 K.~D.~Kokkotas and J.~Ruoff,
 Astron.\ Astrophys.\  {\bf 366} (2001) 565
 [arXiv:gr-qc/0011093].


\bibitem{Glass}
E~.N.~Glass and L. Lindblom,
Astrophys.J.Suppl. {\bf 53}, 93-103 (1983);  errata, {\bf 71}, 173 (1989).

\bibitem{Gerlach68}
U.H.~Gerlach, Phys.\ Rev.\ {\bf 172} (1968) 1325.

\bibitem{Schertler:2000xq}
  K.~Schertler, C.~Greiner, J.~Schaffner-Bielich and M.~H.~Thoma,
  Nucl.\ Phys.\  A {\bf 677} (2000) 463
  [arXiv:astro-ph/0001467].



\bibitem{Alford:2004pf}
  M.~Alford, M.~Braby, M.~W.~Paris and S.~Reddy,
  Astrophys.\ J.\  {\bf 629} (2005) 969
  [arXiv:nucl-th/0411016].

\bibitem{Andersen:2002jz}
  J.~O.~Andersen and M.~Strickland,
  Phys.\ Rev.\  D {\bf 66} (2002) 105001
  [arXiv:hep-ph/0206196].



\bibitem{Paerels:2009pz}
  F.~Paerels {\it et al.},
  arXiv:0904.0435 [astro-ph.HE].

\bibitem{Ozel:2006bv}
  F.~Ozel,
  Nature {\bf 441} (2006) 1115.

\bibitem{Alford:2006vz}
  M.~Alford, D.~Blaschke, A.~Drago, T.~Klahn, G.~Pagliara and J.~Schaffner-Bielich,
  Nature {\bf 445} (2007) E7
  [arXiv:astro-ph/0606524].


\bibitem{Lattimer:2006xb}
  J.~M.~Lattimer and M.~Prakash,
  Phys.\ Rept.\  {\bf 442} (2007) 109
  [arXiv:astro-ph/0612440].


\bibitem{URL}
The results may be obtained from the web page of one of the authors,
http://hep.itp.tuwien.ac.at/$\sim$paulrom/


\bibitem{Vermaseren:2000nd}
  J.~A.~M.~Vermaseren,
  arXiv:math-ph/0010025.


\bibitem{Huber:2005yg}
  T.~Huber and D.~Maitre,
  Comput.\ Phys.\ Commun.\  {\bf 175}, 122 (2006)
  [arXiv:hep-ph/0507094].




\end{thebibliography}
\end{document}